\PassOptionsToPackage{table,xcdraw}{xcolor}
\documentclass[11pt,a4paper]{article}
\pdfoutput=1
\usepackage{jheppub}
\usepackage{gensymb}
\DeclareSymbolFont{matha}{OML}{txmi}{m}{it}
\DeclareMathSymbol{\varv}{\mathord}{matha}{118}
\usepackage{amsmath}
\usepackage{amssymb}
\usepackage{graphicx}
\usepackage{subfigure}
\usepackage{pstricks}
\usepackage{bm}
\usepackage{pbox}
\usepackage{placeins}
\usepackage{booktabs}
\usepackage[T1]{fontenc}
\usepackage{footnote}
\usepackage{pdfpages}
\usepackage{hhline}
\usepackage{multirow}
\usepackage{multicol}
\usepackage{enumitem}
\usepackage[toc,page]{appendix}
\allowdisplaybreaks
\usepackage{comment}
\usepackage{float}
\usepackage{slashed}
\usepackage{xcolor}
\usepackage{textcomp}
\usepackage{multirow}
\usepackage[numbers]{natbib}
\usepackage{notoccite}

\usepackage{cleveref}
\usepackage{extarrows}
\newcommand\ddfrac[2]{\frac{\displaystyle #1}{\displaystyle #2}}
\usepackage{physics}
\usepackage{comment}
\usepackage{dsfont}
\usepackage{blindtext}
\usepackage{csquotes}
\usepackage[none]{hyphenat}
\DeclareUnicodeCharacter{0304}{ }
\usepackage{rotating}
\usepackage{tabularx}

\usepackage[many]{tcolorbox}

\renewcommand{\thetable}{\Roman{table}}

\definecolor{MyDarkBlue}{rgb}{0.1, 0.1, 0.8} 
\definecolor{MyLightBlue}{rgb}{0.22,0.51,0.9}
\definecolor{MyGreen}{rgb}{0.0, 0.5, 0.0}
\definecolor{BrickRed}{rgb}{0.8, 0.25, 0.33}

\RequirePackage{hyperref}
\hypersetup{colorlinks=true, citecolor=MyGreen,linkcolor=BrickRed, urlcolor=MyLightBlue}

\title{\bf Flavor Physics Constraints on Left-Right Symmetric Models with Universal Seesaw}
\author[]{Ritu Dcruz}
\affiliation[]{Department of Physics, Oklahoma State University, Stillwater, OK 74078, USA}
\emailAdd{rdcruz@okstate.edu}
\abstract{We study the phenomenological constraints on the parameter space in the parity symmetric scenario of a class of left-right symmetric models in which the fermion masses are generated through a universal seesaw mechanism. The model, motivated by the axionless solution to the strong $\mathcal{CP}$ problem, has a simple Higgs sector consisting of left- and right-handed doublets. The fermion masses are then generated through their mixing with heavy vector-like fermions, which leads to flavor changing neutral currents arising at tree-level and also introduces non-unitarity in the charged current interactions. These new contributions lead to flavor and flavor universality violating processes and forbidden decays, which are used to derive constraints on the parameter space of the model. We also argue that although the model has the potential to resolve flavor anomalies, it fails to do so in the case of B-anomalies $R_{K^{(*)}}$ and anomalous magnetic moment of muon $a_\mu$.}
\begin{document}
\maketitle

\begin{sloppypar}

\section{Introduction}
The left-right symmetric models~(LRSMs) are simple extensions of the standard model~(SM), initially proposed as a solution to the parity asymmetry observed at low energy scales, involving the symmetry group $SU(3)_c \times SU(2)_L \times SU(2)_R \times U(1)_{B-L}$. The LRSMs, based on the concept that the $V-A$ structure of weak interaction may be a low-energy phenomenon that vanishes at very high energy scales of the order of a few TeV, overcome this by putting both left- and right-handed particles on equal footing thereby restoring the symmetry~\cite{ Pati:1974yy, Mohapatra:1974gc, Senjanovic:1975rk}. Such models have several attractive features. Right-handed neutrinos arise naturally, generating light neutrino masses through seesaw mechanism~\cite{Minkowski:1977sc,GellMann:1980vs, Mohapatra:1979ia, Yanagida:1980xy, Schechter:1980gr,Glashow:1979nm, Mohapatra:2004zh}.  These models also give a physical interpretation of the hypercharge as a quantity arising from the $B-L$ quantum number. 

The class of LRSMs discussed in this paper is motivated by the axionless solution to the strong $\mathcal{CP}$ problem~\cite{Babu:1989rb} since parity is a good symmetry broken spontaneously \cite{Beg:1978mt}. The fermion masses in these models are induced through a universal seesaw mechanism \cite{ Berezhiani:1983hm, Davidson:1987mh, Babu:1988mw} by introducing heavy vector-like fermionic~(VLF) singlet partners. The scalar sector of the model is minimal, containing only two Higgs doublets. The light fermion masses are quadratically dependent on Yukawa couplings ($\mathcal{Y}_i$), allowing for the values of $\mathcal{Y}_i$ required to explain fermion mass hierarchy to be in the range $\mathcal{Y}_i=(10^{-3}-1)$ as opposed to $\mathcal{Y}_i=(10^{-6}-1)$ in SM or standard LRSMs. Another exciting feature of the LRSMs is the flavor-changing neutral current~(FCNC) interactions leading to flavor-violating processes occurring at the tree level due to direct couplings between VLFs and SM fermions. There can also be corrections to the SM charged current interactions. These can result in stringent constraints on the parameter space of the model which are explored in great detail in this work. The model can also potentially resolve the recent flavor anomalies such as $(g-2)_\mu$, $R_{K^{(*)}}$, $R_{D^{(*)}}$, CDF-$W$ mass shift and the Cabibbo anomaly. This model has also been studied recently in the context of low energy experimental signals~\cite{Craig:2020bnv}, gravitational waves~\cite{Graf:2021xku}, neutrino oscillations~\cite{Babu:2020bgz} and, in explaining CKM unitarity problem and CDF $W$-boson mass~\cite{Babu:2022jbn}.

Several measurements of semi-leptonic decays of B mesons have shown significant deviations from their SM predicted values~\cite{Bobeth:2007dw, Bordone:2016gaq}. The most notable ones are the lepton flavor universality~(LFU) violation observed in the neutral current transitions related to the ratio $R_{K^{(*)}}=\text{BR}(B\to K^{(*)}\mu\mu)/\text{BR}(B\to K^{(*)}ee)$ and in charged current mediated $R_{D^{(*)}}=\text{BR}(B\to D^{(*)}\tau\nu_\tau)/\text{BR}(B\to D^{(*)}\ell\nu_\ell)$, $\ell=e, ~\mu$. These are theoretically clean observables since the hadronic uncertainties cancel out in the ratios, making them extremely sensitive to new physics probes. Deviations in the measurements of $R_{D^{(*)}}$ and $R_{K^{(*)}}$ \cite{Aaij:2017vbb, Abdesselam:2019wac, Aaij:2019wad,LHCb:2021trn} from the SM are collectively referred to as B-anomalies. A solution to $R_{D^{(*)}}$ anomaly using this model has been discussed in~\cite{Babu:2018vrl}. The most precise measurement of $R_K$ has resulted in a combined $3.1~\sigma$ discrepancy in $R_{K^{(*)}}$ and related processes. As a natural consequence of parity invariance, a right-handed neutral gauge boson exists that mixes with the SM $Z$ boson giving rise to two neutral gauge boson eigenstates, $Z_1$ and $Z_2$. Both these gauge bosons can mediate tree-level FCNC processes that contribute to $R_{K^{(*)}}$ to occur at the tree level. Hence, the LRSM model can potentially resolve the neutral current B-anomalies as well.
 
Another major hint towards physics beyond the SM comes from the long-standing discrepancy between the experiment and theory in the anomalous magnetic moment~(AMM) of muon, $a_\mu$. The most recent measurement of $a_\mu$ at Fermilab National Accelerator Laboratory~(FNAL) reports a $4.2\,\sigma$ deviation from the SM predicted value of AMM. The model explored in this work can provide chirally enhanced one-loop corrections to AMM of muon mediated by neutral scalar~($h,\,H$) or gauge~($Z_1,\,Z_2$) fields. The possibility of resolving the anomaly in muon AMM is explored briefly in this paper.

In this work, we look at the flavor physics constraints on the model, exploring the allowed parameter space from neutral current gauge interactions, investigate the most stringent constraints for charged current and Higgs interactions, and address the limitations of the model in resolving $R_{K^{(*)}}$ discrepancies as well as the AMM of muon. The rest of the paper is organized as follows. We describe the details of the LRSM model \cite{Babu:2018vrl}, giving the particle content, Higgs potential and gauge boson mass matrix diagonalization in Sec.~\ref{sec:model} and the fermion mass matrix diagonalizations and their interaction Lagrangian in the physical basis in Sec.~\ref{sec:int}. In Secs.~\ref{sec:constraint}, \ref{sec:cc_coupling}, and \ref{sec:constrainthiggs}, we list the different constraints on the model parameters obtained from neutral and charged current gauge interactions, and scalar interactions, respectively, in the parity symmetric scenario of the model. In Sec.~\ref{sec:Banom}, we explore the solution to neutral current B-anomalies arising at tree-level mediated by the neutral gauge bosons, and at one-loop level mediated by the left-handed scalar doublet. Sec.~\ref{sec:AMM} examines the VL-lepton mass-enhanced corrections to the AMM of muon at one-loop level mediated by the neutral gauge as well as scalar fields, and we conclude with Sec.~\ref{conclude}.

\section{Model Description\label{sec:model}}

The particle spectrum of the LRSM with universal seesaw is composed of the usual SM fermions, right-handed neutrinos, and four sets of vector-like singlet fermions, denoted as $(U_a, D_a, E_a, N_a)$, where, the index $a$ runs from 1 to 3 for the three generations of fermions. The SM fermions, along with the right-handed neutrinos, form left- or right-handed doublets, assigned to the gauge group $SU(3)_c \times SU(2)_L \times SU(2)_R \times U(1)_{B-L}$ as follows ($i=1 - 3$ is the family index):
    \begin{equation}
        \begin{aligned}
            \mathcal{Q}_{L,i}\left(3,2,1,+\frac{1}{3}\right)=\begin{pmatrix}
            u_L\\d_L
            \end{pmatrix}_i, && \mathcal{Q}_{R,i}\left(3,1,2,+\frac{1}{3}\right)=\begin{pmatrix}
            u_R\\d_R
            \end{pmatrix}_i,\\
            \psi_{L,i}\left(1,2,1,-1\right)=\begin{pmatrix}
            \nu_L\\e_L
            \end{pmatrix}_i, && \psi_{R,i}\left(1,1,2,-1\right)=\begin{pmatrix}
            \nu_R\\e_R
            \end{pmatrix}_i.\\
        \end{aligned}\label{eq:sm particles}
    \end{equation}

The Higgs sector is comprised simply of left- and right-handed doublets:
    \begin{equation}
        \begin{aligned}
            \chi_L\left(1,2,1,+1\right)=\begin{pmatrix}
            \chi_L^+\\ \chi_L^0
            \end{pmatrix},&& \chi_R\left(1,1,2,+1\right)=\begin{pmatrix}
            \chi_R^+\\ \chi_R^0
        \end{pmatrix},
        \end{aligned}\label{eq:higgs}
    \end{equation}
with the general Higgs potential is given by
    \begin{equation}
        V=-(\mu_L^2\chi_L^{\dagger}\chi_L+\mu_R^2\chi_R^{\dagger}\chi_R)+\frac{\lambda_{1L}}{2}(\chi_L^{\dagger}\chi_L)^2+\frac{\lambda_{1R}}{2}(\chi_R^{\dagger}\chi_R)^2+\lambda_2(\chi_L^{\dagger}\chi_L)(\chi_R^{\dagger}\chi_R).\label{eq:higgs potential}
    \end{equation}
In the parity symmetric limit, $\lambda_{1L}=\lambda_{1R}$, but $\mu_L$ can be allowed to be different from $\mu_R$ since parity can be softly broken. The vacuum expectation values of the neutral fields for which the potential Eq.~\eqref{eq:higgs potential} has a minimum are denoted by
    \begin{equation}
        \begin{aligned}
            \left\langle\chi_L^0\right\rangle=\kappa_L, && \left\langle\chi_R^0\right\rangle=\kappa_R,
        \end{aligned}
    \end{equation}
with $\kappa_L\simeq 174$ GeV. The neutral scalar fields are $\sigma_L=Re(\chi_L^0)/\sqrt{2}$ and  $\sigma_R=Re(\chi_R^0)/\sqrt{2}$ are which mix to give the SM-like Higgs boson $h$ (125 GeV) and a heavy $H$. In the limit of small mixing, the SM Higgs is identified as $\sigma_L$. The remaining fields are \enquote{eaten up} by $W_{L,R}^\pm$ and $Z_{L,R}$ bosons upon symmetry breaking.
The physical Higgs spectrum obtained from the diagonalization of  $\sigma_L-\sigma_R$ mixing matrix,
    \begin{equation}
        \mathcal{M}^2_{\sigma_{L,R}}=\begin{bmatrix}
        2\lambda_{1L}\kappa_L^2 && 2\lambda_2\kappa_L\kappa_R\\
        2\lambda_2\kappa_L\kappa_R && 2\lambda_{1R}\kappa_R^2,
        \end{bmatrix},
    \end{equation}
are as follows:
    \begin{equation}
        \begin{aligned}
            h=\cos\zeta\sigma_L-\sin\zeta\sigma_R, && H=\sin\zeta\sigma_L+\cos\zeta\sigma_R,\\
            M_h^2\simeq 2\lambda_{1L}\left(1-\frac{\lambda_2^2}{\lambda_{1L}\lambda_{1R}}\right)\kappa_L^2, && M_H^2=2\lambda_{1R}\kappa_R^2,
        \end{aligned}\label{eq:higssmix}
    \end{equation}
with the mixing angle $\zeta$ given by 
    \begin{equation}
        \tan 2\zeta=\frac{2\lambda_2\kappa_L\kappa_R}{(\lambda_{1R}\kappa_R^2-\lambda_{1L}\kappa_L^2)}.
    \end{equation}
To generate the fermion masses in the absence of the Higgs bidoublet $(1,2,2,0)$ seen in the standard LRSMs, vector-like fermions (VLFs) are introduced. The gauge quantum numbers of the VLFs are 
    \begin{equation}
        \begin{aligned}
            U_a \left(3,1,1,+\frac{4}{3}\right),&& D_a \left(3,1,1,-\frac{2}{3}\right),&& E_a \left(1,1,1,-2\right),&& N_a \left(1,1,1,0\right),
        \end{aligned}\label{eq:VL particles}
    \end{equation}
The electric charge is given by 
    \begin{equation}
    \begin{aligned}
        Q=&T_{3L}+T_{3R}+\frac{B-L}{2}\text{ with,}\\
        \frac{Y}{2}=&T_{3R}+\frac{B-L}{2},   
    \end{aligned}
   \end{equation}
 thereby giving the hypercharge a definition in terms of the $SU(2)_R$ and $U(1)_{B-L}$ quantum numbers. Defining the covariant derivative as
    \begin{equation}
        D_{\mu}=\partial_{\mu}+i\frac{g_{L,R}}{2}(\vec{\tau}.\overrightarrow{W}_{L,R\mu})+ig_B\frac{B-L}{2}B_{\mu},
    \end{equation}
the interaction of gauge bosons with Higgs field can be derived from the kinetic term of the Lagrangian
    \begin{equation}
        \mathcal{L}^{\text{KE}}_{\text{Higgs}}=(D_{\mu}\chi_L)^{\dagger}(D_{\mu}\chi_L)+(D_{\mu}\chi_R)^{\dagger}(D_{\mu}\chi_R).
    \end{equation}
It can be easily verified that the charged gauge bosons do not mix at tree level, and their masses are 
    \begin{equation}
        \begin{aligned}
            M^2_{W^\pm_{L(R)}}=\frac{1}{2}g_{L(R)}^2\kappa_{L(R)}^2.
        \end{aligned}
    \end{equation}
Of the neutral gauge bosons, the photon field $A_\mu$ remains massless while the two orthogonal fields $Z_L$ and  $Z_R$ mix. In the limit of small $\kappa_L$, these fields are related to the gauge eigenstates by:
    \begin{equation}
        \begin{aligned}
            A^\mu &= \frac{g_Lg_RB^\mu+g_Bg_RW^\mu _{3L}+g_Bg_LW^\mu _{3R}}{\sqrt{g_B^2(g_L^2+g_R^2)+g_L^2g_R^2}},\\
            Z^\mu _R &= \frac{g_BB^\mu-g_RW^\mu_{3R}}{\sqrt{g_R^2+g_B^2}},\\
            Z^\mu _L &= \frac{g_B g_RB^\mu-g_Lg_R\left(1+\frac{g_B^2}{g_R^2}\right)W^\mu_{3L}+g_B^2W^\mu _{3R}}{\sqrt{g_B^2+g_R^2}\sqrt{g_B^2+g_L^2+\frac{g_B^2g_L^2}{g_R^2}}}.
        \end{aligned}
    \end{equation}
 The hypercharge relation implies $g_Y^{-2}=g_R^{-2}+g_B^{-2}$, which can be used to eliminate $g_B$ in terms of $g_Y$ resulting in the $Z_L-Z_R$ mixing matrix  
    \begin{equation}
        \mathcal{M}^2_{Z_L-Z_R}=\frac{1}{2}\begin{pmatrix}
        (g_L^2+g_Y^2)\kappa_L^2 && g_Y^2\sqrt{\frac{g_L^2+g_Y^2}{g_R^2-g_Y^2}}\kappa_L^2\\
        g_Y^2\sqrt{\frac{g_L^2+g_Y^2}{g_R^2-g_Y^2}}\kappa_L^2 && \frac{g_R^4\kappa_R^2+g_Y^4\kappa_L^2}{g_R^2-g_Y^2}
        \end{pmatrix},
    \end{equation}
from which the physical states and masses can be obtained as:
    \begin{equation}
        \begin{aligned}
            Z_1=\cos\xi Z_L-\sin\xi Z_R, & \quad \quad Z_2=\sin\xi Z_L+\cos\xi Z_R,\\
            M^2_{Z_1}\simeq \frac{1}{2}(g_Y^2+g_L^2)\kappa_L^2, & \quad \quad M^2_{Z_2}\simeq \frac{g_R^4}{2(g_R^2-g_Y^2)}\kappa_R^2+\frac{g_Y^4}{2(g_R^2-g_Y^2)}\kappa_L^2,
        \end{aligned}\label{eq:Z1Z2}
    \end{equation}
and the mixing angle is given by $\xi\simeq\frac{g_Y^2}{g_R^4}\frac{\kappa_L^2}{\kappa_R^2}\sqrt{(g_L^2+g_Y^2)(g_R^2-g_Y^2)}$ or more accurately,
    \begin{equation}
    		\tan (-2\xi)=\frac{2g_Y^2\sqrt{(g_L^2+g_Y^2)(g_R^2-g_Y^2)}\kappa_L^2}{g_R^4\kappa_R^2+g_Y^4\kappa_L^2-(g_L^2+g_R^2)(g_R^2-g_Y^2)\kappa_L^2}.
    \end{equation}
The SM $Z$ boson of mass $91.18$ GeV is identified to be $Z_1$ or, in the limit of small mixing angle, $Z_L$. $Z_1$ and $Z$ will be used interchangeably henceforth.

\section{Interactions among Physical Fields \label{sec:int}}

In this section, the diagonalization of fermion mass matrices is studied and the new contributions to fermionic interactions with the bosonic fields are explored.

\subsection{Charged Fermions\label{sec:chargedfermionmass}}

The fermion masses in the model arise from the Yukawa coupling between quarks, leptons, and the Higgs bosons involved in symmetry breaking. In the flavor basis, the Yukawa interactions of the charged fermions and the bare masses of VLFs are given by the Lagrangian
    \begin{equation}
        \begin{aligned}
            \mathcal{L}_{\text{Yuk}}&=
            \mathcal{Y}_U\overline{\mathcal{Q}}_L\widetilde{\chi}_LU_R+\mathcal{Y}'_U\overline{\mathcal{Q}}_R\widetilde{\chi}_RU_L+M_U\overline{U}_LU_R\\&+\mathcal{Y}_D\overline{\mathcal{Q}}_L\chi_LD_R+\mathcal{Y}'_D\overline{\mathcal{Q}}_R\chi_RD_L+M_D\overline{D}_LD_R\\&+\mathcal{Y}_E\overline{\psi}_L\chi_LE_R+\mathcal{Y}'_E\overline{\psi}_R\chi_RE_L+M_E\overline{E}_LE_R+\mathrm{h.c.}
        \end{aligned}
    \end{equation}
with $\widetilde{\chi}_{L,R}=i\tau_2\chi^*_{L,R}$. Under parity symmetry, $\mathcal{Y}=\mathcal{Y}'$ and $M=M^\dagger$. The above Lagrangian gives $6\times 6$ mass matrices for up-type quarks $(u,U)$, down-type quarks $(d,D)$ and charged leptons $(e,E)$ which are written in the form:
    \begin{equation}
        \mathcal{M}_{U,D,E}=\begin{pmatrix}
        0&\mathcal{Y}_{U,D,E}\kappa_L\\
        \mathcal{Y}'^\dagger_{U,D,E}\kappa_R & M_{U,D,E}
        \end{pmatrix}.
    \end{equation}
Matrices of the form $\mathcal{M}$ can be block-diagonalized by a bi-unitary transform with two unitary matrices parametrized by $\rho_L$ and $\rho_R$. 
\begin{equation}
    \mathcal{U}_{X}=\begin{pmatrix}
        \mathds{1}-\frac{1}{2}\rho^\dagger_X\rho_X&&\rho^\dagger_X\\
        -\rho_X&&\mathds{1}-\frac{1}{2}\rho_X\rho^\dagger_X
        \end{pmatrix}, \,\, X=\{L,R\}.
\end{equation}
$\rho_L$ is necessarily small while we assume $\rho_R\ll1$ as well, to simplify the analysis. 
    \begin{equation}
       \mathcal{M}_{\text{diag}}  =\begin{pmatrix}
        \mathds{1}-\frac{1}{2}\rho^\dagger_L\rho_L&&-\rho^\dagger_L\\
        \rho_L&&\mathds{1}-\frac{1}{2}\rho_L\rho^\dagger_L
        \end{pmatrix}\begin{pmatrix}
        0& \mathcal{Y}\kappa_L\\
        \mathcal{Y}'^\dagger\kappa_R&M
        \end{pmatrix}\begin{pmatrix}
        \mathds{1}-\frac{1}{2}\rho^\dagger_R\rho_R&&\rho^\dagger_R\\
        -\rho_R&&\mathds{1}-\frac{1}{2}\rho_R\rho^\dagger_R
        \end{pmatrix},
    \end{equation}
where the matrices $\mathcal{U}_{L,R}$ are unitary up to $\mathcal{O}(\rho^2)$, and the parameters are related to the masses and Yukawa interactions by
    \begin{equation}
        \begin{aligned}
            \rho_L=\kappa_LM^{-1\dagger}\mathcal{Y}^\dagger,\\  
            \rho_R=\kappa_R M^{-1}\mathcal{Y}'^\dagger, 
        \end{aligned}
    \end{equation}
while the mass eigenvalues are
    \begin{equation}
        \begin{aligned}
            \hat{m}=-\kappa_L\kappa_R\mathcal{Y}M^{-1}\mathcal{Y}'^\dagger,&&\textrm{and}&&\hat{M}=M+\frac{1}{2}(\kappa_R^2\mathcal{Y}'^\dagger \mathcal{Y}'M^{-1\dagger}+\kappa_L^2M^{-1\dagger}\mathcal{Y}^\dagger \mathcal{Y}).
        \end{aligned}
    \end{equation}
$M$ is assumed to be diagonal while $\hat{m}$ needs to be diagonalized by a subsequent bi-unitary transform such that $m_f=V_{L_f}\hat{m}V^\dagger_{R_f}$. Now, it is possible to write the interactions of charged fermions with gauge and scalar bosons in the mass basis. In the following sections, $f (\ell)$ stands for mass eigenstate of charged fermions (leptons), and $F$, $U$, $D$, and $E$ represent charged VLF.

\subsubsection{Neutral Current\label{sec:NC}}

The tree-level interactions of charged fermions in their mass basis with the neutral gauge bosons are:
\begin{equation}
\begin{aligned}
-g_{Z_L}^{-1} \mathcal{L}_{Z_L}&=\overline{f}_L\gamma^\mu Z_{L_\mu}\left(A_{L}-(A_{L}-B_{L})V_{L_f}\rho_{L_F}^\dagger \rho_{L_F} V^\dagger_{L_f}\right)f_L\\
&+\overline{f}_R\gamma^\mu Z_{L_\mu}\left(A_{L}'\right)f_R\\
&+\overline{f}_L\gamma^\mu Z_{L_\mu}\left((A_{L}-B_{L})V_{L_f}\rho_{L_F}^\dagger\right)F_L\\
&+\overline{F}_L\gamma^\mu Z_{L_\mu}\left((A_{L}-B_{L})\rho_{_LF} V^\dagger_{L_f}\right)f_L\\
& +\overline{F}_{L,R}\gamma^\mu Z_{L_\mu}\left(B_{L}\right)F_{L,R},
\end{aligned}\label{eq:LZL}
\end{equation}

\begin{equation}
\begin{aligned}
-g_{Z_R}^{-1} \mathcal{L}_{Z_R}&=\overline{f}_L\gamma^\mu Z_{R_\mu}\left(A_{R}-(A_{R}-B_{R})V_{L_f}\rho_{L_F}^\dagger \rho_{L_F} V^\dagger_{L_f}\right)f_L\\
&+\overline{f}_L\gamma^\mu Z_{R_\mu}\left((A_{R}-B_{R})V_{L_f}\rho_{L_F}^\dagger\right)F_L\\
& +\overline{F}_L\gamma^\mu Z_{R_\mu}\left((A_{R}-B_{R})\rho_{L_F} V^\dagger_{L_f}\right)f_L\\
&+\{L\to R\text{ with }A_R\to A_R' \}\\
&+\overline{F}_{L,R}\gamma^\mu Z_{R_\mu}\left(B_{R}\right)F_{L,R},
\end{aligned}\label{eq:LZR}
\end{equation}
where,
\begin{equation}
    \begin{aligned}
g_{Z_{L,R}}=\frac{g_{L,R}^2}{\sqrt{g_{L,R}^2\pm g_Y^2}}~~~,&&~~~~~B_{L,R}=-\frac{g_Y^2}{g_{L,R}^2}\frac{Y_{F}}{2},\\ 
A_{L,R}=T_{3L,3R}-\frac{g_Y^2}{g_{L,R}^2}\frac{Y_{f_{L,R}}}{2},&& ~~~~~~~~~~
A'_{L,R}=-\frac{g_Y^2}{g_{L,R}^2}\frac{Y_{f_{R,L}}}{2}.
\end{aligned}\label{eq:ABex}
\end{equation}
$Y_{f_{L,R}}$ are the hypercharges of SM fermions and $Y_F$ is the hypercharge of the VLF. Under parity symmetry, $g_R=g_L$, $V_{R_f}=V_{L_f}$ and $\rho_R=\frac{\kappa_R}{\kappa_L}\rho_L$. The SM interaction of charged fermions to $Z$ boson can be obtained from $\mathcal{L}_{Z_L}$ as the VLF decouples from SM fermions. The above expressions can be converted into the mass basis of the neutral gauge bosons using the relations in Eq.~\eqref{eq:Z1Z2}, as shown in Eqs.~\eqref{eq:LZ1} and \eqref{eq:LZ2} with the couplings given in Appendix.~\ref{app:Lag}.

\subsubsection{Charged Current}

The charged current interactions of quarks are given by
    \begin{equation}
        \begin{aligned}
            -\frac{\sqrt{2}}{g_L}\mathcal{L}_{W_X}&=\overline{u}_X\gamma^\mu W^+_{X\mu}\Bigl(V_{X_u}V_{X_d}^\dagger-\frac{1}{2}(V_{X_u}\rho_{X_D}^\dagger \rho_{X_D} V_{X_d}^\dagger \\ 
            &+V_{X_u}\rho_{X_U}^\dagger \rho_{X_U} V_{X_d}^\dagger)\Bigr)d_X
            +\overline{u}_X\gamma\mu W_{X\mu}^+ V_{X_u}\rho_{X_D}^\dagger D_X\\ 
            &+\overline{U}_X\gamma\mu W_{X\mu}^+ \rho_{X_U} V_{X_d}^\dagger d_X+\overline{U}_X\gamma^\mu W^+_{X\mu} \rho_{X_U}\rho_{X_D}^\dagger D_X+ \text{h.c.} ,
        \end{aligned}\label{eq:W_X current}
    \end{equation}
with $X=\{L,R\}$ for $\{W_L,\,W_R\}$. Since the charged current interactions of leptons involve neutrinos, these are discussed following the diagonalization of the neutrino mass matrix in Sec.~\ref{sec:neutrino_diag}.

\subsubsection{Higgs Current}

The interactions of the charged fermions with the scalar fields in the flavor basis are given by
    \begin{equation}
    	\begin{aligned}    	\mathcal{L}_{\sigma_L}&=\overline{f}_L \frac{\sigma_L}{\sqrt{2}} \left(-V_{L_f}\mathcal{Y}_F\rho_{R_F}V_{R_f}^\dagger+\frac{1}{2}V_{L_f}\rho_{L_F}^\dagger\rho_{L_F}\mathcal{Y}_F\rho_{R_F} V_{Rf}^\dagger\right)f_R\\
    		&+\overline{f}_L \frac{\sigma_L}{\sqrt{2}}\Bigl(V_{L_f}\mathcal{Y}_F-\frac{1}{2}(V_{L_f}\mathcal{Y}_F\rho_{R_F}\rho_{R_F}^\dagger+V_{L_f}\rho_{L_F}^\dagger\rho_{L_F}\mathcal{Y}_F)\Bigr)F_R\\
    		&+\overline{F}_L \frac{\sigma_L}{\sqrt{2}}\left(-\rho_{L_F}\mathcal{Y}_F\rho_{R_F}V_{R_f}^\dagger\right)f_R+\overline{F}_L\frac{\sigma_L}{\sqrt{2}}\left(
    		\rho_{L_F}\mathcal{Y}_F- \frac{1}{2}\rho_{L_F}\mathcal{Y}_F\rho_{R_F}\rho_{R_F}^\dagger\right) F_R	+ \mathrm{h.c.}.\label{eq:Lhiggsferm}
    	\end{aligned}
    \end{equation}
The $\sigma_R-f$ interaction can be obtained with the transformation $ 	\mathcal{L}_{\sigma_R}=\mathcal{L}_{\sigma_L}(L\leftrightarrow R, \mathcal{Y}\to \mathcal{Y}').$ Since $\sigma_L$ and $\sigma_R$ mix, the interaction in the mass basis can be obtained by
    \begin{equation}
    	\begin{aligned}
    		\mathcal{L}_{h}&=\cos\zeta\mathcal{L}_{\sigma_L}-\sin\zeta\mathcal{L}_{\sigma_R},\\					\mathcal{L}_H&=\sin\zeta\mathcal{L}_{\sigma_L}+\cos\zeta\mathcal{L}_{\sigma_R}.
    	\end{aligned}
    \end{equation}
The Eq.~\eqref{eq:Lhiggsferm} reduces to SM interaction in the limit of $\zeta$ and the NP contributions tending to zero. This may be easily deduced from the interaction Lagrangian
    \begin{equation}
    	\begin{aligned}
    		\mathcal{L}_{h}&\supset \overline{f}_L \cos\zeta\frac{h}{\sqrt{2}} \left(\frac{m_f}{\kappa_L}-\frac{1}{2}V_{Lf}\rho_{LF}^\dagger\rho_{LF}V_{Lf}^\dagger \frac{m_f}{\kappa_L}\right)f_R\\
    		&-\overline{f}_R \sin\zeta\frac{h}{\sqrt{2}} \left(\frac{m_f^\dagger}{\kappa_R}-\frac{1}{2}V_{Rf}\rho_{RF}^\dagger\rho_{RF}V_{Rf}^\dagger \frac{m_f^\dagger}{\kappa_R}\right)f_L+\text{h.c.} .\\
    	\end{aligned}\label{eq:higgs1}
    \end{equation}
For completeness, the corresponding interaction of heavy Higgs is 
    \begin{equation}
    	\begin{aligned}
    		\mathcal{L}_{H}&\supset \overline{f}_L \sin\zeta\frac{H}{\sqrt{2}} \left(\frac{m_f}{\kappa_L}-\frac{1}{2}V_{Lf}\rho_{LF}^\dagger\rho_{LF}V_{Lf}^\dagger \frac{m_f}{\kappa_L}\right)f_R\\
    		&+\overline{f}_R \cos\zeta\frac{H}{\sqrt{2}} \left(\frac{m_f^\dagger}{\kappa_R}-\frac{1}{2}V_{Rf}\rho_{RF}^\dagger\rho_{RF}V_{Rf}^\dagger \frac{m_f^\dagger}{\kappa_R}\right)f_L+\text{h.c.} .\\
    	\end{aligned}\label{eq:higgs2}
    \end{equation}

\subsection{Neutrinos\label{sec:neutrino_diag}}

The light neutrinos have Yukawa couplings with the heavy VL-neutrinos and the Higgs bosons. The VL-neutrinos being electrically neutral, can have both Dirac ($M_N$) and Majorana ($\mu'_L$ and $\mu'_R$) mass terms. The Yukawa interactions of neutrinos are given by 
    \begin{equation}
    	\begin{aligned}
    		\mathcal{L}^\nu_{\text{Yuk}}=&Y_\nu \overline{\psi}_L\widetilde{\chi}_LN_R+\widetilde{Y}_\nu \overline{\psi}_L\widetilde{\chi}_L(N_L)_R^c+Y'_\nu \overline{\psi}_R\widetilde{\chi}_RN_L+\widetilde{Y}'_\nu \overline{\psi}_R\widetilde{\chi}_R(N_R)_L^c \\
    		&+ M_N \overline{N_L}N_R+\mu'_L N_L^TCN_L+\mu'_R N_R^TCN_R+ \text{h.c.} .
    	\end{aligned}
    \end{equation}
 We assume both Dirac ($M_N$) and Majorana ($\mu_L$ and $\mu_R$) mass terms for vector-like neutrinos. Under parity symmetry, $N_L \leftrightarrow N_R$, $\psi_L \leftrightarrow \psi_R$, and $\chi_L \leftrightarrow \chi_R$, which makes $Y_\nu=Y'_\nu$, $\widetilde{Y}_\nu=\widetilde{Y}'_\nu$, $\mu_L'=\mu_R'$ and $M_N=M_N^\dagger$. If $\mu_L' \simeq \mu_R' \gg M_N \gg \kappa_R \gg \kappa_L$, the model will give rise to light sterile neutrinos potentially of sub-MeV or eV range and the Majorana mass terms can be of the order of $10^{10}$~GeV.  We shall assume that $\mu_L' \simeq \mu_R' \simeq \kappa_R$, in which case the sterile neutrinos will have masses of the order of $\kappa_R$.   The heavy singlet neutrino mass matrix can be block diagonalized to a physical mass basis $(N_1, N_2) $. Then, the neutrino mass matrix sandwiched between $(\nu^T,\nu^{cT},N_1^T,N_2^{T}) C$ and $(\nu,\nu^{c},N_1,N_2)^T$ is
    \begin{equation}
     \mathcal{M}_N=   \begin{pmatrix}
        0&&0&&\widetilde{Y}^*\kappa_L&&Y^*\kappa_L\\
        0&&0&&Y'\kappa_R&&\widetilde{Y}'\kappa_R\\
        \widetilde{Y}^\dagger\kappa_L&&Y'^T\kappa_R&&\mu_L&&0\\
        Y^\dagger\kappa_L&&\widetilde{Y}'^T\kappa_R&&0&&\mu^*_R\\
        \end{pmatrix},
    \end{equation}
where all the fields are taken to be left-handed, and $\mu_R^\dagger=\mu_R^*$. The mass matrix can be reduced to a $2\times2$ form assuming the mass hierarchy mentioned above, and it being a complex symmetric matrix, can be block diagonalized with a unitary transformation:
    \begin{equation}
        \begin{pmatrix}
            \mathds{1}-\frac{1}{2}\rho^\dagger\rho&&\rho^\dagger\\
            -\rho&&\mathds{1}-\frac{1}{2}\rho\rho^\dagger
            \end{pmatrix}\begin{pmatrix}
            0&&\Upsilon\\
            \Upsilon^T&&M
            \end{pmatrix}\begin{pmatrix}
            \mathds{1}-\frac{1}{2}\rho^T\rho^*&&-\rho^T\\
            \rho^*&&\mathds{1}-\frac{1}{2}\rho^*\rho^T
        \end{pmatrix}.
    \end{equation}

Here $M$ is block diagonal and $\Upsilon$ is a $6\times6$ matrix containing both $\kappa_L$ and $\kappa_R$ couplings. Note that the parameter $\rho$ is different from the ones in Sec.~\ref{sec:chargedfermionmass}.
    \begin{equation}
        \rho=-(\Upsilon M^{-1})^\dagger,
    \end{equation}
    and
    \begin{equation}
        \begin{aligned}
            \hat{\mathbf{m}}=-2\Upsilon M^{-1}\Upsilon^T, &&\hat{\mathbf{M}}=M+\frac{1}{2}\left(\Upsilon^T\Upsilon^*M^{-1\dagger}+M^{-1\dagger}\Upsilon^\dagger\Upsilon\right),
        \end{aligned}
    \end{equation}
where, $\hat{\mathbf{m}}$ and $\hat{\mathbf{M}}$ are $6\times 6$ matrices. $\hat{\mathbf{m}}$, which represents the mixing between the doublet neutrinos, may be written as
    \begin{equation}
        \begin{pmatrix}
        M_L&&M_D\\
        M_D^T&&M_R
        \end{pmatrix},
    \end{equation}
with
    \begin{equation}
    	\begin{aligned}
        	M_L&=-2\kappa_L^2\widetilde{Y}^*\mu_L^{-1}\widetilde{Y}^\dagger-2\kappa_L^2Y^*\mu_R^{*-1}Y^\dagger,
        	\\ 
        	M_D&=-2\kappa_L\kappa_R\widetilde{Y}^*\mu_L^{-1}Y^{'T}-2\kappa_L\kappa_RY^*\mu_R^{*- 1}\widetilde{Y}^{'T},
        	\\ 
        	M_D^T&=-2\kappa_L\kappa_RY'\mu_L^{-1}\widetilde{Y}^\dagger-2\kappa_L\kappa_R\widetilde{Y}'\mu_R^{*- 1}Y^\dagger,
        	\\ 
        	M_R&=-2\kappa_R^2Y'\mu_L^{-1}Y^{'T}-2\kappa_R^2\widetilde{Y}'\mu_R^{*-1}\widetilde{Y}^{'T},
    	\end{aligned}
    \end{equation}
and can be diagonalized again by the transformation:
    \begin{equation}
        \begin{pmatrix}
        \mathds{1}-\frac{1}{2}\delta^\dagger\delta&&\delta^\dagger\\
        -\delta&&\mathds{1}-\frac{1}{2}\delta\delta^\dagger
        \end{pmatrix}\begin{pmatrix}
        M_L&&M_D\\
        M_D^T&&M_R
        \end{pmatrix}\begin{pmatrix}
        \mathds{1}-\frac{1}{2}\delta^T\delta^*&&-\delta^T\\
        \delta^*&&\mathds{1}-\frac{1}{2}\delta^*\delta^T
        \end{pmatrix}.
    \end{equation}
Here, the mixing parameter is given by
    \begin{equation}
    \delta=-(M_D M_R^{-1})^\dagger,
    \end{equation}
and the masses of the light neutrinos are
    \begin{equation}
     \begin{aligned}
            m_{\nu_L}&=M_L-M_DM_R^{-1}M_D^T,\\ m_{(\nu^c)_L}&=M_R+\frac{1}{2}\left(M_R^{-1\dagger}M_D^\dagger M_D+M_D^TM_D^*M_R^{-1\dagger}\right).
       \end{aligned}
    \end{equation}
To obtain the mass basis $(\nu_1,\nu_2,\nu_3)$ of neutrinos, we need a subsequent unitary transformation equivalent to that of the PMNS matrix.  It may be worth noting that if the lepton number violating interactions of neutrinos were ignored, the neutrino mass matrix would reduce to the same form as the charged fermions and can be diagonalized using the same bi-unitary transformation. Up to the second order, the neutral current interactions of the light doublet neutrinos, under parity symmetry, are
    \begin{equation}
    	\begin{aligned}
       	g_{Z_L}^{-1}\mathcal{L}_{Z_L}&\supset \overline{\nu}_L\gamma^\mu Z_L \Bigl( A_L(1-\rho^T\rho^*-\delta^T\delta^\dagger)\Bigr)\nu_L+(\overline{\nu}^c)_L\gamma^\mu Z_L \Bigl(  A_L\delta^*\delta^T\Bigr)(\nu^c)_L,\\ 
        g_{Z_R}^{-1}\mathcal{L}_{Z_R}&\supset \overline{\nu}_L\gamma^\mu Z_R \Bigl(A'_R\left(1-\delta^T\delta^*-\rho^T\rho^*\right) +A_R\delta^T\delta^*\Bigr)\nu_L\\ 
        		&+(\overline{\nu}^c)_L\gamma^\mu Z_R \Bigl( A'_R\delta^*\delta^T +A_R\left(1-\delta^*\delta^T-\rho^*\rho^T\right)\Bigr)(\nu^c)_L,
    	\end{aligned}
    \end{equation}
following the same notations used in Eq.~\eqref{eq:ABex}. The interaction of leptons with $W_L$ is
    \begin{equation}
        \begin{aligned}
            -\frac{\sqrt{2}}{g_L}\mathcal{L}_{W_L}&\supset \overline{\nu_{L_l}}\gamma^\mu W^+_{L\mu} \left(1-V_{L_l}\Bigl(\frac{1}{2}\rho_{L_L}^\dagger\rho_{L_L}-\frac{1}{2}\delta^*\delta^T-\frac{1}{2}\rho^T\rho^*\Bigr)V_{L_l}^\dagger\right) \ell_L\\
            &+(\overline{\nu}^c)_{L}\gamma^\mu W^+_{L\mu}\left(-\delta^*+\frac{1}{2}\delta^*\rho_{L_L}^\dagger\rho_{L_L}-\frac{1}{2}\rho^T\rho^*\right)V_L^\dagger \ell_L+ \text{h.c.}.
        \end{aligned}\label{eq:nulcharged}
    \end{equation}
Here, we assume that $V_{L_l}$ rotation associates $\nu_{L_l}$ with the corresponding charged lepton. Similarly, the $W_R$-lepton interaction Lagrangian is
    \begin{equation}
    	\begin{aligned}
    	\frac{\sqrt{2}}{g_R}\mathcal{L}_{W_R}&\supset \overline{\nu_{L_l}}\gamma^\mu W^+_{R\mu} V_{L_l}\left(\delta^T-\frac{1}{2}\delta^T\rho_{R_L}^T\rho_{R_L}-\frac{1}{2}\rho^T\rho^*\right)V_{R_l}^* (l_R)^c\\
    		&+(\overline{\nu}^c)_{L}\gamma^\mu W^+_{L\mu}\left(1-\frac{1}{2}\rho_{R_L}^T\rho_{R_L}^*-\frac{1}{2}\delta^*\delta^T-\frac{1}{2}\rho^T\rho^*\right)V_R^* (l_R)^c+ \text{h.c.}.
    	\end{aligned}
    \end{equation}

\section{Constraints on Neutral Current Couplings \label{sec:constraint}}

With the features of the interaction Lagrangian detailed above and the plethora of experimental signals available, we can study the bounds on various NP parameters of the model. This section explore different constraints on couplings to $Z_1$ and $Z_2$. In the charged lepton sector, we consider the flavor-conserving and -violating two-body decays of $Z_1$ and three-body decays of charged leptons. The most stringent limit comes from the three-body decay of muon. In the quark sector, we study the different meson decay processes and mass differences from neutral meson mixing. Since both $Z_1$ and $Z_2$ contributes to FCNCs, which are absent in the SM, we focus on the tree-level neutral current interactions of SM fermions. The $Z_1$ interaction to SM fermions can be rewritten, under parity symmetry, as
    \begin{equation}
       -\mathcal{L}_{Z_1}\supset \frac{g_L}{\cos\theta_W}\overline{f}_L\gamma^\mu Z_{1\mu}(C_{L_1}+\widetilde{C}_{L_1})f_L+\frac{g_L}{\cos\theta_W}\overline{f}_R\gamma^\mu Z_{1\mu}(C_{R_1}+\widetilde{C}_{R_1})f_R.\label{eq:LZ1}
    \end{equation}
Similarly, the corresponding $Z_2$ Lagrangian is
    \begin{equation}
        -\mathcal{L}_{Z_2}\supset \frac{g_L}{\cos\theta_W}\overline{f}_L\gamma^\mu Z_{2\mu}(C_{L_2}+\widetilde{C}_{L_2})f_L+\frac{g_L}{\cos\theta_W}\overline{f}_R\gamma^\mu Z_{2\mu}(C_{R_2}+\widetilde{C}_{R_2})f_R,\label{eq:LZ2}
    \end{equation}
where, $\widetilde{C}$ contains the mixing with VLFs. These expressions can be found in Appendix.~\ref{app:Lag}. For the analysis, we assume $g_L=g_R$, $M_{Z_1}=M_Z=91.18$ GeV and $M_{Z_2}=5$ TeV, consistent with the current experimental limits~\cite{ATLAS:2019erb}. A shorthand notation is used for the new couplings:
\begin{equation}
    \begin{aligned}
    (V_{L_f}\rho_{L_F}^\dagger\rho_{L_F}V_{L_f}^\dagger)_{ij}&=R_{ij},\\
    (V_{R_f}\rho_{R_F}^\dagger\rho_{R_F}V_{R_f}^\dagger)_{ij}&=R'_{ij}.
    \end{aligned}
\end{equation}
Then, under parity symmetry,
    \begin{equation}
    	R'_{ij}=\frac{\kappa_R^2}{\kappa_L^2}R_{ij}.\label{eq:Rparity}
    \end{equation}  
In the limit of small mixing between the neutral gauge bosons, $\sin\xi=\xi$, and $\cos\xi$ is set to 1. We also assume $R_{ij}$ is real for simplicity. For a clear understanding of how the constraints affect Yukawa couplings at different vector-like fermion masses, the results are given in terms of $(V_{L_f}\mathcal{Y}_F\mathcal{Y}_F^\dagger V_{L_f}^\dagger)_{ij}$. Unless otherwise stated, the experimental bounds used in the analyses are obtained from PDG (Ref.~\cite{Zyla:2020zbs}). The results are tabulated in the following sections. 

\subsection{\texorpdfstring{$Z$ Decays}{Z decays}}

The constraints on the couplings to charged leptons from $Z~(\equiv Z_1)$ decay are reported here. As seen from Sec.~\ref{sec:NC} the couplings are modified by the presence of vector-like leptons as well as the mixing between the two neutral gauge bosons. In obtaining the branching ratios these contributions are included in the total decay width of $Z$. For $Z \to \ell_i^+ \ell_i^-$ the new contribution to the appropriate couplings ($R_{ii}$) is turned on in both the $\Gamma(Z \to \ell_i^+ \ell_i^-)$ and $\Gamma_{total}$.  We made use of the PDG results of partial decay widths of $Z$-boson to achieve the precision needed to compare theoretical calculations with the experimental results. In computing the branching ratios in the case of $Z \to \ell_i^+\ell_j^-$ decays, $\Gamma_{total}= 2.4952\text{ GeV}$ was used.

\subsubsection{\texorpdfstring{$Z\rightarrow \ell_i^+\ell_i^-$}{Ztolili}}

Here, we study the flavor-conserving decays of SM-like $Z_1$ boson to charged leptons. There are new contributions to the diagonal couplings of $Z_1$ which will be constrained from the $Z$ decay modes. These decay modes are more constraining because of the interference of SM and NP terms as opposed to the case in the next subsection where $Z$ decays to $\ell_i \ell_j$. The decay rate can be obtained from the equation
    \begin{equation}
        \Gamma=\frac{g_L^2M_{Z_1}}{24\pi\cos^2\theta_W}\left(|C_{L_1}^{ii}+\widetilde{C}_{L_1}^{ii}|^2+|C_{R_1}^{ii}+\widetilde{C}_{R_1}^{ii}|^2\right)\label{eq:zlili}
    \end{equation}
 \begin{table}[]
    \footnotesize
    	{\renewcommand{\arraystretch}{1.5}
    		\begin{center}
    			\begin{tabular}{|c|c|c|}
    				\hline
    				\textbf{Process} & \textbf{Exp. Bound} & \textbf{Constraint} \\
    				\hline
    				$Z\rightarrow e^+ e^-$ & $(3.3632\pm 0.0042)\%$ & $|(V_{L_l}\mathcal{Y}_E\mathcal{Y}_E^\dagger V_{L_l}^\dagger)_{ee}|\leq 5.47\times 10^{-2}~\left(\frac{M_L}{1\text{ TeV}}\right)^2 $\\
    				    				\hline
    				$Z\rightarrow \mu^+ \mu^-$ & $(3.3662\pm 0.0066)\%$& $|(V_{L_l}\mathcal{Y}_E\mathcal{Y}_E^\dagger V_{L_l}^\dagger)_{\mu\mu}|\leq 6.59\times 10^{-2}~\left(\frac{M_L}{1\text{ TeV}}\right)^2 $\\
    				\hline
    				$Z\rightarrow \tau^+ \tau^-$ & $(3.3696\pm 0.0083)\%$& $|(V_{L_l}\mathcal{Y}_E\mathcal{Y}_E^\dagger V_{L_l}^\dagger)_{\tau\tau}|\leq 5.98\times 10^{-2}~\left(\frac{M_L}{1\text{ TeV}}\right)^2$\\
        \hline
    			\end{tabular}
    			\caption{Constraints from flavour conserving $Z$ decays to charged leptons. These are the allowed regions in the $2\,\sigma $ range as a function of vector-like lepton mass $M_L~(\text{TeV})$. These constraints were obtained by setting the total decay width of $Z$ as a function of the NP contributions to the appropriate diagonal couplings and fixing the rest of the decays to its SM values.
    			}
    			\label{tab:Ztolili}
    	\end{center}}
    \end{table}
The experimental bounds of the decay modes and the corresponding constraints on the new diagonal couplings are given in Table~\ref{tab:Ztolili}. The constraints are obtained by allowing $2\,\sigma $ deviation from the central value of the experimental results quoted.
   
\subsubsection{\texorpdfstring{$Z\rightarrow \ell_i^+\ell_j^-$}{Ztolilj}}

The presence of FCNCs lead to tree-level flavor violating decay modes of $Z$ boson to charged leptons, which are studied in this section. The constraints on the off-diagonal couplings contributing to the decay modes are given in Table~\ref{tab:Ztolilj}, with the following expression for decay rate:
    \begin{equation}
        \Gamma=\frac{g_L^2M_{Z_1}}{24\pi\cos^2\theta_W}(|\widetilde{C}_{L_1}^{ij}|^2+|\widetilde{C}_{R_1}^{ij}|^2).\label{eq:zlilj}
    \end{equation}

    \begin{table}[]
        \footnotesize
        {\renewcommand{\arraystretch}{1.5}%
          \begin{center}
            \begin{tabular}{|c|c|c|}
        \hline
              \textbf{Process} & \textbf{Exp. Bound} & \textbf{Constraint}\\
        \hline
        $Z\rightarrow e^\pm \mu^\mp$ &$<7.5\times 10^{-7}$ &$|(V_{L_l}\mathcal{Y}_E\mathcal{Y}_E^\dagger V_{L_l}^\dagger)_{e\mu}|< 0.11~\left(\frac{M_L}{1\text{ TeV}}\right)^2$ \\
        \hline
        $Z\rightarrow e^\pm \tau^\mp$ &$<5.0\times 10^{-6}$ & $|(V_{L_l}\mathcal{Y}_E\mathcal{Y}_E^\dagger V_{L_l}^\dagger)_{e\tau}|< 0.275~\left(\frac{M_L}{1\text{ TeV}}\right)^2$ \\
        \hline
        $Z\rightarrow \mu^\pm \tau^\mp$  &$<6.5\times 10^{-6}$&  $|(V_{L_l}\mathcal{Y}_E\mathcal{Y}_E^\dagger V_{L_l}^\dagger)_{\mu\tau}|< 0.313~\left(\frac{M_L}{1\text{ TeV}}\right)^2$ \\
        \hline
            \end{tabular}
        \caption{Constraints from flavor violating $Z$ decays to charged leptons as a function of vector-like lepton mass $M_L$ (TeV). $\Gamma_{total}$ was taken to be 2.4952 GeV.}
        \label{tab:Ztolilj}
          \end{center}}
    \end{table}

\subsubsection{Lepton Flavour Universality Violation}

The lepton flavor universality violations of $Z$ decays are studied here. The ratio of the decays to $\ell^+_i\ell^-_i$ pairs is calculated using Eq.~\eqref{eq:zlili} and taking a Taylor expansion to the first order in both the $R_{ii}$ parameters involved successively. In the Taylor series, the product of two parameters is ignored so that the quantity constrained is of the form $\left(1+\text{constant}\times (R_{ii}-R_{jj})\right)$ for 	$\Gamma(Z\rightarrow \ell^+_i\ell^-_i )/\Gamma(Z\rightarrow \ell^+_j\ell^-_j)$. The constraints are listed in Table~\ref{tab:Lfuv}. 
\begin{table}[]
\footnotesize
	{\renewcommand{\arraystretch}{2}
		\begin{center}
			\begin{tabular}{|c|c|c|}
				\hline
				\textbf{Process} & \textbf{Exp. Bound} & \textbf{Constraint$\big(\frac{M_L}{1\text{ TeV} }\big)^2$} \\
				\hline
				$\ddfrac{\Gamma(Z\rightarrow\mu^+ \mu^- )}{\Gamma(Z\rightarrow e^+ e^-)}$ & $(1.0001\pm 0.0024)$ & $-6.0\times 10^{-2}\leq (V_{L_l}\mathcal{Y}_E\mathcal{Y}_E^\dagger V_{L_l}^\dagger)_{\mu\mu-ee}\leq 5.73\times 10^{-2}$ \\
				\hline
				$\ddfrac{\Gamma(Z\rightarrow \tau^+ \tau^-)}{\Gamma(Z\rightarrow e^+ e^-)}$& $(1.002\pm 0.0032)$& $-1.03\times 10^{-2}\leq (V_{L_l}\mathcal{Y}_E\mathcal{Y}_E^\dagger V_{L_l}^\dagger)_{\tau\tau-ee} \leq 5.39\times 10^{-2}$ \\
				\hline
				$\ddfrac{\Gamma(Z\rightarrow\tau^+ \tau^-)}{ \Gamma(Z\rightarrow\mu^+ \mu^-)}$ & $(1.001\pm 0.0026)$& $-7.55\times 10^{-2}\leq (V_{L_l}\mathcal{Y}_E\mathcal{Y}_E^\dagger V_{L_l}^\dagger)_{\tau\tau-\mu\mu}\leq 5.14\times 10^{-2}$ \\
				\hline
					\end{tabular}
			\caption{Constraints from lepton flavour universality violations in $Z$ decays. A $2\,\sigma $ range is allowed in constraining the parameters. The results are quoted as a function of vector-like lepton mass $M_L$(TeV).}
			\label{tab:Lfuv}
	\end{center}}
\end{table}

\subsection{3 Body Decay of Charged Leptons}

In this section, we examine the decay of leptons to three charged leptons involving one flavor conserving vertex. These decay amplitudes are proportional to $R_{ij}$ since the flavor conserving vertex is SM-like coupling. The processes where both the vertices are flavor violating are not considered here since their amplitudes are proportional to $R_{ij}^2$ which are much less constrained. In computing the decay rates for the processes with one of the vertices being SM-like, the NP contribution to that vertex is ignored (although, the mixing between $Z_1$ and $Z_2$ is retained). The decay rate for $\ell_i^-\rightarrow \ell_j^-\ell_k^-\ell_k^+$ is given by
    \begin{equation}
        \Gamma=\frac{1}{(1+\delta_{jk})}\frac{g_L^4m_i^5 }{1536\cos^4\theta_W \pi ^3}\frac{\mathcal{C}}{M_{Z_1}^4M_{Z_2}^4},
    \end{equation}
where,
    \begin{equation}
        \begin{aligned}
            \mathcal{C}=&M_{Z_2}^4\Bigl(|C_{L_1}^{kk}|^2+|C_{R_1}^{kk}|^2\Bigr)\Bigl(|\widetilde{C}_{L_1}^{ji}|^2+|\widetilde{C}_{R_1}^{ji}|^2\Bigr)\\ 
            &+M_{Z_1}^4\left(|\widetilde{C}_{L_2}^{kk}|^2\Bigl(|C_{L_2}^{ji}|^2+\frac{7}{10}|C_{R_2}^{ji}|^2\Bigr)+|\widetilde{C}_{R_2}^{kk}|^2\Bigl(|C_{R_2}^{ji}|^2+\frac{7}{10}|
            C_{L_2}^{ji}|^2\Bigr)\right)\\ 
            &+M_{Z_1}^2M_{Z_2}^2\Bigl(\Bigl(\widetilde{C}_{L_1}^{ji}\widetilde{C}_{L_2}^{*ji}+\widetilde{C}_{R_1}^{ji}\widetilde{C}_{R_2}^{*ji}\Bigr)\Bigl(C_{L_1}^{kk}C_{L_2}^{*kk}+C_{R_1}^{kk}C_{R_2}^{*kk}\Bigr)\\ 
            &+C_{L_1}^{*kk}C_{L_2}^{kk}\Bigl(\widetilde{C}_{L_1}^{*ji}\widetilde{C}_{L_2}^{ji}+\frac{7}{10}\widetilde{C}_{R_1}^{*ji}\widetilde{C}_{R_2}^{ji}\Bigr)+C_{R_1}^{*kk}C_{R_2}^{kk}\Bigl(\widetilde{C}_{R_1}^{*ji}\widetilde{C}_{R_2}^{ji}+\frac{7}{10}\widetilde{C}_{L_1}^{*ji}\widetilde{C}_{L_2}^{ji}\Bigr)\Bigr).
        \end{aligned}
    \end{equation}
        \begin{figure}[]
    	\centering
    	\includegraphics[scale=0.2]{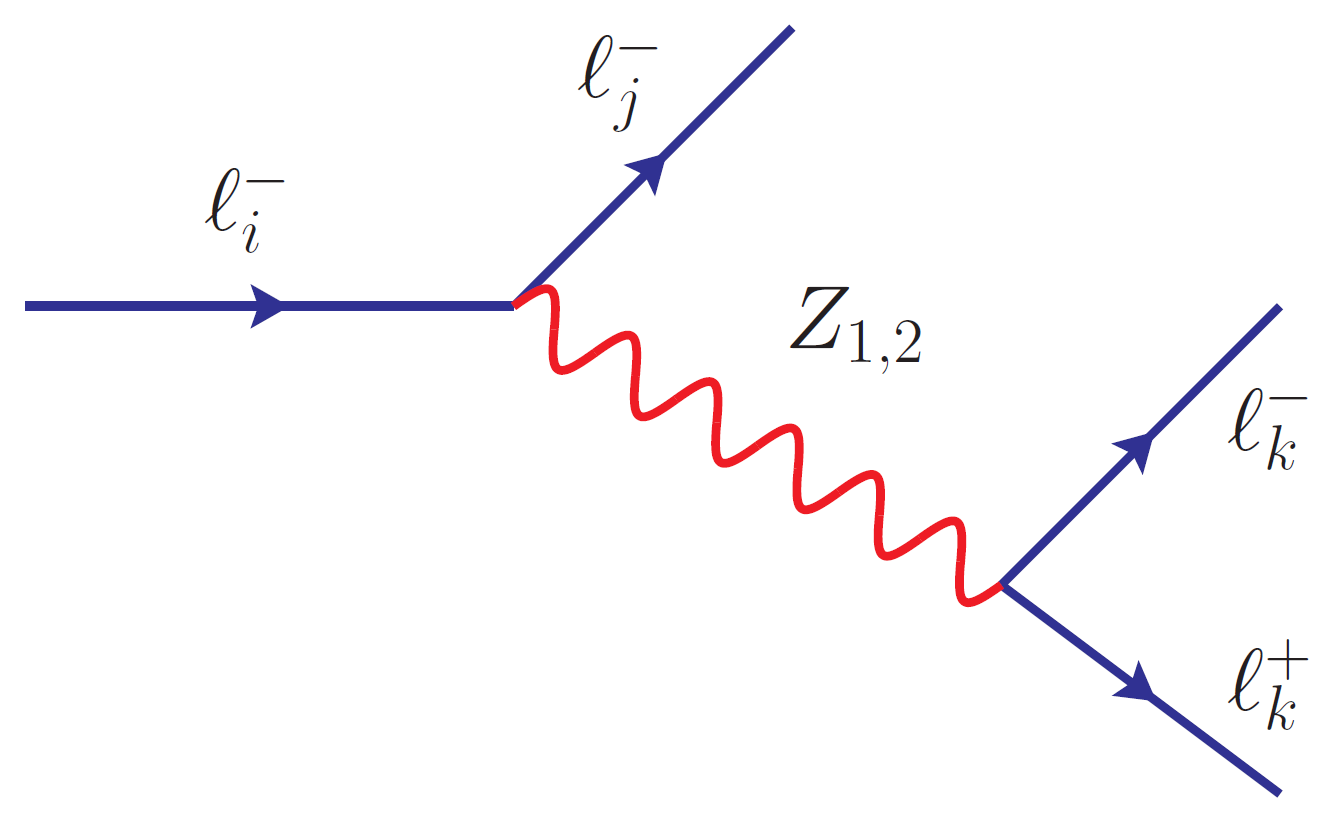}
    	\caption{Tree-level diagram for $\mu\rightarrow 3e$. }
    	\label{mutoegammafig}
    \end{figure}
The $(1+\delta_{ij})$ factor takes care of identical particles in the final state. The constraints on the off-diagonal couplings of neutral bosons to charged fermions are given in Table~\ref{3body}.

    \begin{table}
        \footnotesize
        {\renewcommand{\arraystretch}{1.5}
        \begin{center}
        \begin{tabular}{|c|c|c|c|}
            \hline
            \textbf{Process} & \textbf{Exp. Bound} & \textbf{Constraint}\\
            \hline
            $\mu^-\rightarrow e^-e^+e^-$ & $<1.0\times 10^{-12}$ & $|(V_{L_l}\mathcal{Y}_E\mathcal{Y}_E^\dagger V_{L_l}^\dagger)_{e\mu}|<9.51\times 10^{-5}~\left(\frac{M_L}{1\text{ TeV}}\right)^2$\\
            \hline
             $\tau^-\rightarrow e^-e^+e^-$ & $<2.7\times 10^{-8}$ & $|(V_{L_l}\mathcal{Y}_E\mathcal{Y}_E^\dagger V_{L_l}^\dagger)_{e\tau}|< 3.69\times 10^{-2}~\left(\frac{M_L}{1\text{ TeV}}\right)^2$\\
            \hline
            $\tau^-\rightarrow e^-\mu^+\mu^-$ & $<2.7\times 10^{-8}$ & $|(V_{L_l}\mathcal{Y}_E\mathcal{Y}_E^\dagger V_{L_l}^\dagger)_{e\tau}|< 2.61\times 10^{-2}~\left(\frac{M_L}{1\text{ TeV}}\right)^2$\\
            \hline
            $\tau^-\rightarrow \mu^-e^+e^-$ & $<1.8\times 10^{-8}$ & $|(V_{L_l}\mathcal{Y}_E\mathcal{Y}_E^\dagger V_{L_l}^\dagger)_{\mu\tau}|< 2.14\times 10^{-2}~\left(\frac{M_L}{1\text{ TeV}}\right)^2$\\
            \hline
            $\tau^-\rightarrow \mu^-\mu^+\mu^-$ & $<2.1\times 10^{-8}$ & $|(V_{L_l}\mathcal{Y}_E\mathcal{Y}_E^\dagger V_{L_l}^\dagger)_{\mu\tau}|<3.27\times 10^{-2}~\left(\frac{M_L}{1\text{ TeV}}\right)^2$\\
            \hline
        \end{tabular}
        \caption{Constraints from three body decays of charged leptons. New contributions to diagonal couplings are ignored in this analysis but the mixing between $Z_1$ and $Z_2$ are not omitted. The results are quoted as a function of vector-like lepton mass $M_L$ (TeV).
        }
        \label{3body}
        \end{center}}
    \end{table}

\subsection{Radiative Decays of Charged Leptons\label{l1tol2gamma}}
    \begin{figure}[]
    	\centering
    	\includegraphics[scale=.25]{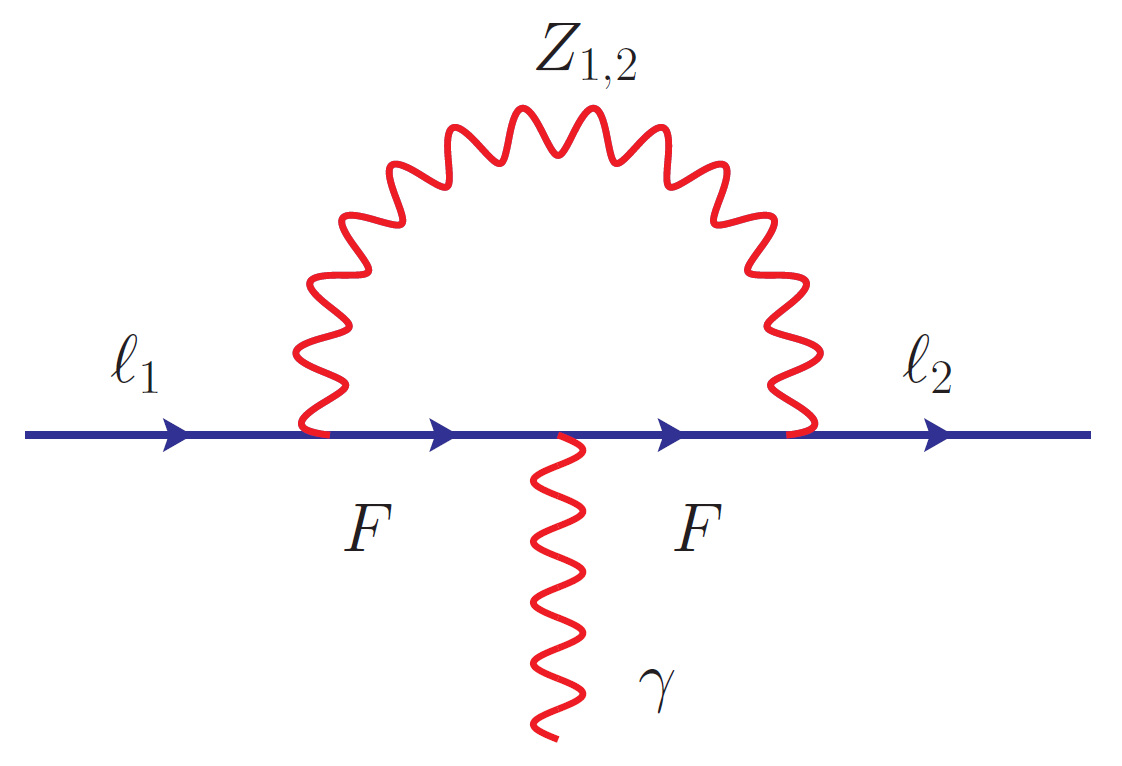}
    	\caption{One-loop $\mu \to e \gamma$ mediated by neutral gauge bosons. $F$ stands for SM as well as vector-like charged leptons.}
    	\label{mutoegammafigure}
    \end{figure}
The radiative decays of the form $\ell_1 \to \ell_2 \gamma$ arise from one-loop diagrams as shown in Fig.~\ref{mutoegammafigure} for $\mu \to e \gamma$. The gauge interaction leading to such decays with VL-lepton $F$ in the internal line is given by
    \begin{equation}
        \mathcal{L}_{gauge}=\sum_{j=1}^{2}\sum_{i=1}^{2}\overline{F}\gamma^\mu \left(\mathcal{C}_{L_i}^{Fj}P_L+\mathcal{C}_{R_i}^{Fj}P_R\right) \ell_j Z_{i_\mu}+ \text{h.c.} .\label{Lmutoegamma}
    \end{equation}
The appropriate coefficients can be obtained from Eqs.~\eqref{eq:LZL}, and \eqref{eq:LZR} converted to the mass basis of neutral gauge bosons, imposing parity symmetry: \begin{equation}
\begin{aligned}
      \mathcal{C}_{L_1}^{Fj}=&~g_L\cos\theta_W\left(\sin\xi\frac{1}{2\sqrt{\cos 2\theta_W}}\frac{g_Y^2}{ g_L^2}-\cos\xi\left(\frac{1}{2}+\frac{g_Y^2}{2g_L^2}\right)\rho_{L_F}V^\dagger_{L_j}\right),\\
    \mathcal{C}_{R_1}^{Fj}=&~g_L\cos\theta_W\left( \sin\xi\frac{1}{2\sqrt{\cos 2\theta_W}}\frac{\kappa_R}{\kappa_L}\rho_{L_F}V^\dagger_{L_j}\right),\\
     \mathcal{C}_{L(R)_2}^{Fj}=&-\mathcal{C}_{L(R)_1}^{Fj}(\xi\to\xi+\pi/2).
\end{aligned}\label{eq:coeff}
\end{equation}

        \begin{table}[]
    \footnotesize
    	{\renewcommand{\arraystretch}{1.5}%
    		\begin{center}
    			\begin{tabular}{|c|c|c|}
    				\hline
    				\textbf{Process} & \textbf{Exp. Bound} & \textbf{Constraint}\\
    				\hline
    				$\mu^-\rightarrow e^-\gamma$ & $<4.2\times 10^{-13}$ & $|(V_{L_l}\mathcal{Y}_E\mathcal{Y}_E^\dagger V_{L_l}^\dagger)_{e\mu}|< 2.41\times 10^{-6}~\left(\frac{M_L}{1\text{ TeV}}\right)^2$\\
    				\hline
    				$\tau^-\rightarrow e^-\gamma$ & $<3.3\times 10^{-8}$ & $|(V_{L_l}\mathcal{Y}_E\mathcal{Y}_E^\dagger V_{L_l}^\dagger)_{e\tau}|< 2.69\times 10^{-2}~\left(\frac{M_L}{1\text{ TeV}}\right)^2$\\
    				\hline
    				$\tau^-\rightarrow \mu^-\gamma$ & $<4.2\times 10^{-8}$ & $|(V_{L_l}\mathcal{Y}_E\mathcal{Y}_E^\dagger V_{L_l}^\dagger)_{\mu\tau}|< 3.54\times 10^{-2}~\left(\frac{M_L}{1\text{ TeV}}\right)^2$\\
    				\hline
    			\end{tabular}
    			\caption{Constraints from $\ell_1 \to \ell_2 \gamma$ processes. The vector-like lepton mass enhancement appearing in these diagrams is taken to be $=1.0$ TeV. The results are quoted as a function of vector-like lepton mass $M_L$(TeV).}\label{tab:raddecay}
    	\end{center}}
    \end{table}
The partial width for $\ell_1\to \ell_2\gamma$ is~\cite{Lavoura:2003xp}
    \begin{equation}
    	\Gamma=\frac{(m_1^2-m_2^2)^3}{16\pi m_1^3}\left(|\sigma_L|^2+|\sigma_R|^2\right).
    \end{equation}
Using the Eqs.~\eqref{Lmutoegamma} and \eqref{eq:coeff}, the coefficients $\sigma_{L, R}$ for $\mu \to e \gamma$ mediated by the VL-lepton $F$ are the following, where the subscript $i=1,2$ corresponds to $Z_{1,2}$: 
    \begin{equation}
                 \sigma_{L,R}=\sum_{i=1}^{2}\sigma_{L_i,R_i},
    \end{equation}
where,
    \begin{equation}
     	\begin{aligned}
     		\sigma_{L_i}=&-\left(\mathcal{C}_{R_i}^{eF}\mathcal{C}_{R_i}^{F\mu}y_1+\mathcal{C}_{L_i}^{eF}\mathcal{C}_{L_i}^{F\mu}y_2+\mathcal{C}_{R_i}^{eF}\mathcal{C}_{L_i}^{F\mu}y_3+\mathcal{C}_{L_i}^{eF}\mathcal{C}_{R_i}^{F\mu}y_4\right),\\
     		\sigma_{R_i}=&-\left(\mathcal{C}_{R_i}^{eF}\mathcal{C}_{R_i}^{F\mu}y_2+\mathcal{C}_{L_i}^{eF}\mathcal{C}_{L_i}^{F\mu}y_1+\mathcal{C}_{R_i}^{eF}\mathcal{C}_{L_i}^{F\mu}y_4+\mathcal{C}_{L_i}^{eF}\mathcal{C}_{R_i}^{F\mu}y_3\right).       
     	\end{aligned}\label{mutoegamma}
    \end{equation}
The factors $y_i$, with $c= \dfrac{i}{96m_{Z_i}^2 (m_{Z_i}^2-m_F^2)^4 \pi^2}$, are
\begin{equation}
    \begin{aligned}
       y_1=&c\frac{m_\mu}{2}\Bigl((m_{Z_i}^2-m_F^2)\left\{-8m_{Z_i}^6+30m_{Z_i}^4m_F^2-9m_{Z_i}^2m_F^4+5m_F^6\right.\\
       &\left.+m_e^2(2m_{Z_i}^4+5m_{Z_i}^2m_F^2-mF^4)\right\}
       +6m_{Z_i}^4m_F^2(m_e^2+3m_F^2)\ln{\frac{m_F^2}{m_{Z_i}^2}}\Bigr),  \\
       y_2=&y_1(m_e\leftrightarrow m_\mu),\\
       y_3=&c\frac{m_F}{2}\Bigl((m_{Z_i}^2-m_F^2)\left\{(m_\mu^2+m_e^2)(m_F^4-2m_{Z_i}^4-5m_{Z_i}^2m_F^2)+6(4m_{Z_i}^6-3m_{Z_i}^4m_F^2-m_F^6)\right\}\\
       &-6m_{Z_i}^4m_F^2(m_\mu^2+m_e^2-6m_{Z_i}^2+6m_F^2)\ln{\frac{m_F^2}{m_{Z_i}^2}}\Bigr),\\
       y_4=&-c\,m_em_\mu m_F\left(11m_{Z_i}^6-18m_{Z_i}^4m_F^2+9m_{Z_i}^2m_F^4-2m_F^6+6m_{Z_i}^6\ln{\frac{m_F^2}{m_{Z_i}^2}}\right).
    \end{aligned}
\end{equation}
 The dominant contribution to radiative decays arise from the chirally enhanced VL-lepton mediated diagrams. In obtaining the constraints, the mass of the mediator VLF is set to 1 TeV. The results are given in Table~\ref{tab:raddecay}. 


\subsection{Mass difference of Neutral Mesons\label{sec:mass diff}}
\begin{figure}[]
    	\centering
    	\includegraphics[scale=0.2]{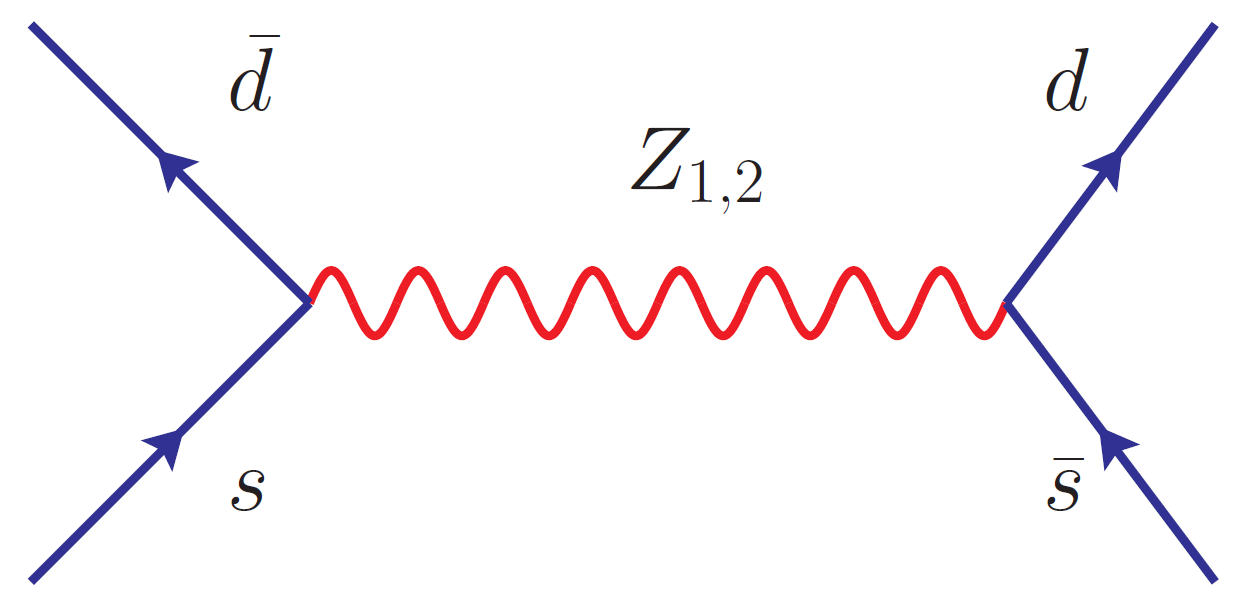}
    	\caption{Tree-level diagram contributing to kaon mixing. }
    	\label{fig:kaon_mixing_tree}
    \end{figure}

The effective Lagrangian for $\Delta F=2$ processes mediated by neutral gauge bosons, as shown in Fig.~\ref{fig:kaon_mixing_tree}, is of the form
    \begin{equation}
        \mathcal{L}=\sum_{k=1}^{2}\left(C_{L_k}(\Lambda)\,\overline{q}_{L_i}^\alpha\gamma^\mu q_{L_j}^\alpha Z_{k_\mu} +C_{R_k}(\Lambda)\,\overline{q}_{R_i}^\alpha\gamma^\mu q_{R_j}^\alpha Z_{k_\mu} +\frac{1}{2}M^2_{Z_k} Z_{k_\mu} Z_k^\mu\right)
    \end{equation}
where i and j are flavor indices, Greek indices stand for color, and $C(\Lambda)$ are the coefficients at high-scale. Upon integrating out the $Z_1$ and $Z_2$ masses, the Lagrangian becomes
    \begin{equation}
      \begin{aligned}
         \mathcal{H}=&\sum_{k=1}^{2}\frac{1}{2M^2_{Z_k}}\Bigl(C^2_{L_k}(\Lambda)\,\overline{q}_{L_i}^\alpha\gamma_\mu q_{L_j}^\alpha \,\overline{q}_{L_i}^\beta\gamma^\mu q_{L_j}^\beta  +C_{R_k}^2(\Lambda)\,\overline{q}_{R_i}^\alpha\gamma_\mu q_{R_j}^\alpha \,\overline{q}_{R_i}^\beta\gamma^\mu q_{R_j}^\beta \\ 
        &\qquad\qquad\qquad+2 C_{L_k}(\Lambda)\, C_{R_k}(\Lambda)\,\overline{q}_{L_i}^\alpha\gamma_\mu q_{L_j}^\alpha \, \overline{q}_{R_i}^\beta\gamma^\mu q_{R_j}^\beta \Bigr).\label{eq:Heffmeson}
      \end{aligned}  
    \end{equation}
   \begin{table}[]
    \footnotesize
    	{\renewcommand{\arraystretch}{1.5}%
    		\begin{center}
    			\begin{tabular}{|c|c|c|c|c|}
    				\hline
    				\textbf{Process} & \textbf{$\Delta \text{M}_{\text{exp}}$(GeV)} &\textbf{$\Delta \text{M}_{\text{SM}}$} & \textbf{Constraint $~\left(\frac{M_F}{1\text{ TeV}}\right)^2$}\\
    				\hline
    				\hline
    				$K-\overline{K}$ & $(3.484\pm0.006)\times10^{-15}$ & $2.364\times 10^{-15}$ \text{GeV}& $|(V_{L_f}\mathcal{Y}_D\mathcal{Y}_D^\dagger V_{L_f}^\dagger)_{ds}| \leq 2.12\times10^{-4}$ \\
    				\hline
    				$D-\overline{D}$ & $(6.2586^{+2.70}_{-2.90})\times10^{-15}$ & $3.87\times 10^{-15}$ GeV&$ |(V_{L_f}\mathcal{Y}_U\mathcal{Y}_U^\dagger V_{L_f}^\dagger)_{uc}|\leq 9.63\times 10^{-5}$  \\
    				\hline
    				$B-\overline{B}$ & $(3.334\pm0.013)\times 10^{-13}$ &$(0.543 \pm 0.029) \text{ps}^{-1}$\cite{Lenz:2019lvd}& $|(V_{L_f}\mathcal{Y}_D\mathcal{Y}_D^\dagger V_{L_f}^\dagger)_{db}|\leq 3.76\times 10^{-4}$ \\
    				\hline
    				$B_s-\overline{B}_s$ & $(1.1688\pm0.0014)\times 10^{-11}$ & $(18.77\pm 0.86)\text{ps}^{-1}$\cite{Lenz:2019lvd} & $|(V_{L_f}\mathcal{Y}_D\mathcal{Y}_D^\dagger V_{L_f}^\dagger)_{sb}| \leq 4.35\times10^{-4}$ \\
    				\hline 
    			\end{tabular}
    			\caption{Constraints from neutral meson mass differences. The results are quoted as a function of vector-like quark mass $M_F$ (TeV). $B_q-\overline{B_q}$ calculations are done using Eq.~\eqref{B_q} while the other two are evaluated using Eq.~\eqref{K,D}. More details can be found in Appendix.~{\ref{1}}.}
    			\label{mixing}
    	\end{center}}
    \end{table}
The mass difference can be evaluated as $\Delta M = 2\text{Re}\bra{\phi}\mathcal{H}\ket{\overline{\phi}}$, $\phi$ being the meson state. In general, the NP operators contributing to $\Delta F =2$ processes are~\cite{Bona:2007vz}
    \begin{equation}
    	\begin{aligned}
    		\mathcal{O}_1^{\,q_i\,q_j}&=\overline{q}_{Lj}^\alpha\gamma_\mu q_{Li}^\alpha \,\overline{q}_{Lj}^\beta\gamma^\mu q_{Li}^\beta\,, \\ 
    		\mathcal{O}_2^{\,q_i\,q_j}&=\overline{q}_{Rj}^\alpha q_{Li}^\alpha \,\overline{q}_{Rj}^\beta q_{Li}^\beta\,,\\ 
    		\mathcal{O}_3^{\,q_i\,q_j}&=\overline{q}_{Rj}^\alpha q_{Li}^\beta \,\overline{q}_{Rj}^\beta q_{Li}^\alpha\,,\\ 
    		\mathcal{O}_4^{\,q_i\,q_j}&=\overline{q}_{Rj}^\alpha q_{Li}^\alpha \,\overline{q}_{Lj}^\beta q_{Ri}^\beta\,,\\
    		\mathcal{O}_5^{\,q_i\,q_j}&=\overline{q}_{Rj}^\alpha q_{Li}^\beta\, \overline{q}_{Lj}^\beta q_{Ri}^\alpha\,,\\
    	\end{aligned}\label{eq:operators}
    \end{equation}
and $\widetilde{\mathcal{O}}_{1,2,3}^{\,q_i\,q_j}$, obtained by the exchange $L \leftrightarrow R$. Using Eqs.~\eqref{eq:Heffmeson} and \eqref{eq:operators}, the effective Hamiltonian for meson mixing mass difference can be written as
    \begin{equation}
        \mathcal{H}_\text{{eff}}=C_1(M_Z)\mathcal{O}_1 + \widetilde{C}_1(M_Z)\mathcal{\widetilde{O}}_1 -4 C_5(M_Z)\mathcal{O}_5.
    \end{equation}
Here, the operator $\mathcal{O}_5$ is obtained by Fierz transformation~\cite{ Nishi:2004st} of $\overline{q}_{Li}^\alpha\gamma_\mu q_{Lj}^\alpha \overline{q}_{Ri}^\beta\gamma^\mu q_{Rj}^\beta$ in Eq.~\eqref{eq:Heffmeson}.
The Wilson coefficients~(WCs) at $M_{Z_{1(2)}}$ scale (or any NP scale, $\Lambda$) need to be evolved down to hadronic scale $m_b=4.6 $ GeV for bottom mesons, $\mu_D=2.8 $ GeV for charmed mesons and $\mu_K=2 $ GeV  for Kaons, which are the renormalization scales used in lattice computation of matrix elements~\cite{Bona:2007vz}. Renormalization group evolution of $C_5$ induces $C_4$, causing the corresponding operator to appear in the expression of $\Delta M$ at the hadronic scale. This is given by the analytical formula:
    \begin{equation} \bra{\overline{\phi}}\mathcal{H}_\text{eff}\ket{\phi}_i=\sum_{j=1}^5\sum_{r=1}^5(b_j^{(r,i)}+\eta c_j^{(r,i)})\eta^{a_j}C_i(\Lambda))\bra{\overline{\phi}}\mathcal{Q}_r\ket{\phi},\label{K,D}
    \end{equation}
where, $\eta = \alpha_s (\Lambda)/\alpha_s (m_t)$, and $a_j$, $b_j^{(r,i)}$ and $c_j^{(r,i)}$ are the magic numbers. The relevant constants and magic numbers are provided in Appendix.~\ref{app:Bqmix}.

We follow a different approach for obtaining the mass differences in the case of $B_q$-mesons~\cite{DiLuzio:2019jyq, He:2006bk}, although the expression in Eq.~\eqref{K,D} is equally valid. The SM predicts~\cite{Buchalla:1995vs}
    \begin{equation}
        \Delta M_q = \frac{2G_F^2}{4\pi^2}M_W^2\lambda_q^2\hat{\eta}_BS_0(x_t)\frac{\langle \mathcal{O}_1\rangle}{2M_{B_q}},
    \end{equation}
where, $q=\{s, d\}$, $\lambda_q=V_{tb}V_{tq}^*$, $S_0$ is the Inami-Lim function~\cite{Inami:1980fz}, $x_t=(\overline{m}_t(\overline{m}_t)/M_W)^2$, $\hat{\eta}_B=0.84$ and $\langle \mathcal{O}_1\rangle=\frac{2}{3}M_{B_q}^2f^2_{b_q}B_1(m_b)$. Using this, the SM+NP contribution to mass mixing normalized to the SM one is given by
    \begin{equation}
    	\begin{aligned}
    		\frac{\Delta M_q^\text{SM+NP}}{\Delta M_q^\text{SM}}=&\bigg|1+\frac{\sqrt{2}G_FM_{Z_1}^2}{\mathtt{C}^{\Delta_{SM}}}\sum_{k=1,2}\bigg[\frac{2}{3}\bigg(\frac{\widetilde{C}_{L_k}^{{ij}^2}+\widetilde{C}_{R_k}^{{ij}^2}}{M_{Z_k}^2}\eta_k^{6/23}\bigg)\\
    		&-\frac{B_5}{B_1}\bigg(\frac{2M^2_{B_q}}{3(m_b+m_q)^2}+1\bigg)\bigg(\frac{\widetilde{C}_{L_k}^{ij}\widetilde{C}_{R_k}^{ij}}{M_{Z_k}^2}\eta_k^{3/23}\bigg)\\
    		&+\frac{1}{3}\frac{B_4}{B_1}\bigg(\frac{2 M^2_{B_q}}{(m_b+m_q)^2}+\frac{1}{3}\bigg)\bigg(\frac{\widetilde{C}_{L_k}^{ij}\widetilde{C}_{R_k}^{ij}}{M_{Z_k}^2}\left(\eta_k^{3/23}-\eta_k^{-24/23}\right)\bigg)\bigg]\bigg|,
    	\end{aligned} \label{B_q}
    \end{equation}
with $\mathtt{C}^{\Delta_\text{SM}}=\frac{G_F^2}{12\pi}\lambda_q^2M_W^2S_0(x_t)\hat{\eta}_B$. The analysis for $B_q$ mixings is done with a $2\,\sigma$ deviation in the SM predictions and $1\,\sigma$ in experimental measurements while $2\,\sigma$ deviation in the experimental measurements is allowed for $K$ and $D$ mass differences. The constraints obtained are tabulated in Table~\ref{mixing} along with the experimental bounds and the SM predictions used.

\subsection{Charged Leptonic Decays of Mesons }
Neutral mesons decaying to charged leptons of the form $\phi\to\ell_i^+\ell_i^-$ are analyzed here, where $\phi$ represents meson and $\ell$ stands for lepton. The decay rate, with $f_\phi$ being the form factor, is given by
    \begin{equation}
        \Gamma=\frac{g_L^4}{32\pi\cos^4\theta_W}f_\phi^2m_\ell^2m_\phi\sqrt{1-\frac{4m_\ell^2}{m_\phi^2}}(|\mathcal{C}_L|^2+|\mathcal{C}_R|^2),
    \end{equation}
where, $\mathcal{C}_X=\dfrac{C_{X_1}^{\ell\ell}}{M_{Z_1}^2}(\widetilde{C}_{L_1}^{ij}-\widetilde{C}_{R_1}^{ij})+ \dfrac{C_{X_2}^{\ell\ell}}{M_{Z_2}^2}(\widetilde{C}_{L_2}^{ij}-\widetilde{C}_{R_2}^{ij}),\text{ and } X=\{L,R\}.$
    \begin{figure}[]
    	\centering
    	\includegraphics[scale=0.25]{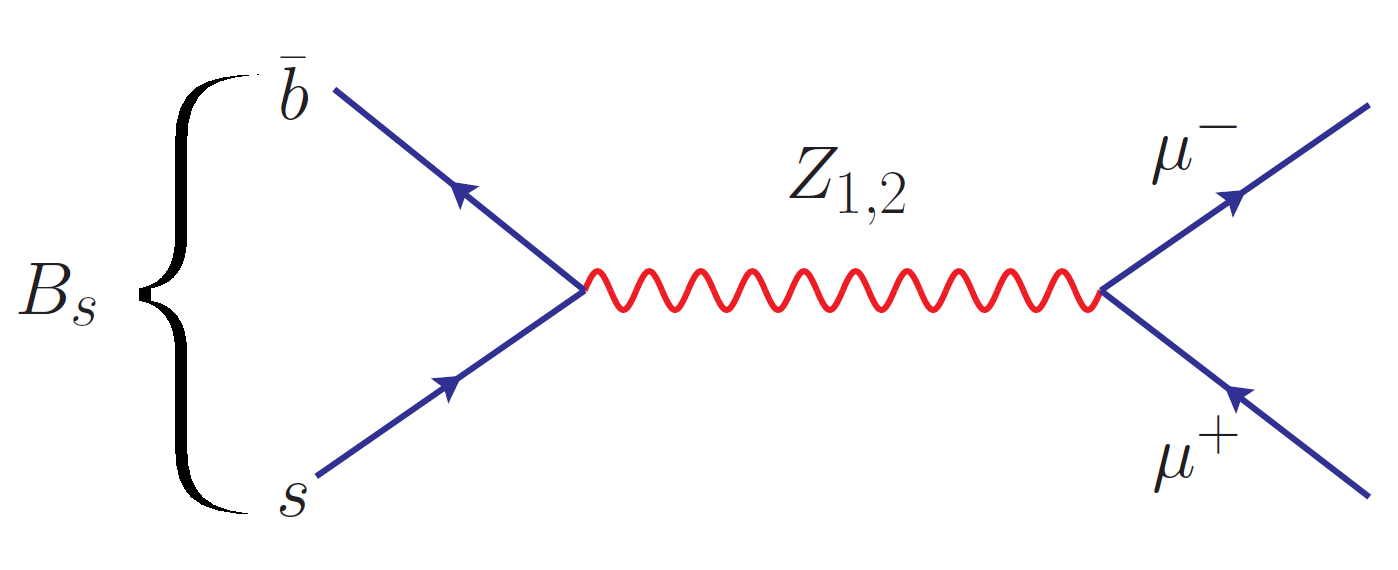}
    	\caption{Tree-level diagram for $B_s\to\mu^+\mu^-$. }
    	\label{leptonicmesondecay}
    \end{figure}
The new contributions to $Z_i\,\overline{\ell}\ell$ couplings are ignored; i.e., $R_{\ell\ell}=0$ and the allowed parameter space of $R_{ij}$ from the quark sector is determined. For $B_q \to \mu^+ \mu^-$ decays, the constraints are obtained for the central value of the experimental measurements. The constraints obtained are listed in  Table~\ref{fphi} along with the experimental bounds and the form factors used. We see that the constraints are small when electrons are the final particles owing to the small mass while muon final states give stronger constraints on account of the stringent experimental bounds. 

    \begin{table}[]
    \small
    	{\renewcommand{\arraystretch}{1.5}%
    		\begin{center}
    			\begin{tabular}{|c|c|c|c|c|}
    				\hline
    				\textbf{Process} & \textbf{Exp. Bound} &\textbf{Constraint~$\left(\frac{M_F}{1\text{ TeV}}\right)^2$
    					} & \textbf{$f_\phi$ (GeV)}\cite{Melikhov:2000yu}\\
    				\hline
    				$D\rightarrow e^+e^-$ & $< 7.9\times 10^{-8}$ & $|(V_{L_f}\mathcal{Y}_U\mathcal{Y}_U^\dagger V_{L_f}^\dagger)_{uc}|< 35.96$ &$0.200$ \\
    				\hline
    				$D\rightarrow \mu^+\mu^-$ & $< 6.2\times 10^{-9}$ & $|(V_{L_f}\mathcal{Y}_U\mathcal{Y}_U^\dagger V_{L_f}^\dagger)_{uc}|< 4.88\times 10^{-2}$ & $0.200$ \\
    				\hline 
    				$B\rightarrow e^+e^-$ & $< 2.5\times 10^{-9}$ & $|(V_{L_f}\mathcal{Y}_D\mathcal{Y}_D^\dagger V_{L_f}^\dagger)_{bd}|< 2.21$ & $0.180$ \\
    				\hline
    				$B\rightarrow \mu^+\mu^-$ & $< 1.5\times 10^{-10}$ & $|(V_{L_f}\mathcal{Y}_D\mathcal{Y}_D^\dagger V_{L_f}^\dagger)_{bd}|\leq 2.62\times10^{-3}$ & $0.180$\\
    				\hline
    				$B\rightarrow \tau^+\tau^-$ & $< 1.6\times 10^{-3}$ & $|(V_{L_f}\mathcal{Y}_D\mathcal{Y}_D^\dagger V_{L_f}^\dagger)_{bd}|< 0.59$ & $0.180$ \\
    				\hline
    				$B_s\rightarrow e^+e^-$ & $< 9.4\times 10^{-9}$ & $|(V_{L_f}\mathcal{Y}_D\mathcal{Y}_D^\dagger V_{L_f}^\dagger)_{bs}|< 3.83$ & $0.200$\\
    				\hline
    				$B_s\rightarrow \mu^+\mu^-$ & $(3.45\pm 0.29)\times 10^{-9}$
        & $|(V_{L_f}\mathcal{Y}_D\mathcal{Y}_D^\dagger V_{L_f}^\dagger)_{bs}|\leq 1.21\times10^{-2}$ & $0.200$\\
    				\hline
    				$B_s\rightarrow \tau^+\tau^-$ & $< 5.2\times 10^{-3}$ & $|(V_{L_f}\mathcal{Y}_D\mathcal{Y}_D^\dagger V_{L_f}^\dagger)_{bs}|< 0.947$ & $0.200$ \\
    				\hline
    				\end{tabular}
    			\caption{Constraints from flavor conserving charged leptonic decays of neutral mesons. The results are quoted as a function of vector-like quark mass $M_F$. The experimental values of rare $B$ decays are obtained from HFLAV Collaboration~\cite{HFLAV:2022esi}.
    			}
    			\label{fphi}
    	\end{center}}
    \end{table}


\subsection{Semi-Leptonic Meson Decay}
Here, we explore the tree-level lepton flavor conserving decays of mesons of the for $\phi_H\to phi_L \ell^+\ell^-$. We assume that total decay width is comprised of only the tree-level contribution from the NP model parameters, so the results are less constraining than what would be obtained with a more thorough investigation including the SM interference terms. The decay width for such processes is given by 
    \begin{equation}
        \Gamma=\frac{g^4_L}{64\pi^3\cos^4\theta_W m_H}|f_+(0)|^2\mathcal{F}\,(|\mathcal{C}_L|^2+|\mathcal{C}_R|^2),\label{eq:semilepmesondecay}
    \end{equation}
where, $\mathcal{F}=\dfrac{m_V^2}{48m_H^2}\left(-2m_H^6+9m_H^4m_V^2-6m_H^2m_V^4-6(m_V^3-m_H^2m_V)^2 \ln{\frac{m_V^2-m_H^2}{m_V^2}}\right)$, $m_H$ is the parent meson mass and $m_V$ is the corresponding vector meson mass. The input parameters used are given in Appendix.~\ref{app1}. The coefficients $\mathcal{C}_{L,R}$ are given by
    \begin{equation}
        \mathcal{C}_X=\frac{C_{X_1}^{\ell\ell}(\widetilde{C}_{L_1}^{ij}+\widetilde{C}_{R_1}^{ij})}{M_{Z_1}^2}+ \frac{C_{X_2}^{l\ell\ell}(\widetilde{C}_{L_2}^{ij}+\widetilde{C}_{R_2}^{ij})}{M_{Z_2}^2}, \quad X=\{L, R\}.
    \end{equation}
    \begin{figure}[]
    	\centering
    	\includegraphics[scale=0.25]{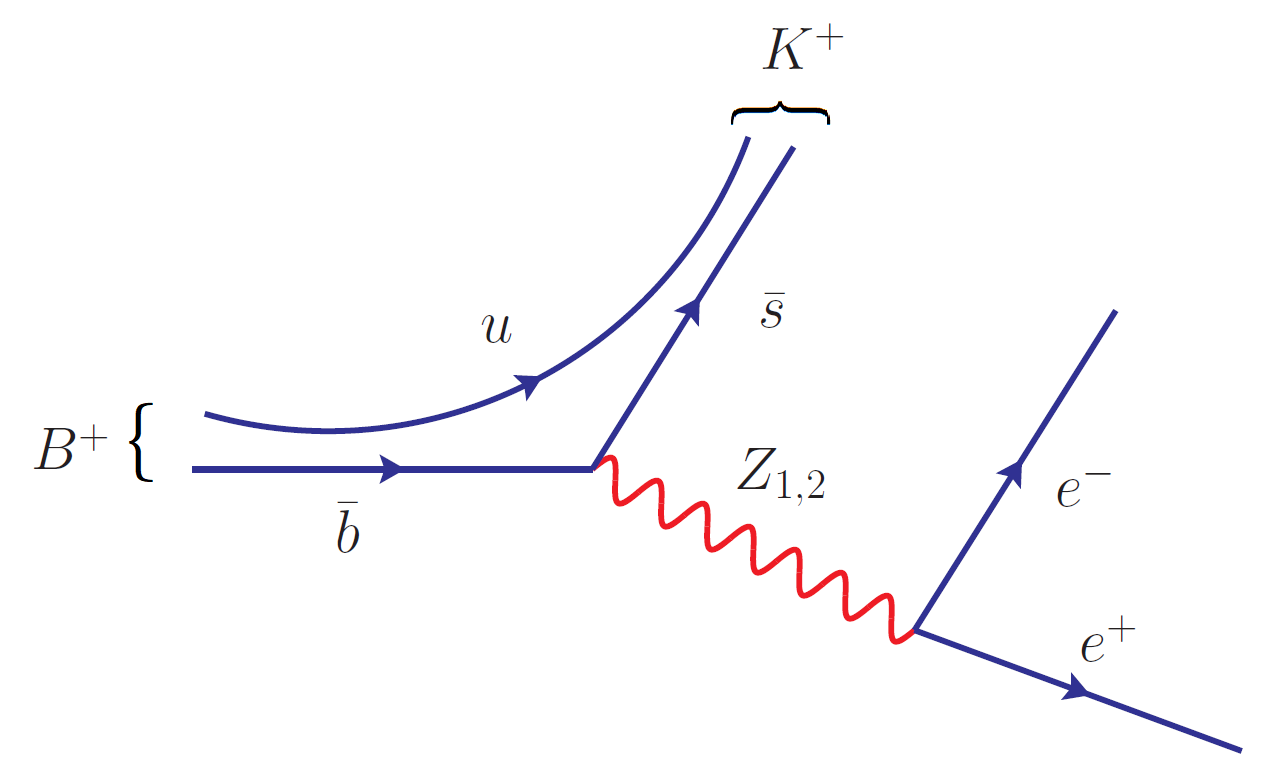}
    	\caption{Tree-level diagram for $B^+\to K^+e^+e^-$. }
    	\label{semileptonicmesondecay}
    \end{figure} 
    \begin{table}[]
    \footnotesize
    	{\renewcommand{\arraystretch}{1.5}%
    		\begin{center}
    			\begin{tabular}{|c|c|c|c|}
    				\hline
    				\textbf{Process} & \textbf{Exp. Bound} &\textbf{Constraint~$\left(\frac{M_F}{1\text{ TeV}}\right)^2$}\\
    				\hline
    				$K^+\to \pi ^+ e^+ e^-$ & $(3\pm 0.09)\times 10^{-7}$ & $|(V_{L_f}\mathcal{Y}_D\mathcal{Y}_D^\dagger V_{L_f}^\dagger)_{sd}|\leq 3.83\times10^{-2}$\\
    				\hline
    				$K^+\to \pi ^+ \mu^+ \mu^-$ & $(9.4\pm 0.6)\times 10^{-8}$ & $|(V_{L_f}\mathcal{Y}_D\mathcal{Y}_D^\dagger V_{L_f}^\dagger)_{sd}|\leq 2.15\times10^{-2}$\\
    								\hline
    				$K^+\to \pi ^+ \nu \overline{\nu}$ & $(1.7\pm 1.1)\times 10^{-10}$ & $|(V_{L_f}\mathcal{Y}_D\mathcal{Y}_D^\dagger V_{L_f}^\dagger)_{sd}|\leq 3.66\times10^{-4}$\\
    				\hline 
    				\hline
    				$B^+\to \pi ^+ \ell^+ \ell^-$ & $< 4.9\times 10^{-8}$ & $|(V_{L_f}\mathcal{Y}_D\mathcal{Y}_D^\dagger V_{L_f}^\dagger)_{bd}|< 6.49\times10^{-3}$\\
    				\hline
    				$B^+\to \pi ^+ e^+ e^- $ & $< 8.0\times 10^{-8}$ &$|(V_{L_f}\mathcal{Y}_D\mathcal{Y}_D^\dagger V_{L_f}^\dagger)_{bd}|< 1.18\times10^{-2}$\\
    				\hline
    				$B^+\to \pi ^+ \mu ^+ \mu ^-$ & $(1.78 \pm 0.23)\times 10^{-8}$ & $|(V_{L_f}\mathcal{Y}_D\mathcal{Y}_D^\dagger V_{L_f}^\dagger)_{bd}|\leq 6.23\times10^{-3}$\\
    				\hline
    				$B\to \pi \ell^+ \ell^-$ & $< 5.3\times 10^{-8}$ & $|(V_{L_f}\mathcal{Y}_D\mathcal{Y}_D^\dagger V_{L_f}^\dagger)_{bd}|< 9.6\times10^{-3}$\\
    				\hline
    				$B\to \pi e^+ e^-$ & $8.4\times 10^{-8}$ & $|(V_{L_f}\mathcal{Y}_D\mathcal{Y}_D^\dagger V_{L_f}^\dagger)_{bd}|< 1.71\times10^{-2}$\\
    				\hline
    				$B\to \pi \mu^+ \mu^-$ & $< 6.9\times 10^{-8}$ & $|(V_{L_f}\mathcal{Y}_D\mathcal{Y}_D^\dagger V_{L_f}^\dagger)_{bd}|< 1.55\times10^{-2}$\\
    				\hline
    				$B^+\to \pi ^+ \nu \overline{\nu}$ & $< 1.4\times 10^{-5}$ & $|(V_{L_f}\mathcal{Y}_D\mathcal{Y}_D^\dagger V_{L_f}^\dagger)_{bd}|< 6.24\times10^{-2}$\\
    				\hline
    				$B\to \pi \nu \overline{\nu}$ & $< 9\times 10^{-6}$ & $|(V_{L_f}\mathcal{Y}_D\mathcal{Y}_D^\dagger V_{L_f}^\dagger)_{bd}|< 7.06\times10^{-2}$\\
    				\hline
    				\hline
    				$B^+\to K ^+ \ell^+ \ell^-$ & $(4.63 \pm 0.19)\times 10^{-7}$ & $|(V_{L_f}\mathcal{Y}_D\mathcal{Y}_D^\dagger V_{L_f}^\dagger)_{bs}|\leq 1.72\times10^{-2}$\\
    				\hline
    				$B^+\to K ^+ e^+ e^-$ & $(5.61 \pm 0.56)\times 10^{-7}$ &$|(V_{L_f}\mathcal{Y}_D\mathcal{Y}_D^\dagger V_{L_f}^\dagger)_{bs}|\leq 2.83\times10^{-2}$\\
    				\hline
    				$B^+\to K ^+ \mu^+ \mu^-$ & $(4.50 \pm 0.21)\times 10^{-7}$ & $|(V_{L_f}\mathcal{Y}_D\mathcal{Y}_D^\dagger V_{L_f}^\dagger)_{bs}|\leq 2.42\times10^{-2}$\\
    				\hline
    				$B^+\to K ^+ \tau^+ \tau^-$ & $< 2.2\times 10^{-3}$ & $|(V_{L_f}\mathcal{Y}_D\mathcal{Y}_D^\dagger V_{L_f}^\dagger)_{bs}|< 1.62$\\
    				\hline
    				$B\to K \ell^+ \ell^-$ & $(3.28\pm0.32)\times 10^{-7}$ & $|(V_{L_f}\mathcal{Y}_D\mathcal{Y}_D^\dagger V_{L_f}^\dagger)_{bs}|\leq 1.52\times10^{-2}$\\
    				\hline
    				$B\to K e^+ e^-$ & $(2.49\pm0.72)\times 10^{-7}$ & $|(V_{L_f}\mathcal{Y}_D\mathcal{Y}_D^\dagger V_{L_f}^\dagger)_{bs}|\leq 2.16\times10^{-2}$\\
    				\hline
    				$B\to K \mu^+ \mu^-$ & $(3.41\pm 0.34)\times 10^{-7}$ & $|(V_{L_f}\mathcal{Y}_D\mathcal{Y}_D^\dagger V_{L_f}^\dagger)_{bs}|\leq 2.20\times10^{-2}$\\
    				\hline								$B^+\to K ^+ \nu \overline{\nu}$ & $<1.6\times 10^{-5}$ & $|(V_{L_f}\mathcal{Y}_D\mathcal{Y}_D^\dagger V_{L_f}^\dagger)_{bs}|< 5.51\times10^{-2}$\\
    				\hline    
    				$B\to K \nu \overline{\nu}$ & $< 2.6\times 10^{-5}$ & $|(V_{L_f}\mathcal{Y}_D\mathcal{Y}_D^\dagger V_{L_f}^\dagger)_{bs}|< 7.03\times10^{-2}$\\
    				\hline
    				\hline
    				$D^+\to \pi^+ e^+ e^-$ & $< 1.1\times 10^{-6}$ & $|(V_{L_f}\mathcal{Y}_U\mathcal{Y}_U^\dagger V_{L_f}^\dagger)_{uc}|< 0.33$ \\
    				\hline
    				$D^+\to \pi^+ \mu^+ \mu^-$ & $< 7.3\times 10^{-8}$ & $|(V_{L_f}\mathcal{Y}_U\mathcal{Y}_U^\dagger V_{L_f}^\dagger)_{uc}|< 8.54\times10^{-2}$ \\
    				\hline
    				$D\to \pi e^+ e^-$ & $< 4\times 10^{-6}$ & $|(V_{L_f}\mathcal{Y}_U\mathcal{Y}_U^\dagger V_{L_f}^\dagger)_{uc}|< 1.44$ \\
    				\hline
    				$D\to \pi \mu^+ \mu^-$ & $< 1.8\times 10^{-4}$ & $|(V_{L_f}\mathcal{Y}_U\mathcal{Y}_U^\dagger V_{L_f}^\dagger)_{uc}|< 9.67$ \\
    				\hline
    				$D_s\to K^+ \mu^+ \mu^-$ & $< 2.1\times 10^{-5}$ & $|(V_{L_f}\mathcal{Y}_U\mathcal{Y}_U^\dagger V_{L_f}^\dagger)_{uc}|< 1.67$ \\
    				\hline
    				$D_s\to K^+ e^+e^-$ & $< 3.7\times 10^{-6}$ & $|(V_{L_f}\mathcal{Y}_U\mathcal{Y}_U^\dagger V_{L_f}^\dagger)_{uc}|< 0.99$ \\
    				\hline
     			\end{tabular}
    			\caption{Constraints from flavor conserving charged leptonic decays of heavy mesons, quoted as a function of vector-like quark mass $M_F$. The experimental values of rare $B$ decays are obtained from HFLAV Collaboration~\cite{HFLAV:2022esi}.}
    			\label{H to L}
    	\end{center}}
    \end{table}
The NP contribution to diagonal couplings from the lepton sector is set to zero since the quark vertex has the leading contribution. When the final state particles involve neutrinos, only the SM neutrino vertex is taken into account so that the $C_{X_2}$ coefficients vanish and the three flavors of neutrinos are summed together. As in the previous section, the constraints are obtained for the central values of the experimental measurements. They are tabulated in Table~\ref{H to L}.

\section{Constraints on Charged Current Couplings\label{sec:cc_coupling}}

In this section, we find the constraints on the theory parameters modifying the charged current couplings. The major constraints arise from lepton flavor universality violating processes. We also look at $W_L$ decay to leptons and $\ell_1 \to\ell_2 \gamma$ processes. For simplicity, the part of Eq.~\eqref{eq:nulcharged} relevant to the processes being studied is rewritten as
    \begin{equation}
    		-\mathcal{L}_{W_L}=\overline{\nu}_{L} \gamma^\mu W^+_{L\mu}\frac{g_L}{\sqrt{2}} (1-f_{\nu \ell}) \ell_L+...+ \text{h.c.}. \label{Wfcoupling}
    \end{equation}

\subsection{\texorpdfstring{$W_L$ Decay to Leptons}{WLdecaytoleptons}}
    \begin{table}[]
    \footnotesize
    {\renewcommand{\arraystretch}{1.5}
        \begin{center}
            \begin{tabular}{|c|c|c|}
            \hline
            \textbf{Process} & \textbf{Exp. Bound}~\cite{CMS:2022mhs} & \textbf{Constraint} \\
            \hline
            $W_L^+\rightarrow e^+ \nu$ & $(10.83\pm 0.10)\%$ & $-7.03\times 10^{-3}\leq f_{\nu e}\leq 1.36\times 10^{-2}$ \\
            \hline
            $W_L^+\rightarrow \mu^+ \nu$ & $(10.94\pm 0.08)\%$& $-1.06\times 10^{-2}\leq f_{\nu \mu } \leq 5.84\times 10^{-3}$ \\
            \hline
            $W_L^+\rightarrow \tau^+ \nu$ & $(10.77\pm 0.21)\%$& $-1.52 \times 10^{-2}\leq f_{\nu \tau}\leq 2.82 \times 10^{-2}$ \\
            \hline
            \end{tabular}
        \caption{Constraints from leptonic decays of $W_L$. $f$'s are the new contribution to the charged current couplings as defined in Eq.~\eqref{Wfcoupling}. A  $2\,\sigma$ deviation  is  allowed  in obtaining the constraints. 
        }\label{Wdecaylep}
        \end{center}}
    \end{table}
Here, we obtain the constraints on the coupling of $W_L$ to leptons by studying its different decay modes. The decay rate of such processes is given by
    \begin{equation}
        \Gamma=\frac{g_L^2M_W}{48\pi}|1-f_{\nu \ell}|^2.\label{5.2}
    \end{equation}
To accommodate the deviation in the total decay width of $W_L$ due to NP contribution, the branching ratio is computed by turning on the relevant coupling in each case. Since $\Gamma(W_L\to q_i q_j)_\text{SM}=6\Gamma(W_L\to \nu_\ell \ell)_\text{SM}$ owing to the apparent unitarity of the CKM matrix, the total decay width of $W_L$ maybe written as 
    \begin{equation}
        \Gamma_{W_L}=8\Gamma(W_L\to \nu_\ell \ell)_\text{SM}+\frac{g_L^2M_W}{48\pi}|1-f_{\nu \ell}|^2.
    \end{equation}
Then, the branching ratio of $W_L^+ \to e^+\nu$, for example, would be $\Gamma(f_{\nu \ell}=f_{\nu e})/\Gamma_{W_L}(f_{\nu \ell}=f_{\nu e})$. We allow a $2\,\sigma$ deviation above and below the central value in obtaining the constraints. The constraints from $W_L$ decays are given in Table~\ref{Wdecaylep}.
    
\subsection{Radiative Decays of Charged Leptons}

    \begin{figure}[]
        \centering
      \includegraphics[width=0.37\linewidth]{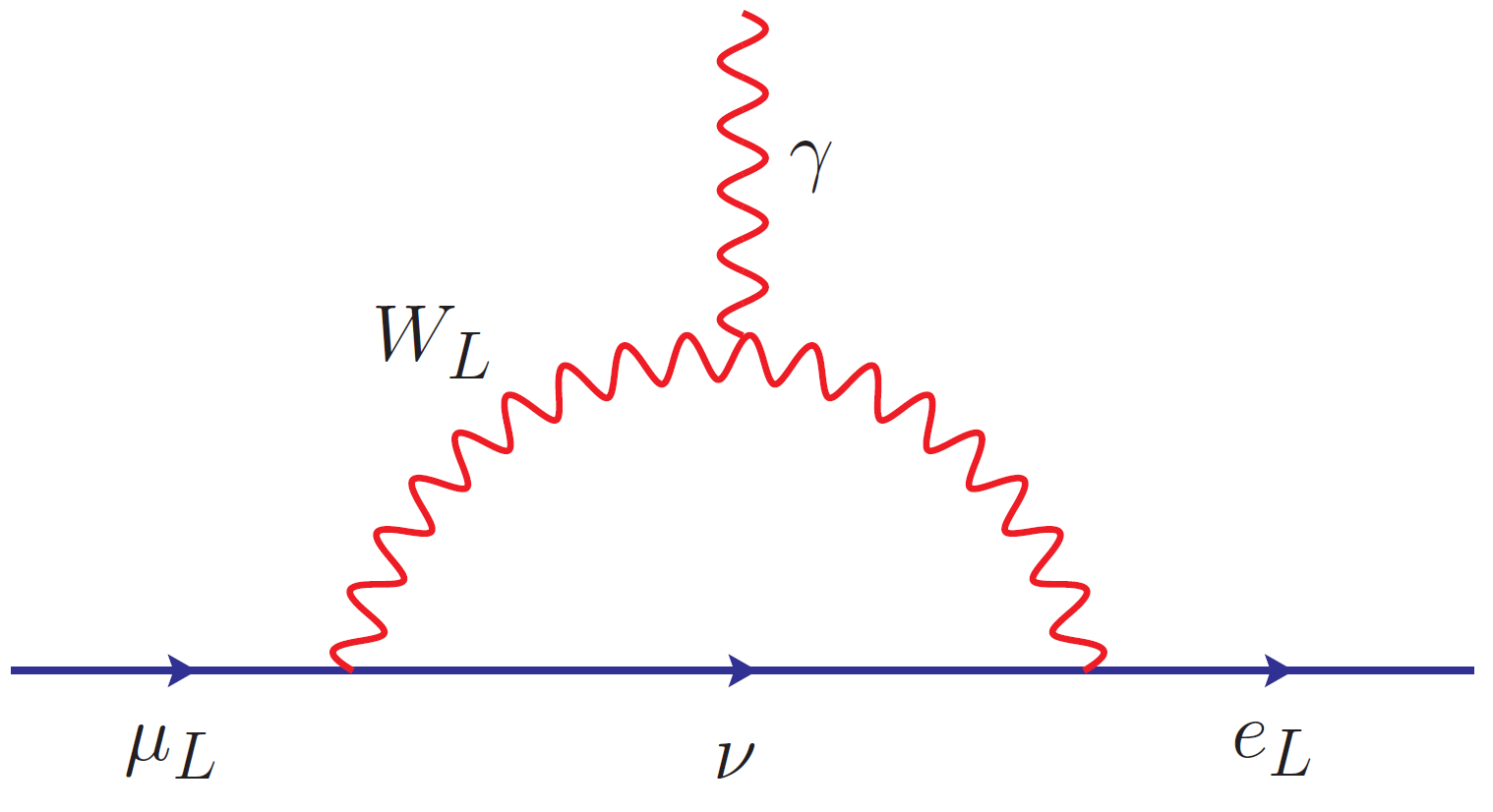} 
        \label{mutoegamma_WL} 
        \caption{One-loop diagram of $\mu\to e \gamma$ mediated by $W_L$. Keeping one of the vertices to be SM-like will eliminate $\nu_{\tau}$ from appearing in this process.}
    \end{figure}
Since the charged current couplings can be flavor-changing in this model, $\ell_1 \to \ell_2 \gamma$ processes can occur at one-loop level without a neutrino mass insertion in the internal fermion line. The gauge interaction relevant for the decay is
    \begin{equation}
    	\mathcal{L}=\frac{g_L}{\sqrt{2}}\overline{e}\gamma^\sigma W_{L\sigma}^-\nu_e+\frac{g_L}{\sqrt{2}}\overline{\mu}\gamma^\sigma W_{L\sigma}^-\nu_\mu +\frac{g_L}{\sqrt{2}}f_{e\nu_\mu}\overline{e}\gamma^\sigma W_{L\sigma}^-\nu_\mu + \frac{g_L}{\sqrt{2}}f_{\mu\nu_e}\overline{\mu}\gamma^\sigma W_{L\sigma}^-\nu_e+... + \text{h.c.}
    \end{equation}
and the decay rate is computed along the same lines as Sec.~\ref{l1tol2gamma}. The internal fermion mediators are considered to be SM neutrinos such that one of the vertices is a SM vertex. The constraints from these processes are listed in Table~\ref{l1l2gammacharged}.

    \begin{table}[]
    \footnotesize
    	{\renewcommand{\arraystretch}{1.5}%
    		\begin{center}
    			\begin{tabular}{|c|c|c|}
    				\hline
    				\textbf{Process} & \textbf{Exp. Bound} & \textbf{Constraint}\\
    				\hline
    				$\mu^-\rightarrow e^-\gamma$ & $<4.2\times 10^{-13}$ & $|f_{e\nu}+f_{\mu\nu}^*|< 2.59\times 10^{-6}$\\
    				\hline
    				$\tau^-\rightarrow e^-\gamma$ & $<3.3\times 10^{-8}$ &  $|f_{e\nu}+f_{\tau\nu}^*|< 1.73\times 10^{-3}$\\
    				\hline
    				$\tau^-\rightarrow \mu^-\gamma$ & $<4.2\times 10^{-8}$ & $|f_{\mu\nu}+f_{\tau\nu}^*|< 1.95\times 10^{-3}$\\
    				\hline
    			\end{tabular}
    			\caption{Constraints from $l_1 \to l_2 \gamma$ processes. The $f$'s are the new contributions to charged current interaction.}
    			\label{l1l2gammacharged}
    	\end{center}}
    \end{table}

\subsection{Lepton Flavour Universality Tests}

In this section, we have analyzed the constraints arising from various lepton flavor universality~(LFU) tests involving charged current. The ratio of leptonic decays of $W$ boson, predicted to be unity in the SM~\cite{Kniehl:2000rb}, can be a direct test of lepton universality. The theoretical uncertainties for these processes appear from the mass of the final state leptons, which are small compared to the experimental uncertainties, and the other theoretical parameters cancel out in the ratio. Using Eq.~\eqref{5.2}, the ratios of decay rates given in Table~\ref{lfu1} take the form $1-2(f_{\nu \ell_i}-f_{\nu \ell_j})$ to the first order in both the parameters, where, $i$ is lepton flavor in the numerator, and $j$ is the one in the denominator. This formulation is also valid for $\Gamma(\tau^-\to\mu^- \overline{\nu}_\mu \nu_\tau)/\Gamma(\tau^-\to e^- \overline{\nu}_e \nu_\tau)$ and, for $\Gamma(K\to\pi\mu \nu)/\Gamma(K\to\pi e \nu)$ given in Table~\ref{lfu2}, which, including the radiative corrections, is predicted to be $0.6591(31)$~\cite{Wanke:2007km} in SM. In computing the constraints for such process that have non-negligible radiative corrections, we adopt $\dfrac{\rm R_\text{SM+NP}}{\rm R_\text{SM}}\leq \dfrac{\rm R_\text{Exp}
}{\rm R_\text{SM}}$ where, ${\rm R}$ is the ratio of the branching fractions.
    \begin{table}[]
    \footnotesize
    	{\renewcommand{\arraystretch}{1.5}%
    		\begin{center}
    			\begin{tabular}{|c|c|c|}
    				\hline
    				\textbf{Process}
        &\textbf{Exp. Bound} & \textbf{Constraint} \\
    				\hline		
    				$\Gamma(W_L\to\mu \nu)/\Gamma(W_L\to e \nu)$ & $(1.002\pm 0.006)$ &$			
    				-0.007\leq f_{\nu \mu}-f_{\nu e}\leq 0.005$
    				\\
    				\hline
    				$\Gamma(W_L\to\tau \nu)/\Gamma(W_L\to e \nu)$ & $(1.015\pm 0.020)$ &$			
    				-0.027\leq f_{\nu \tau}-f_{\nu e}\leq 0.012$ \\
    				\hline
    				$\Gamma(W_L\to\tau \nu)/\Gamma(W_L\to\mu \nu)$  &$(1.002\pm 0.020)$ &$			
        				-0.021\leq f_{\nu \tau}-f_{\nu \mu}\leq 0.019$  \\
    				\hline
    	 			\end{tabular}
    			\caption{Lepton flavor universality violation in $W$ decays. The $f$'s are the new contributions to charged current interactions. A  $2\,\sigma$ deviation is  allowed in the experimental bound for obtaining the constraints.}
    		\label{lfu1}
    	\end{center}}
    \end{table}
    
Another set of LFU-violating constraints come from the purely leptonic decays of charged mesons, $\phi$. The ratios of such decays will be of the form
    \begin{equation}
    	{\rm R}=\frac{(m_i^3-m_i m_\phi^2)^2\left(1+\delta_{RC}-2(f_{\nu j}-f_{\nu i})\right)}{(m_j^3-m_\phi^2m_j)^2}
    \end{equation}
where, $i$ is lepton in the numerator, $j$ is the one in the denominator and $\delta_{RC}$ is the radiative correction~\cite{Cirigliano:2007xi,BESIII:2021anh,Bryman:2019bjg,Mescia:2007ku}. The constraints from these are given in Table~\ref{lfuphi}. Yet another set of constraints come from the three-body decays of charged leptons. The theoretical prediction to the lowest order for 
    \begin{equation}
    	\frac{\Gamma(\tau^-\to e^- \overline{\nu}_e \nu_\tau)}{\Gamma(\mu^-\to e^- \overline{\nu}_e \nu_\mu)}=\frac{m_\tau^2}{m_\mu^5}\left(1-2(f_{\nu\tau}-f_{\nu\mu})\right).
    \end{equation}
Finally, we also look at ratios of the form $\Gamma(\tau\to M \nu_\tau)/\Gamma(M \to \ell \nu_\ell)$ which are given by~\cite{deGouvea:2015euy}
    \begin{equation}
    	{\rm R}_{\tau/M}=\frac{\Gamma(\tau\to M \nu_\tau)}{\Gamma(M \to \ell \nu_\ell)}=\frac{1}{2}\frac{m_\tau^3}{m_Mm_\ell^2}\frac{(1-m_M^2/m_\tau^2)^2}{(1-m_\ell^2/m_M^2)^2}\left(1+\delta_{\ell}^M-2(f_{\nu\tau}-f_{\nu \ell})\right),
    \end{equation}
    where, $\delta_\ell^M$ represents radiative corrections~\cite{Decker:1993py}. The results obtained from these sets of LFU-violating decays are given in Table~\ref{lfu2}.
    \begin{table}[]
    \footnotesize
    	{\renewcommand{\arraystretch}{1.6}%
    		\begin{center}
    			\begin{tabular}{|c|c|c|c|}
    				\hline
    				\textbf{Process} &\textbf{SM} & \textbf{Exp. Bound} & \textbf{Constraint} \\
    				\hline
    				$\dfrac{\Gamma(K\to e \nu)}{\Gamma(K\to \mu \nu)}$ &$2.477\times 10^{-5}$& $(2.488\pm 0.009)\times 10^{-5}$ & $
    				-0.002\leq f_{\nu \mu}-f_{\nu e}\leq 0.006$\\
    				\hline  
    				$ \dfrac{\Gamma(\pi\to e \nu)}{\Gamma(\pi\to\mu \nu)}$ &$(1.2352\pm 0.0002)\times 10^{-4}$& $(1.2327\pm 0.0023)\times 10^{-4}$ & $
    				-0.003\leq f_{\nu \mu}-f_{\nu e}\leq 0.001$\\
    				\hline  
    				$ \dfrac{\Gamma(D_s\to \tau \nu)}{\Gamma(D_s\to\mu \nu)}$&9.75 & $(10.73\pm 0.69)$ & $
    				-0.12\leq f_{\nu \tau}-f_{\nu \mu}\leq 0.020$\\
    				\hline 
    				\end{tabular}
    				\caption{LFUV from leptonic decays of mesons. The $f$'s are the new contributions to charged current interaction. A  $2\,\sigma$ deviation is  allowed in the experimental bound and the central value of SM prediction is used for obtaining the constraints.}
    			\label{lfuphi}
    	\end{center}}
    \end{table}
    \begin{table}[]
    	{\renewcommand{\arraystretch}{2}%
    		\begin{center}
    			\begin{tabular}{|c|c|c|c|}
    				\hline
    				\textbf{Process} & \textbf{SM}&\textbf{Exp. Bound} & \textbf{Constraint}~($\times 10^{-3} $)\\
    				\hline		
    				$\dfrac{\Gamma(\tau^-\to\mu^- \overline{\nu}_\mu \nu_\tau)}{\Gamma(\tau^-\to e^- \overline{\nu}_e \nu_\tau)}$&$0.9726$ & $(0.9764\pm 0.0028)$ & $			
    				-4.73\leq f_{\nu \mu}-f_{\nu e}\leq 1.03$
    				\\
    				\hline
    				$\dfrac{\Gamma(\tau^-\to e^- \overline{\nu}_e \nu_\tau)}{\Gamma(\mu^-\to e^- \overline{\nu}_e \nu_\mu)}$ &$1.345\times 10^{6}$& $(1.349\pm 0.004)\times 10^6$ &$			
    				-4.42\leq f_{\nu \tau}-f_{\nu \mu}\leq 1.52$ \\
    				\hline
    				$\dfrac{\Gamma(\tau^+\to\pi^+ \nu_\tau)}{\Gamma(\pi^+\to\mu^+ \nu_\mu)}$ &9771& $(9704\pm 56)$ &$			
    				-3.07\leq f_{\nu \tau}-f_{\nu \mu}\leq 8.41$  \\
    				\hline
    				$\dfrac{\Gamma(\tau^+\to K^+ \nu_\tau)}{\Gamma(K^+\to e^+ \nu_e)}$& $1.94\times 10^7$& $(1.89\pm 0.03)\times 10^{7}$ &$			
    				-24.1\leq f_{\nu \tau}-f_{\nu e}\leq 8.13$  \\
    				\hline $\dfrac{\Gamma(\tau^+\to K^+ \nu_\tau)}{\Gamma(K^+\to \mu^+ \nu_\mu)}$&480 & $(469\pm 7)$ &$			
    				-7.33\leq f_{\nu \tau}-f_{\nu e}\leq 22.1$  \\
    				\hline 
        $ \dfrac{\Gamma(K\to\pi\mu \nu)}{\Gamma(K\to\pi e \nu)}$ &$0.6591\pm 0.0031$& $(0.6608\pm 0.0029)$ & $
    				-5.69\leq f_{\nu \mu}-f_{\nu e}\leq 3.11$\\
    				\hline 
    			\end{tabular}
    			\caption{Constraints from other LFUV processes. The $f$'s are the new contributions to charged current interaction.  A  $2\,\sigma$ deviation is  allowed in the experimental bound and the central value of SM prediction is used for obtaining the constraints.}
    		\label{lfu2}
    	\end{center}}
    \end{table}

\subsection{Mass Difference of Neutral Mesons\label{sec:boxWLR}}

\begin{figure}[]
    	\centering
    	\includegraphics[scale=0.3]{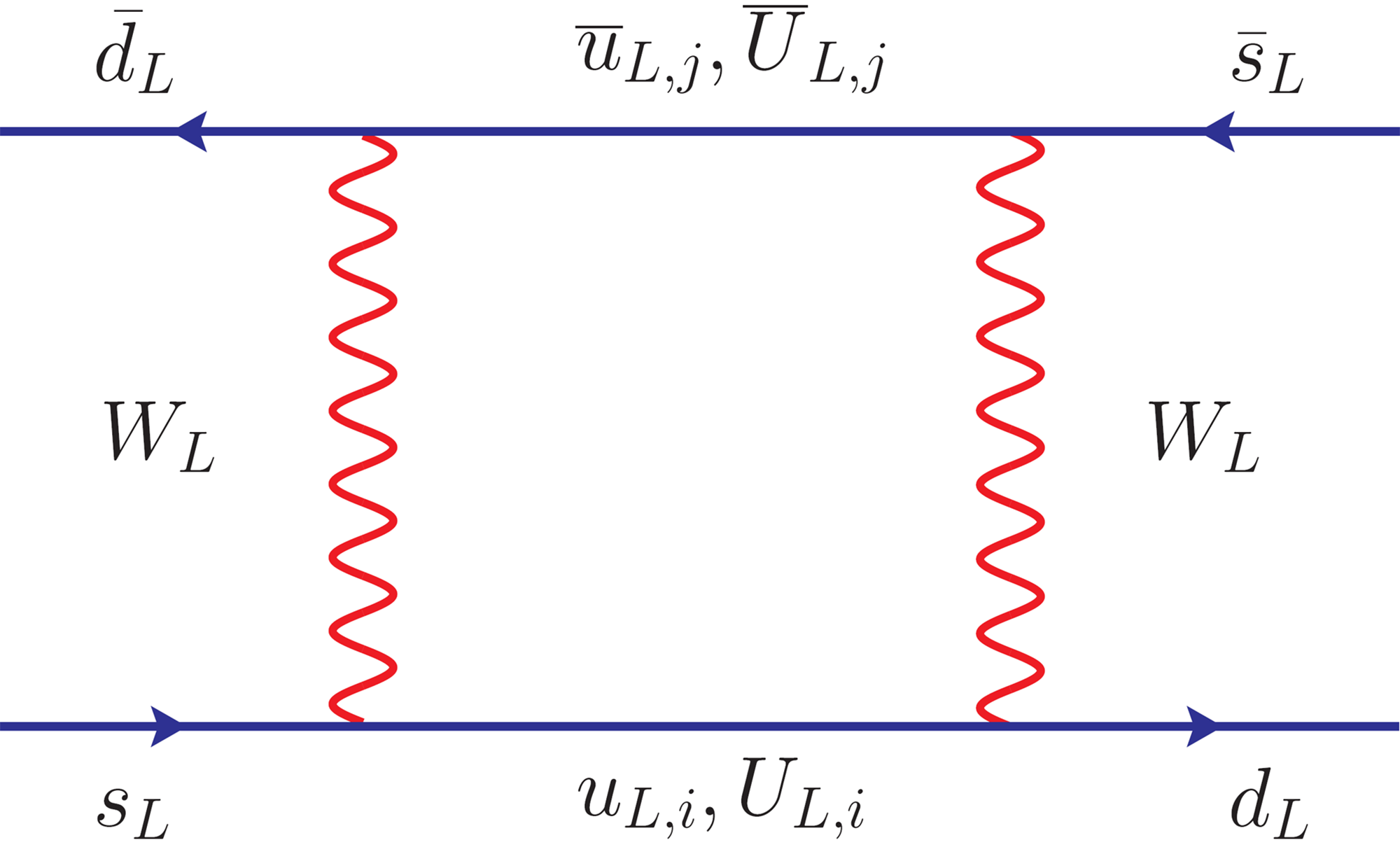}~~~~\includegraphics[scale=0.3]{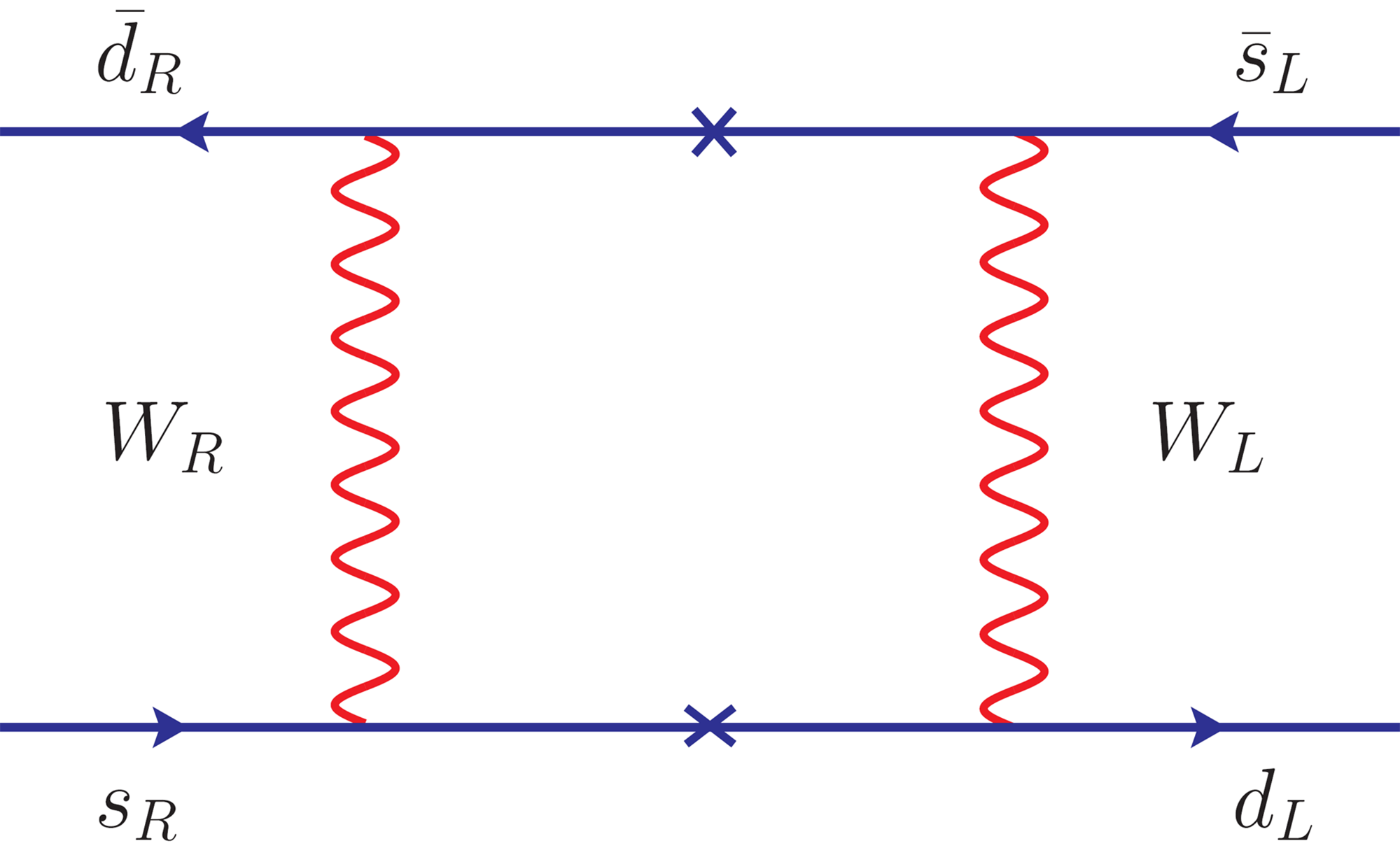}~~~~\includegraphics[scale=0.3]{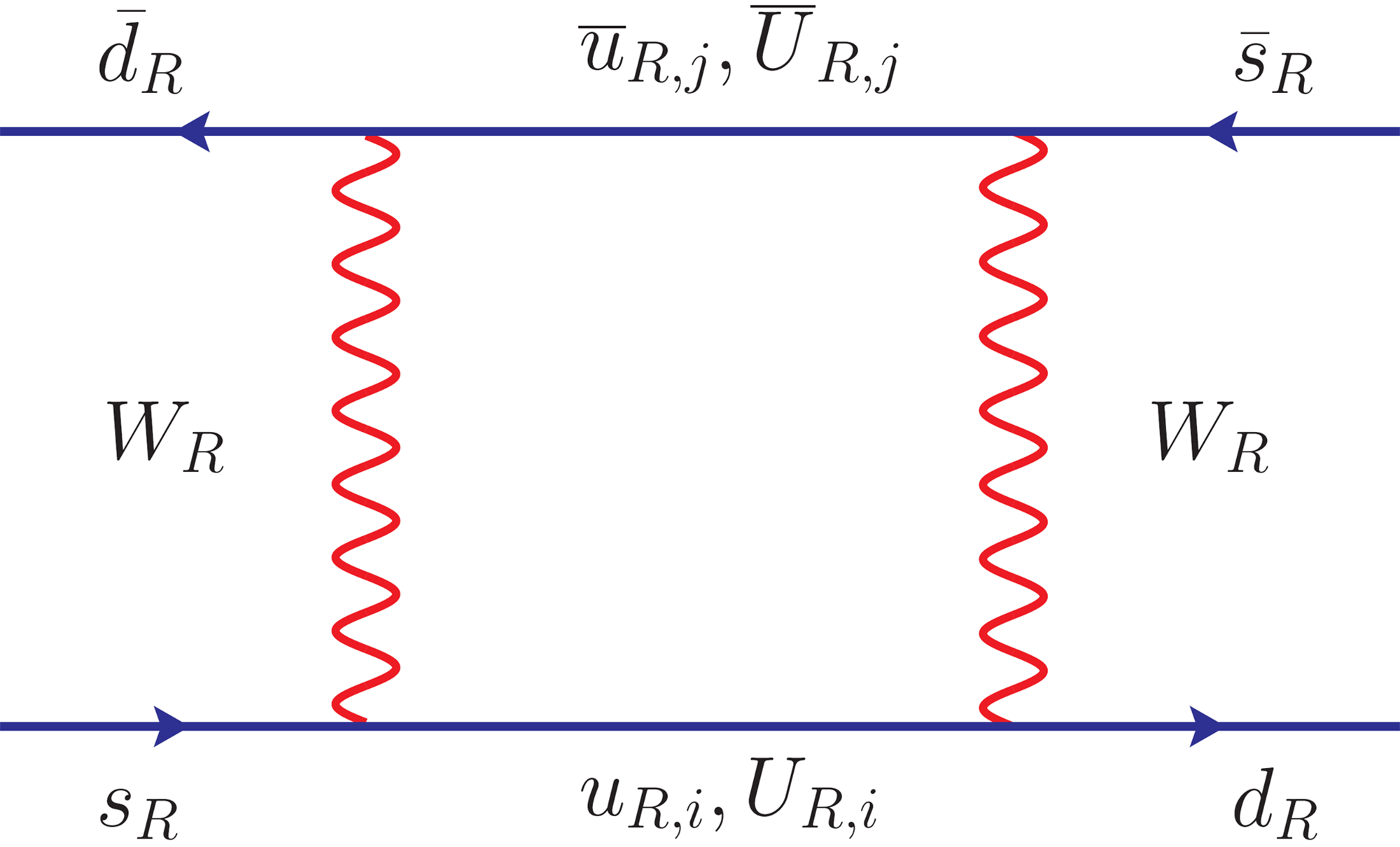}\\
    ~~~~~~~~(a)~~~~~~~~~~~~~~~~~~~~~~~~~~~~~~~~~~~~~~~~(b)~~~~~~~~~~~~~~~~~~~~~~~~~~~~~~~~~~~~~~~~~~(c)
    	\caption{Box diagrams contributing to kaon mixing. The $F$ in the internal line represents both SM and vector-like up-type quarks. The last diagram is qualitatively insignificant compared to the other two. }
    	\label{kaon_mixing}
    \end{figure}

In SM, neutral meson mixing occurs at one-loop level through the familiar box diagram involving  $W_L$ boson. The presence of $W_R$ boson and heavy vector-like quarks can have additional effects on these processes which can give significant constraints on the NP parameters including the mass of $W_R$. The diagrams contributing to kaon mixing are shown in Fig.~\ref{kaon_mixing}. We only consider the first two diagrams since the contribution from the one with two $W_R$'s is extremely small.

The $W_L-W_L$ box diagram contributes~\cite{Ecker:1985vv}
    \begin{equation}
        \mathcal{H}_{LL}=\frac{G_F^2M_{L}^2}{4\pi^2}(\bar{d}\gamma^\mu P_Ls)^2 \sum_{i,j}\lambda_i^{LL}\lambda_j^{LL}\big\{(1+\frac{x_ix_j}{4})I_2(x_i, x_j; 1)-2x_ix_jI_1(x_1,x_j;1)\big\}
    \end{equation}
while the contribution from $W_L-W_R$ diagram is
    \begin{equation}
        \mathcal{H}_{LR}=\frac{G_F^2M_{L}^2\beta}{2\pi^2}\bar{d} P_Ls\,\bar{d} P_Rs\sum_{i,j}\lambda_i^{LR}\lambda_j^{RL}\sqrt{x_ix_j}\big\{(4+\beta x_ix_j)I_1(x_i, x_j; \beta)-(1+\beta)I_2(x_1,x_j;\beta)\big\}
    \end{equation}
where,
    \begin{equation}
        \begin{aligned}
        x_i=m_i^2/M_{L}^2, &\qquad\qquad\beta=M_{L}^2/M_{R}^2\\
        I_1(x_i,x_j;\beta)=&\frac{x_i\ln{x_i}}{(1-x_i)(1-x_i\beta)(x_i-x_j)}+(i\leftrightarrow j)-\frac{\beta\ln{\beta}}{(1-\beta)(1-x_i\beta)(1-x_j\beta)},\\
        I_2(x_i,x_j;\beta)=&\frac{x_i^2\ln{x_i}}{(1-x_i)(1-x_i\beta)(x_i-x_j)}+(i\leftrightarrow j)-\frac{\ln{\beta}}{(1-\beta)(1-x_i\beta)(1-x_j\beta)}.
        \end{aligned}
    \end{equation}
In the above equations, we need to implement the GIM cancellation as well as the simplification $m_u\rightarrow 0$, and $m_c^2/M_{L,R}^2$ is kept to the first order \cite{Mohapatra:1983ae}. The expressions for $\lambda_{i,j}$ are given in Appendix.~\ref{app3}. Among the meson mixing processes, $K-\overline{K}$ gave significant constraints on the NP parameters. The theoretical bound on the mass of $W_R$ boson was found to be $M_{W_R}\gtrsim  2.4 \text{ TeV}$. The indirect $\mathcal{CP}$ violation in the meson system could improve this bound; however, it depends heavily on the complex phase associated with the right-handed quark mixing matrix under parity symmetry. The correction to $\epsilon_K$~\cite{Craig:2020bnv}, compared to the theoretical uncertainty of the SM prediction~\cite{Nierste:2017kxd}, yields a conservative limit of $M_{W_R} \gtrsim 10.8$ TeV.
\section{Constraints on Higgs Couplings\label{sec:constrainthiggs}}

In this section, we tabulate the constraints on the 
Higgs couplings to fermions.  Since there is a mixing between the light Higgs $h$ and heavy Higgs $H$, the parts of the Lagrangian given in Eq.~\eqref{eq:higgs1} and Eq.~\eqref{eq:higgs2} are relevant for the processes considered here. Similar to the parity symmetric approximation done in the case of $Z_i$ couplings, the new contributions to Higgs couplings may be written in terms of 
    \begin{equation}
     \left.\begin{aligned}
            \mathcal{R}&=V_{L_f}\rho_{L_F}^\dagger\rho_{L_F}V_{L_f}^\dagger m_f,\\
        \mathcal{R}'&=V_{R_f}\rho_{R_F}^\dagger\rho_{_RF}V_{R_f}^\dagger m^\dagger_f,
           \end{aligned}
     \right\}
     \qquad \mathcal{R}'=\frac{\kappa_R^2}{\kappa_L^2}\mathcal{R}.
    \end{equation}
Using this, the Higgs interactions to charged fermions are
    \begin{equation}
    	\begin{aligned}
    		\mathcal{L}_{h}&=\overline{f} h \left(\left\{ \frac{\cos\zeta}{\kappa_L}K_1-\frac{\sin\zeta}{\kappa_R}K_2^\dagger\right\}P_R+\left\{\frac{\cos\zeta}{\kappa_L}K_1^\dagger-\frac{\sin\zeta}{\kappa_R}K_2\right\}P_L\right) f+...,\\
    		\mathcal{L}_{H}&=\overline{f} H \left(\left\{  \frac{\sin\zeta}{\kappa_L}K_1+\frac{\cos\zeta}{\kappa_R}K_2^\dagger\right\}P_R+\left\{  \frac{\sin\zeta}{\kappa_L}K_1^\dagger+\frac{\cos\zeta}{\kappa_R}K_2\right\}P_L\right) f+...,\\
    	\end{aligned}
    \end{equation}
with,
    \begin{equation}
        \begin{aligned}
            K_1=\Bigl(m_f-\frac{1}{2}\mathcal{R}\Bigr), &&
            K_2=\Bigl(m_f-\frac{1}{2}\frac{\kappa_R^2}{\kappa_L^2} \mathcal{R}\Bigr).
        \end{aligned}
    \end{equation}
The mixing angle is set to zero, with the mass of the heavy Higgs being $\simeq 6.6 \text{ TeV}$ when $M_{Z_2}=5\text{ TeV}$. The first set of constraints comes from the expected deviation in the charged leptonic decay of Higgs from SM,
$\mu_{\ell^+\ell^-}=BR(h \to \ell^+\ell^-)/BR(h \to \ell^+\ell^-)_\text{SM}$ with
    \begin{equation}
        \Gamma(h \to \ell^+\ell^-)=\frac{\sqrt{m_h^2-4m_\ell^2}}{16\pi m_h^2}\Bigg(\left(|C_{L_h}|^2+|C_{R_h}|^2\right)(m_h^2-2m_\ell^2)-2m_\ell^2\left(C_{L_h}C_{R_h}^*+C_{L_h}^*C_{R_h}\right)\Bigg),
    \end{equation}
    and the SM contribution being
\begin{equation}
    \Gamma(h \to \ell^+\ell^-)_\text{SM}=\dfrac{m_h m_\ell^2}{8\pi\kappa_L^2}\Bigl(1-\dfrac{4m_\ell^2}{m_h^2}\Bigr)^{3/2},
\end{equation} 
where, the coefficient of $P_{L(R)}$ is $C_{L(R)}$. The constraints from various expected results are tabulated in Table~\ref{tab:htollnew}.
    \begin{table}[]
    \footnotesize
    	{\renewcommand{\arraystretch}{1.5}%
    		\begin{center}
    			\begin{tabular}{|c|c|c|}
    				\hline
    				\textbf{Process} & \textbf{Exp. Bound} & \textbf{Constraint} \\
    				\hline
   $ \mu_{\mu^+\mu^-}$&$1.21\pm 0.35$&$|(V_{L_l}\mathcal{Y}_E\mathcal{Y}_E^\dagger V_{L_l}^\dagger)_{\mu\mu}|\leq18.86~\left(\frac{M_L}{1\text{ TeV}}\right)^2$\\
    				\hline 
        $\mu_{\tau^+\tau^-}$&$0.91\pm0.09$&$|(V_{L_l}\mathcal{Y}_E\mathcal{Y}_E^\dagger V_{L_l}^\dagger)_{\tau\tau}|\leq9.61~\left(\frac{M_L}{1\text{ TeV}}\right)^2$\\
    				\hline
    			\end{tabular}
    			\caption{Constraint from SM-like Higgs decay to charged leptons. The results are quoted as a function of vector-like lepton mass $M_L$ (TeV).}
       \label{tab:htollnew}
    	\end{center}
     }
    \end{table}
Since the SM-like Higgs can also induce flavor change, we study the decay of the top quark to the up-type quarks and Higgs. The decay rate is given by
    \begin{equation}
        \begin{aligned}
            \Gamma(t\to h q)&=\frac{\sqrt{m_h^4-2m_h^2(m_q^2+m_t^2)+(m_q^2-m_t^2)^2}}{32\pi m_t^3}\times\\
            &\Bigg(\left(|C_{L_h}|^2+|C_{R_h}|^2\right)(m_t^2+m_q^2-m_h^2)+2m_qm_t\left(C_{L_h}C_{R_h}^*+C_{L_h}^*C_{R_h}\right)\Bigg).
        \end{aligned}
    \end{equation}
The constraints arising from the top decays are quoted in Table~\ref{tab:topdecays}.

    \begin{table}[]
    \footnotesize
    {\renewcommand{\arraystretch}{1.5}
    \begin{center}
        \begin{tabular}{|c|c|c|c|}
        \hline
        \textbf{Process} & \textbf{Exp. Bound} & \textbf{Constraint}\\
        \hline
        $t \to h c$ & $<7.3\times 10^{-4}$ & $|(V_{L_f}\mathcal{Y}_F\mathcal{Y}_F^\dagger V_{L_f}^\dagger)_{ct}|<2.39\,\left(\frac{M_F}{1 \text{TeV}}\right)^2$\\
        \hline
         $t\to h u$ & $<1.9\times 10^{-4}$ & $|(V_{L_f}\mathcal{Y}_F\mathcal{Y}_F^\dagger V_{L_f}^\dagger)_{ut}|<1.24\,\left(\frac{M_F}{1 \text{TeV}}\right)^2$\\
        \hline
        \end{tabular}
    \caption{Constrains from top decaying to Higgs and up-type quark. The results are quoted as a function of vector-like quark mass $M_F$ (TeV). }
    \label{tab:topdecays}
    \end{center}}
    \end{table}
\begin{figure}
    \centering
    \includegraphics[scale=0.5]{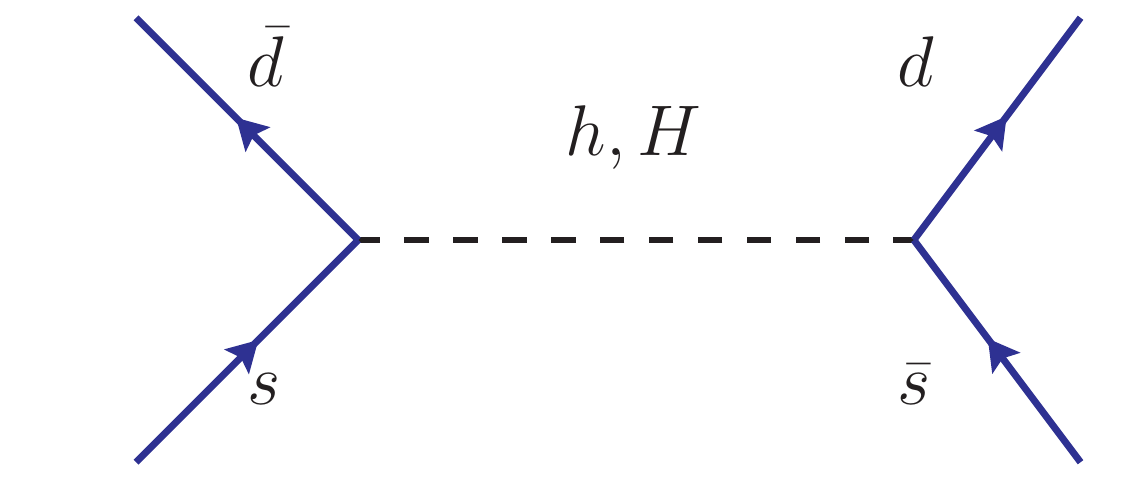}
    \caption{Tree-level diagram of scalar mediated $K-\overline{K}$ mixing.}
    \label{fig:kaon_higgs}
\end{figure}

Another set of constraints comes from the tree-level meson mixing processes mediated by $\{h,H\}$ as shown in Fig.~\ref{fig:kaon_higgs}. The major constraints arise from $K-\overline{K}$ and $B-\overline{B}$ mixing mass difference. The effective Hamiltonian leading to meson mixing may be written as
    \begin{equation}
        \mathcal{H}_\text{eff}=\sum_{k=h,H}\Bigg(\frac{C_{L_k}^2(\Lambda)}{2m_k^2}\mathcal{O}_2 +\frac{C_{R_k}^2(\Lambda)}{2m_k^2}\widetilde{\mathcal{O}}_2 + \frac{C_{L_k}(\Lambda)C_{R_k}(\Lambda)}{m_k^2}\mathcal{O}_4\Bigg),
    \end{equation}
with reference to the operators given in Sec.~\ref{sec:mass diff}. Both the quantities in Table~\ref{KandB} are computed using magic numbers given in Appendix.~\ref{app1}. The $\Delta M$'s obtained this way were constrained using the central value of the experimental results. In the case of $B-\overline{B}$, we also included an analysis with $\Delta M^\text{SM+NP}/\Delta M^\text{SM}$, since there is an attempt to reduce theoretical uncertainties of the SM prediction of $\Delta M_B$~\cite{DiLuzio:2019jyq}. This gives a more stringent constraint quoted in parenthesis in Table~\ref{KandB}.

    \begin{table}[]
    \footnotesize
    {\renewcommand{\arraystretch}{1.5}%
    \begin{center}
        \begin{tabular}{|c|c|c|c|}
            \hline
            \textbf{Process} & \textbf{Exp. Bound (GeV)} & \textbf{Constraint}\\
            \hline
            $K-\overline{K}$ & $(3.484\pm 0.006) \times 10^{-15}$& $ |(V_{L_f}\mathcal{Y}_F\mathcal{Y}_F^\dagger V_{L_f}^\dagger)_{ds}|\leq 1.06(\frac{M_F}{1 \text{TeV}})^2$\\
            \hline
            $B-\overline{B}$ &$(3.334\pm0.013)\times 10^{-13} $ & $ |(V_{L_f}\mathcal{Y}_F\mathcal{Y}_F^\dagger V_{L_f}^\dagger)_{db}|\leq 0.214(\frac{M_F}{1 \text{TeV}})^2 \left(4.21\times 10^{-2}(\frac{M_F}{1 \text{TeV}})^2\right)$\\
            				\hline
        \end{tabular}
    \caption{Constraints from mass differences of neutral mesons. The results are quoted as a function of vector-like quark mass $M_F$(TeV). The NP contributions were constrained against the central value of experimental results. The constraint given in the parenthesis is obtained using $\Delta M^\text{SM+NP}/\Delta M^\text{SM}$.}
    \label{KandB}
    \end{center}}
    \end{table}

\section{Neutral Current B-anomalies \label{sec:Banom}}

The experimental results on several $b \to s \mu^+ \mu^-$ processes showed\footnote{The recent LHCb measurement is consistent with the SM predictions~\cite{LHCb:2022qnv}.} significant deviations from the Standard Model predictions. Over the years the discrepancies have been observed in branching ratios of $B\to K^{(*)} \mu^+\mu^-$, $B_s\to \phi \mu^+\mu^-$, and $B_s\to \mu^+\mu^-$, and in the angular distribution of $B\to K^*\mu^+\mu^-$. In particular, there was a combined $3.1~\sigma$ discrepancy with the measurements of the lepton flavor universality~(LFU) ratio $R_{K^{(*)}}=\Gamma(B\to K^{(*)}\mu^+\mu^-)/\Gamma(B\to K^{(*)}e^+e^-)$. The SM calculations of these observables are devoid of any hadronic uncertainties as they cancel out in the ratio leading to clean and highly accurate predictions~\cite{Bordone:2016gaq}:
    \begin{equation}
        \begin{aligned}
        R_K^{SM}&=1.0004^{+0.0008 }_{-0.0007},\\
        R_{K^*}^{SM}&=\begin{cases}
        0.906\pm 0.028 \quad q^2\in [0.045, 1.1]~\mathrm{GeV}^2,\\
        1.00\pm 0.01 \quad\quad q^2\in [1.1, 6]~\mathrm{GeV}^2,
        \end{cases}
        \end{aligned}
    \end{equation}
    where, $q$ denotes the dilepton mass. The most precise measurement of these ratios  so far has been by the LHCb:
    \begin{equation}
        \begin{aligned}
            R_K&=0.846^{+0.042 +0.013}_{-0.039 -0.012}\quad q^2\in [1.1, 6]~\mbox{GeV}^2,~\text{\cite{LHCb:2021trn}}\\
            R_{K^*}&=\begin{cases}
            0.66^{+0.11}_{-0.07}\pm 0.024 \quad q^2\in [0.045, 1.1]~\mathrm{GeV}^2,\\
            0.69^{+0.113}_{-0.069}\pm 0.047 \quad q^2\in [1.1, 6]~\mathrm{GeV}^2.\\ 
            \end{cases}~\text{\cite{Aaij:2017vbb}}\\ 
        \end{aligned}
    \end{equation}
$R_{K^*}$ measurements have also been done by Belle~\cite{Abdesselam:2019wac}
    \begin{equation}
        \begin{aligned}
            R_{K^*}&=\begin{cases}
            0.52^{+0.36}_{-0.26}\pm 0.05 \quad q^2\in [0.045, 1.1]~\mathrm{GeV}^2,\\
            0.96^{+0.45}_{-0.29}\pm 0.11 \quad q^2\in [1.1, 6]~\mathrm{GeV}^2,\\
            0.90^{+0.27}_{-0.21}\pm 0.10 \quad q^2\in [0.1, 8]~\mathrm{GeV}^2,\\ 
            1.18^{+0.52}_{-0.32}\pm 0.10 \quad q^2\in [15, 19]~\mathrm{GeV}^2,\\
            0.94^{+0.17}_{-0.14}\pm 0.08 \quad q^2\in [0.045, ]~\mathrm{GeV}^2,\\
            \end{cases}
        \end{aligned}
    \end{equation}
but with considerably larger errors compared to LHCb.  These deviations of experimental measurements from the SM predictions, referred to as the neutral current B-anomalies, could be clear signals of new physics. In this section, we explore the contribution to these processes from the LRSM with universal seesaw in the phenomenologically interesting parity symmetric version.
   
The contributions to neutral current B-anomalies in this model arise from the Lagrangian
\begin{equation}
\mathcal{L}_{Z_i}=\sum_{i=1}^{2}\frac{1}{2}M_{Z_i}^2(Z_{i_\mu})^2+\left(C_{L_i}^{bs}(\overline{b}_L\gamma^\mu s_L)+C_{R_i}^{bs}(\overline{b}_R\gamma^\mu s_R)  +C_{L_i}^{\ell\ell}(\overline{\ell}_L\gamma^\mu \ell_L)+C_{R_i}^{\ell\ell}(\overline{\ell}_R\gamma^\mu \ell_R)\right)Z_{i_\mu},
\end{equation}
where $\ell=\{e,\mu\}$. Integrating out $Z_{1,2}$ at tree-level gives the effective Lagrangian:
    \begin{equation}
        \begin{aligned}
            \mathcal{L}_\text{eff}=&-\sum_{i=1}^{2}\frac{1}{M_{Z_i}^2}\Big[C_{L_i}^{bs}C_{L_i}^{\ell\ell}(\overline{b}_L\gamma^\mu s_L) (\overline{\ell}_L\gamma_\mu \ell_L)+C_{L_i}^{bs}C_{R_i}^{\ell\ell}(\overline{b}_L\gamma^\mu s_L) (\overline{\ell}_R\gamma_\mu \ell_R)\\ 
            &+C_{R_i}^{bs}C_{L_i}^{\ell\ell}(\overline{b}_R\gamma^\mu s_R) (\overline{\ell}_L\gamma_\mu \ell_L)+C_{R_i}^{bs}C_{R_i}^{\ell\ell}(\overline{b}_R\gamma^\mu s_R) (\overline{\ell}_R\gamma_\mu \ell_R)\\
            &+\frac{1}{2}(C^{bs}_{L_i})^2(\overline{b}_{L}\gamma_\mu s_{L})^2 +\frac{1}{2}(C^{bs}_{R_i})^2(\overline{b}_{R}\gamma_\mu s_{R})^2+C^{bs}_{L_i} C^{bs}_{R_i}(\overline{b}_{L}\gamma_\mu s_{L})( \overline{b}_{R}\gamma^\mu s_{R})\Big].
        \end{aligned} \label{Leff}
    \end{equation}
Compared to Eqns.~\eqref{eq:LZ1}, ~\eqref{eq:LZ2} and Appendix.~\ref{app:Lag}, certain changes have been made to keep the notations simple. Firstly, the factor $g_L/\cos\theta_W$ has been absorbed into the coefficients $C_{{L,R}_i}$. Furthermore, $C^{bs}_{{L,R}_i}\equiv\widetilde{C}^{bs}_{{L,R}_i}$ while $C^{\ell\ell}_{{L,R}_i}\equiv \widetilde{C}^{\ell\ell}_{{L,R}_i}$ defined in Appendix.~\ref{app:Lag}. From the two expressions above, it is evident that the couplings relevant to resolving neutral current B-anomalies also contribute to $B_s-\overline{B}_s$ mixing, $B_s\to\mu^+\mu^-$, as well as $Z_i$ decays, whose SM predictions and experimental values are given in Table~\ref{tab:Banom_observables}. Therefore, the main constraints arise from 
\begin{itemize}
   \parskip=0.5em
        \item The $B_s-\overline{B_s}$ mass difference: $\Delta M_s^\text{NP+SM}/\Delta M_s^\text{SM} $,
        \item Lepton flavour universality violation of $Z$ decays: $\Gamma(Z\to \mu^+\mu^-)/\Gamma(Z\to e^+ e^-)$, and the individual branching ratios.
            \end{itemize}
One would also need to consider the following observables in reconciling the B-anomalies:
 \begin{itemize}
   \parskip=0.5em
        \item Muonic decay of $B_s$ meson: $\text{BR}(B_s\to \mu^+ \mu^-)^\text{NP+SM}/\text{BR}(B_s\to\mu^+ \mu^-)^\text{SM}$,
        \item Mixing induced $\mathcal{CP}$ asymmetry given by $A_\mathcal{CP}^\text{mix}(B_s\rightarrow J/\psi \phi)=\sin(\phi_\Delta-2\beta_s)$ \cite{Lenz:2006hd} with the value $-0.021\pm 0.031$~ \cite{HFLAV:2019otj}, where,  $\phi_\Delta$ is defined as $\arg\Bigg(\dfrac{\Delta M_s^\text{NP+SM}}{\Delta M_s^\text{SM}}\Bigg)$ with $\beta_s=0.01843^{+0.00048}_{-0.00034}$~\cite{Charles:2004jd}.
    \end{itemize}

    \begin{table}[]
    \footnotesize
    	{\renewcommand{\arraystretch}{1.5}%
    		\begin{center}
    			\begin{tabular}{|c|c|c|}
    				\hline
    				\textbf{Observable}&\textbf{SM prediction}&\textbf{Experimental Value}\\
    				\hline
    			 $\ddfrac{\Gamma(Z\to \mu^+\mu^-)}{\Gamma(Z\to e^+ e^-)}$&$\simeq 1$&$1.0001\pm 0.0024$.\\
    				 \hline
         $\Delta M_s$&$(18.77\pm 0.86) \text{ps}^{-1}$&$(17.749\pm 0.020) \text{ps}^{-1}$\\
    				   \hline
    				  $BR(Z\to e^+e^-)$&$(3.3663\pm 0.0012)\%$&$(3.3632\pm 0.0042)\%$\\
    				  \hline
    				   $BR(Z\to \mu^+\mu^-)$&$(3.3663\pm 0.0012)\%$&$(3.3662\pm 0.0066)\%$\\
    				   \hline
    				$BR(B_s\to\mu^+\mu^-)$&$(3.65\pm 0.29)\times 10^{-9}$~\cite{Bobeth:2013uxa}&$(3.09^{+0.46~+0.15}_{-0.43~-0.11})\times 10^{-9}$~\cite{LHCb:2021awg,LHCb:2021vsc}\\
    				   \hline
    			\end{tabular}
    		\caption{Observables constraining the model parameters that contribute to resolving neutral current B-anomalies.}\label{tab:Banom_observables}
    	\end{center}}
    \end{table}
Assuming the couplings are real, the constraint  from $B_s-\overline{B}_s$ mixing, allowing $2\,\sigma$ deviation, is 
\begin{equation}
   \frac{ \Delta M_s^\text{NP+SM}}{\Delta M_s^\text{SM}}\leq \frac{\Delta M_s^\text{exp}}{\Delta M_s^\text{SM}}\equiv 0.95\pm 0.04\Rightarrow|R_{bs}|\leq 1.41\times 10^{-5}.\label{eq:cons1}
\end{equation}
The LFU violation of Z decays can arise due to NP in either electron- or muon-sector, or both. For simplicity, we assume that the NP appears only in one of these sectors at a time leading to the following constraints, allowing $2\,\sigma$ deviation:
\begin{equation}
   \frac{ \Gamma(Z\to \mu^+\mu^-)}{\Gamma(Z\to e^+ e^-)} \Rightarrow \begin{cases}
        |R_{\mu\mu}|\leq 1.74\times 10^{-3},\\
        |R_{ee}|\leq 1.81\times 10^{-3}.
    \end{cases}\label{eq:cons2}
\end{equation}
From these constraints, the largest allowed values of the conventional Wilson coefficients are tabulated in Table~\ref{tab:WC} along with the respective LFU ratios. The NP WCs in terms of the model parameters are~\cite{Altmannshofer:2021qrr} 
\begin{equation}
    \begin{aligned}
        C_{9\,(10)}^{\ell}&=\mp\frac{4\sqrt{2}\pi}{8G_F \alpha_\text{em}V_{tb}V_{ts^*}}\sum_{i=1}^2\left(\frac{1}{M_{Z_i}^2}C_{L_i}^{bs}\left(C_{L_i}^{\ell\ell}\pm C_{R_i}^{\ell\ell}\right)\right),\\
               C_{9\,(10)}^{'\ell}&=\mp\frac{4\sqrt{2}\pi}{8G_F \alpha_\text{em}V_{tb}V_{ts^*}}\sum_{i=1}^2\left(\frac{1}{M_{Z_i}^2}C_{R_i}^{bs}\left(C_{L_i}^{\ell\ell}\pm C_{R_i}^{\ell\ell}\right)\right).
             \end{aligned}\label{eq:WC}
    \end{equation}    
We see that the NP in electron-sector can improve the theoretical prediction of $R_{K^{(*)}}$ slightly due to the large right-handed current~($C_9^{'e}=C_{10}^{'e}$)~(see also Refs.~\cite{Geng:2017svp, DAmico:2017mtc,Datta:2019zca}) contribution in the denominator. This, however, is not sufficient to explain all the $b\to s\mu^+\mu^-$ related anomalies which points towards NP in the muonic sector. The NP in the muon-sector in this model worsens the $R_{K^{(*)}}$ prediction due to the large right-handed current $C_9^{'\mu}=C_{10}^{'\mu}$ appearing in the numerator, although it can explain the observed $\text{BR}(B_s\to \mu^+\mu^-)$.   
\begin{figure}
    \centering
    \includegraphics[scale=0.5]{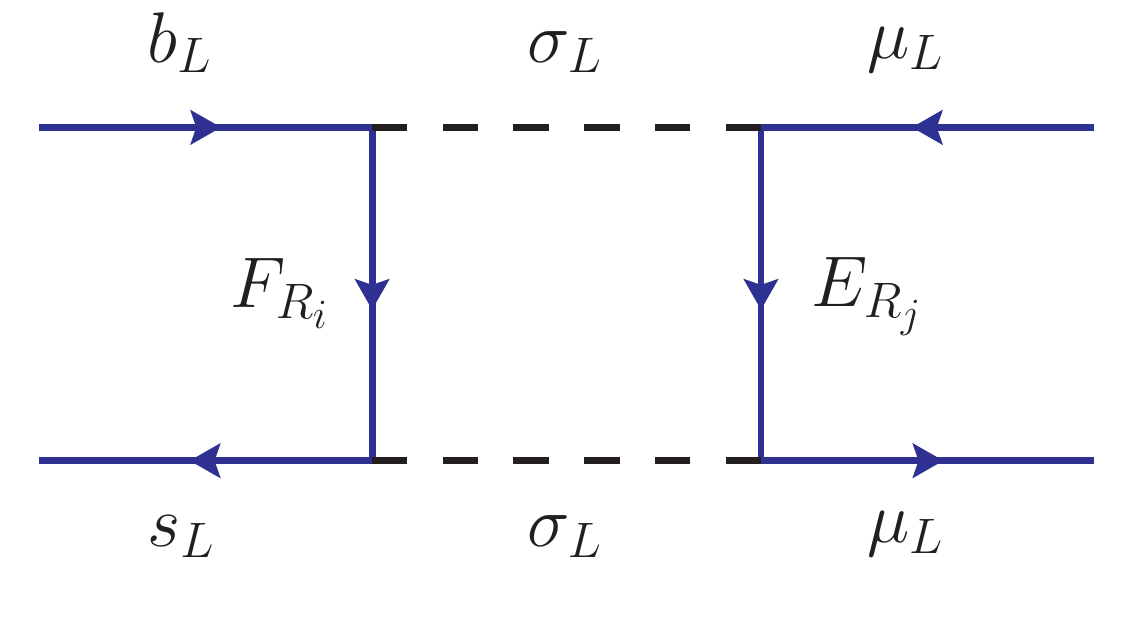}~~~~\includegraphics[scale=0.5]{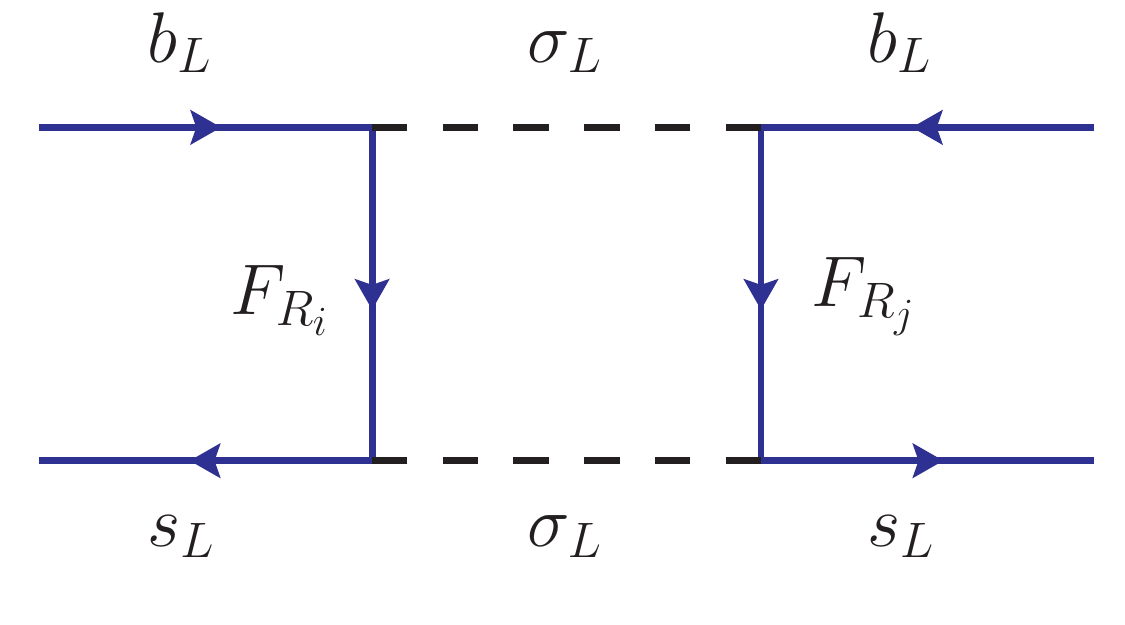}\\
    ~~(a)~~~~~~~~~~~~~~~~~~~~~~~~~~~~~~~~~~~~~~~~~~~~~(b)
    \caption{Box diagram contributing to $C_9$ from $\sigma_L$ mediator~(a) and $B_s-\overline{B}_s$ mixing arising from the same mediator~(b). $F$ represents VL-down type quark whereas $E$ represents VL-lepton.}
    \label{fig:C9box}
\end{figure}    
\begin{table}[]
\footnotesize
    	{\renewcommand{\arraystretch}{0.9}%
    		\begin{center}
\begin{tabular}{|c|c|c|}
\hline
     \textbf{Wilson coefficient}   & \textbf{Value} & \textbf{Observable}\\
     \hline
     \hline
\multirow{3}{*}{ $C_9^e=-C_{10}^e$} & \multirow{3}{*}{ $-2.74 \times 10^{-4}$} & \multirow{2}{*}{$R_K=0.95$} \\
 &  &  \\ \cline{3-3} 
 &  & \multirow{2}{*}{$R_{K^*}=0.96$} \\ \cline{1-2}
\multirow{3}{*}{ $C_9^{'e}=C_{10}^{'e}$} & \multirow{3}{*}{ $-9.03 \times 10^{-1}$} &  \\ \cline{3-3} 
 &  & \multirow{2}{*}{$\text{BR}(B_s\to \mu^+ \mu^-)=3.65\times 10^{-9}$} \\
 &  &  \\ \hline\hline
\multirow{3}{*}{$C_9^\mu=-C_{10}^\mu$} & \multirow{3}{*}{$-2.63 \times 10^{-4}$} & \multirow{2}{*}{$R_K=1.04$} \\
 &  &  \\ \cline{3-3} 
 &  & \multirow{2}{*}{$R_{K^*}=1.03$} \\ \cline{1-2}
\multirow{3}{*}{$C_9^{'\mu}=C_{10}^{'\mu}$} & \multirow{3}{*}{$-8.68\times 10^{-1}$} &  \\ \cline{3-3} 
 &  & \multirow{2}{*}{$\text{BR}(B_s\to \mu^+ \mu^-)=2.35\times 10^{-9}$} \\
  &  &  \\ \hline
\end{tabular}\caption{Maximum values of NP Wilson coefficients~(Eq.~\eqref{eq:WC}) allowed by the model parameters consistent with $B_s-\overline{B}_s$ mixing~(Eq.~\eqref{eq:cons1}) and LFUV of $Z$ decays~(Eq.~\eqref{eq:cons2}) and the corresponding extremum values of the observable. Note that NP from only one of the electron or muon sectors is turned on at a time.}
    \label{tab:WC}
    \end{center}}
\end{table}
Since the global fit to the neutral current B-anomalies alludes to a muon-specific left-handed current~\cite{Altmannshofer:2021qrr}, box diagram contribution mediated by left-handed scalar field $\sigma_L$~(assuming no mixing with $\sigma_R$ such that $(m_{\sigma_L}=m_h)\ll (m_{\sigma_R}=m_H)$), as shown in Fig.~\ref{fig:C9box}~(a), leading to $C_9^\mu=-C_{10}^\mu$ was also explored. The NP WC from the box diagram is 
\begin{equation}
    C_9^\mu=-C_{10}^\mu=-\frac{\sqrt{2}}{128\pi G_F \alpha_\text{em}} \frac{1}{V_{tb}V_{ts}^*}\frac{|y_\mu|^2y_sy^*_b}{4m_{\sigma_L}^2}\mathcal{F}(x_i,x_j),
\end{equation}
where, 
\begin{equation}
    \mathcal{F}(x_i,x_j)=\frac{x_i^2\ln{x_i}}{(x_i-1)^2(x_i-x_j)}-\frac{x_j^2\ln{x_j}}{(x_j-1)^2(x_i-x_j)}+\frac{1}{(x_j-1)(x_i-1)},
\end{equation}
with $x_i=\dfrac{m_{F_i}}{m_{\sigma_L}}$ and $x_j=\dfrac{m_{E_j}}{m_{\sigma_L}}$. Here, $y_\mu$, $y_s$ and $y_b$ are Yukawa couplings of $\mu_L$, $s_L$ and $b_L$, respectively, to $\sigma_L$. The coupling of quarks $b$ and $s$ to $\sigma_L$ and down-type VLFs lead to $B_s-\overline{B}_s$ mixing as shown in Fig.~\ref{fig:C9box}~(b). The mass difference~(procedure for the calculation is described in Sec.~\ref{sec:mass diff} and the magic numbers are given in Appendix.~\ref{app:Bqmix}) is given by
\begin{equation}
    \Delta M_s =\frac{0.86}{32\pi^2}\frac{|y_b|^2|y_s|^2}{4m_{\sigma_L}^2}\frac{8}{3}f^2_{B_s}M_{B_s}B_1(\mu_b)\mathcal{F}(x_i,x_j),
\end{equation}
with $x_i=\dfrac{m_{F_i}}{m_{\sigma_L}}$ and $x_j=\dfrac{m_{F_j}}{m_{\sigma_L}}$. The constraint arising from $\Delta M_s$ is extremely severe to allow large values of WCs required to explain the previously observed $R_{K^{(*)}}$. In light of the recent experimental result of $R_{K^{(*)}}$, which is consistent with the SM predictions, our analysis shows that this model is consistent with the flavor constraints arising from $b\to s\mu^+\mu^-$ experiments.

\section{Anomalous Magnetic Moment of Muon\label{sec:AMM}}
\begin{figure}
    \centering
    \includegraphics[scale=0.5]{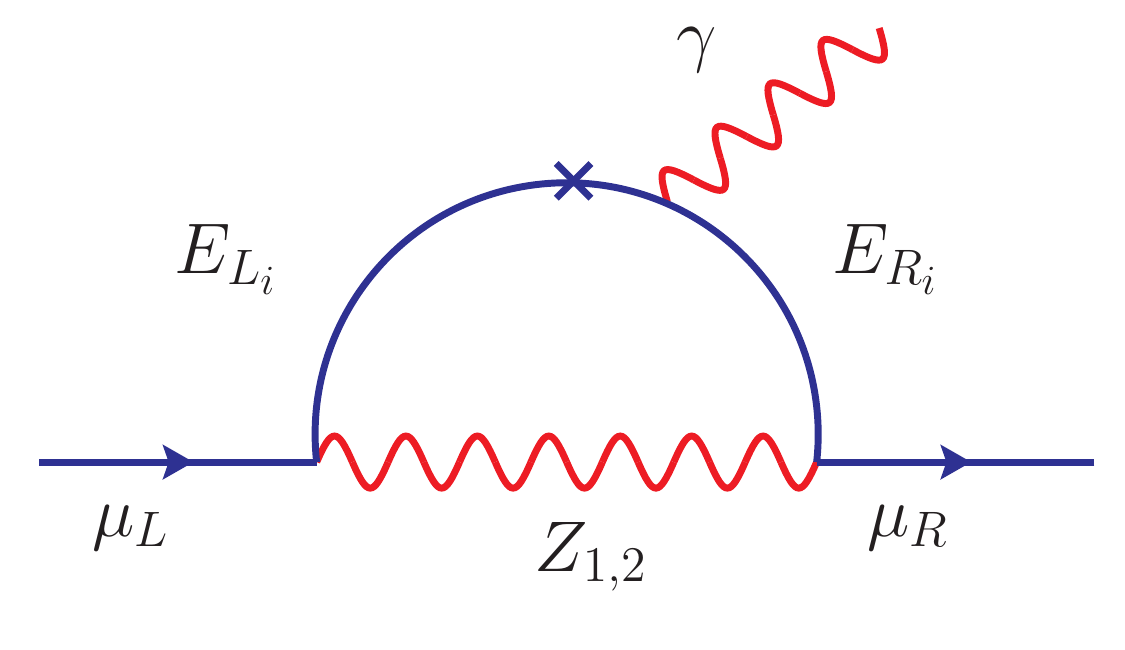}
    \includegraphics[scale=0.5]{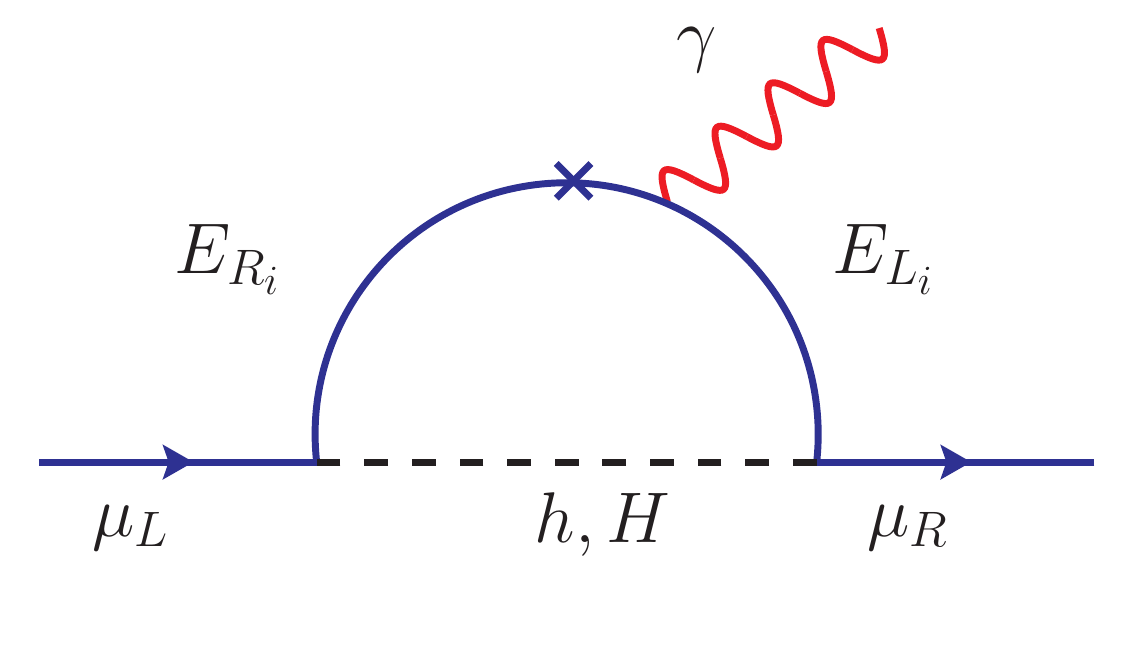}\\
   ~~~(a)~~~~~~~~~~~~~~~~~~~~~~~~~~~~~~~~~~~~~~~~~~~~(b)
    \caption{Diagrams leading to correction to AMM.}
    \label{fig:AMM}
\end{figure}
The anomalous magnetic moment of muon $a_\mu= (g_\mu-2)/2$ is predicted to be $a_\mu^{\textnormal{SM}}=116591810(43)\times 10^{-11}$~\cite{Davier:2019can, Aoyama:2020ynm,Davier:2010nc,Gerardin:2020gpp}. The measurements of $a_\mu$ at Fermilab National Accelerator Laboratory~(FNAL), $a_\mu(\textnormal{FNAL})=116592040(54)\times  10^{-11}$ ~\cite{Muong-2:2021ojo}, which agrees with the previous Brookhaven National Laboratory~(BNL) E821 measurement~\cite{Muong-2:2006rrc,Muong-2:2001kxu} is at odds with the SM prediction. The difference, $\Delta a_\mu = a_\mu ^{\text{exp}}- a_\mu ^{\text{th}} \simeq (251\pm59)\times 10^{-11}$, is a $4.2\,\sigma$ discrepancy. (There is an ambiguity that emerged recently with the results from the BMW collaboration~\cite{Borsanyi:2020mff} which agrees with the experimental measurement within $1.6~\sigma$.) The discrepancy is another major hint towards BSM physics. The neutral scalars $h$ and $H$, and the neutral gauge bosons $Z_1$ and $Z_2$ in the LRSM with universal seesaw can mediate significant chirally enhanced one-loop corrections to $a_\mu$, as shown in Fig.~\ref{fig:AMM}, potentially resolving the anomaly in AMM of muon. 

The corrections to AMM arise from the following Lagrangian:
    \begin{align}
        \mathcal{L}&\supset g_L\cos\theta_W\overline{\mu}\gamma^\mu Z_{1_\mu}\left[\left\{\frac{\sin\xi}{2\sqrt{\cos2\theta_W}}\frac{g_Y^2}{g_L^2}-\cos\xi\left(\frac{1}{2}+\frac{1}{2}\frac{g_Y^2}{g_L^2}\right)\right\}V_L\rho_L^\dagger\, \text{P}_\text{L}
        +\frac{\kappa_R\sin\xi}{2\kappa_L\sqrt{\cos2\theta_W}}V_L\rho_L^\dagger  \,\text{P}_\text{R}\right]F\nonumber\\
        &-g_L\cos\theta_W\overline{\mu}\gamma^\mu Z_{2_\mu}\left[\left\{\frac{\cos\xi}{2\sqrt{\cos2\theta_W}}\frac{g_Y^2}{g_L^2}+\sin\xi\left(\frac{1}{2}+\frac{1}{2}\frac{g_Y^2}{g_L^2}\right)\right\}V_L\rho_L^\dagger\text{P}_\text{L}-\frac{\kappa_R\cos\xi}{2\kappa_L\sqrt{\cos2\theta_W}}V_L\rho_L^\dagger\text{P}_\text{R}\right]F\nonumber\\
        &+\overline{\mu} \,h \left[\frac{\cos\zeta}{\sqrt{2}}\left(\left\{V_{L}\mathcal{Y}-\frac{1}{2}(V_{L}\mathcal{Y}\,\rho_{L}\rho_{L}^\dagger\frac{\kappa_R^2}{\kappa_L^2}+V_{L}\rho_{L}^\dagger\rho_{L}\mathcal{Y})\right\}\text{P}_\text{R}-\left\{V_L\rho^\dagger_{L}\mathcal{Y}^\dagger\rho_{L}^\dagger\frac{\kappa_R}{\kappa_L}\right\}\text{P}_\text{L}\right)\right.\nonumber\\
        &\left.-\frac{\sin\zeta}{\sqrt{2}}\left(\left\{V_{L}\mathcal{Y}-\frac{1}{2}(V_{L}\mathcal{Y}\,\rho_{L}\rho_{L}^\dagger+V_{L}\frac{\kappa_R^2}{\kappa_L^2}\rho_{L}^\dagger\rho_{L}\mathcal{Y})\right\}\text{P}_\text{L}-\left\{V_L\rho^\dagger_{L}\mathcal{Y}^\dagger\rho_{L}^\dagger\frac{\kappa_R}{\kappa_L}\right\}\text{P}_\text{R}\right)\right]F\nonumber\\
        &+\overline{\mu} \,H \left[\frac{\sin\zeta}{\sqrt{2}}\left(\left\{V_{L}\mathcal{Y}-\frac{1}{2}(V_{L}\mathcal{Y}\,\rho_{L}\rho_{L}^\dagger\frac{\kappa_R^2}{\kappa_L^2}+V_{L}\rho_{L}^\dagger\rho_{L}\mathcal{Y})\right\}\text{P}_\text{R}-\left\{V_L\rho^\dagger_{L}\mathcal{Y}^\dagger\rho_{L}^\dagger\frac{\kappa_R}{\kappa_L}\right\}\text{P}_\text{L}\right)\right. \nonumber\\
        &\left.+\frac{\cos\zeta}{\sqrt{2}}\left(\left\{V_{L}\mathcal{Y}-\frac{1}{2}(V_{L}\mathcal{Y}\,\rho_{L}\rho_{L}^\dagger+V_{L}\frac{\kappa_R^2}{\kappa_L^2}\rho_{L}^\dagger\rho_{L}\mathcal{Y})\right\}\text{P}_\text{L}-\left\{V_L\rho^\dagger_{L}\mathcal{Y}^\dagger\rho_{L}^\dagger\frac{\kappa_R}{\kappa_L}\right\}\text{P}_\text{R}\right)\right]F,
    \end{align}
where, $\text{P}_\text{\{L,R\}}$ are left- and right-handed projection operators, and $F$ represents the heavy VL-leptons. Compared to the general form of the interaction Lagrangian
\begin{equation}
\mathcal{L}\supset\sum_{F,X}\overline{\mu}\left[C_V\gamma^\mu+C_A\gamma^\mu\gamma^5\right]F\,X_\mu+\sum_{F,H}\overline{\mu}\left[C_S+C_P\gamma^5\right]F\,H\, ,
\end{equation}
The corrections to AMM~\cite{Leveille:1977rc}, under the assumption $\dfrac{m_\mu}{m_h} \ll 0$, arising from the VL-lepton mass enhancements are: 
\begin{equation}
    \begin{aligned}
        \left[a_\mu\right]_X=&\frac{m_\mu m_F}{4\pi^2}(C_A^2-C_V^2)\mathcal{F}(m_F,m_X)\\
         \left[a_\mu\right]_H=&\frac{m_\mu m_F}{8\pi^2}(C_S^2-C_P^2)\mathcal{G}(m_F,m_H)
    \end{aligned}
\end{equation}
with
\begin{equation}
    \begin{aligned}
        \mathcal{F}(m_F,m_X)=&\frac{\left[m_F^6-4m_X^6+3m_X^4m_F^2+6m_X^4m_F^2\ln{\dfrac{m_X^2}{m_F^2}}\right]}{4m_X^2(m_X^2-m_F^2)^3}\\
    \mathcal{G}(m_F,m_X)=&\frac{\left[m_F^4+3m_H^4-4m_H^2m_F^2+2m_H^4\ln{\dfrac{m_F^2}{m_H^2}}\right]}{2(m_F^2-m_H^2)^3}
    \end{aligned}
\end{equation}
A single-family mixing between muon and the corresponding VLF is not enough to resolve the anomaly in AMM due to the constraint from muon mass. To avoid this, we consider a case where muon mixes with two VL-leptons in a basis where the muon mass is negligible: 
\begin{equation}
\left(\overline{\mu}_L~\overline{E}_{e_L}~\overline{E}_{\mu_L}\right)\begin{pmatrix}
        0&&0&&y_\mu\kappa_L\\
         0&&0&&M_{1}\\
        y_\mu\kappa_R&&M_{1}&&M_{2}
    \end{pmatrix}\begin{pmatrix}
         \mu_R\\ E_{e_R}\\ E_{\mu_R}
    \end{pmatrix},
\end{equation}
such that $y_\mu\kappa_L \ll \{M_{1},\,M_{2}\}$. The mixing matrix $\mathcal{M}_\mu$ is diagonalized by a bi-unitary transform of the form
\begin{equation}
    \text{Diag}(0,\,M_E,\,M_M)=U_L \mathcal{M}_\mu U_R^\dagger
\end{equation}
where,
\begin{equation}
    U_{X=\{L,\,R\}}=\begin{pmatrix}
    c_{X_{1}}&s_{X_{1}}&0\\
    -c_{X_{2}}s_{X_{1}}&c_{X_{1}}c_{X_{2}}&s_{X_{2}}\\
    -s_{X_{1}}s_{X_{2}}&-c_{X_{1}}s_{X_{2}}&c_{X_{2}}
    \end{pmatrix}.
\end{equation}

Here, $c\,(s)$ stand for $\cos\,(\sin)$. The mixing angles and the mass eigenstates are
\begin{equation}
    \begin{aligned}
       L_{1}=\arctan(\frac{-y_\mu\kappa_L }{M_{1}}),&&L_{2}=\frac{1}{2}\arctan\left(-\frac{2M_{2}\sqrt{M_{1}^2+y_\mu^2\kappa_L^2}}{M_{2}^2+(\kappa_R^2-\kappa_L^2)y_\mu^2}\right)\\
      R_{1}=\arctan(\frac{-y_\mu\kappa_R }{M_{1}}),&&R_{2}=\frac{1}{2}\arctan\left(-\frac{2M_{2}\sqrt{M_{1}^2+y_\mu^2\kappa_R^2}}{M_{2}^2-(\kappa_R^2-\kappa_L^2)y_\mu^2}\right)\\       
    \end{aligned}
\end{equation}
\begin{equation}
     M_{E,\,M}=\frac{1}{2}\left(2M_{1}^2+M_{2}^2+(\kappa_L^2+\kappa_R^2)y_\mu^2\mp\sqrt{4M_{2}^2(M_{1}^2+\kappa_L^2y_\mu^2)+\left(M_{2}^2+(\kappa_R^2-\kappa_L^2)y_\mu^2\right)^2}\right)
\end{equation}
The mixing between $Z_{L}-Z_R$ can be ignored such that $Z_{1\,(2)}\equiv Z_{L\,(R)}$, whereas $\sigma_L-\sigma_R$ mixing as of Eq.~\eqref{eq:higssmix} is required to obtain a chiral enhancement. The total correction to AMM is 
\begin{equation}
\begin{aligned}
    a_\mu=&-\frac{m_\mu}{4\pi^2}\Bigg[\frac{g_L^4\tan^2\theta_W}{4(g_L^2-g_Y^2)} c_{L_1}c_{R_1}s_{L_1}s_{R_1}\left\{ c_{L_2}c_{R_2}M_E\mathcal{F}(M_E,m_{Z_R})+s_{L_2}s_{R_2}M_M\mathcal{F}(M_M,m_{Z_R})\right\}\\
    &\qquad~~~~ \frac{y_\mu^2\sin2\zeta}{8} c_{L_1}c_{R_1}\left\{s_{L_2}s_{R_2}M_E\mathcal{G}(M_E,m_h)+c_{L_2}c_{R_2}M_E\mathcal{G}(M_M,m_h)\right.\\
    &\left.\qquad\qquad\qquad\qquad~~~~~~-s_{L_2}s_{R_2}M_E\mathcal{G}(M_E,m_H)-c_{L_2}c_{R_2}M_E\mathcal{G}(M_M,m_H)\right\}\Bigg]
\end{aligned}
\end{equation}
The main contribution to the AMM in this model arises from the scalar sector when there is a large mixing between the neutral scalars.  But this correction is not large enough to explain the $4.2\,\sigma$ discrepancy for scalar and vector-like lepton masses in the experimentally accessible~($\leq\mathcal{O}$~(TeV)) range, although it can be alleviated to some extend. 
\section{Conclusion \label{conclude}}

This paper presents a comprehensive description of the left-right symmetric model with a universal seesaw mechanism for generating fermion masses. The direct coupling of SM fermions with their VLF singlet partners leads to tree-level flavor-changing neutral current interactions and has room for flavor and flavor-universality violating processes, making this a compelling model to explore various flavor anomalies that have come up in recent years. Full tree-level Lagrangian in the physical basis has been explored, allowing us to obtain constraints on the theory parameters from neutral and charged current mediated processes. The parity symmetric version, motivated by the axionless solution to the strong $\mathcal{CP}$ problem, was explored in great detail. The model was applied to find a solution to the neutral current B-anomalies and the AMM of muon. The constraints on the contributions from the model were found to be severe, ruling out the possibility of resolving the two anomalies. The contribution to $R_{K^{(*)}}$ is restricted by the stringent constraint appearing from the mass difference in $B_s-\overline{B}_s$ mixing. Although the model allows VLF mass-enhanced corrections to the AMM of muon, the constraint from muon mass eliminates any substantial correction to survive. We also found that the discrepancies observed in various LFU-violating processes mediated by charged current interactions could also not be simultaneously explained by the parameter space of the model.

\section*{Acknowledgements}
RD would like to thank K.S. Babu for his valuable guidance that led to the completion of this work, and Ajay Kaladharan, Vishnu P.K., and Anil Thapa for useful discussions. This work is supported by the U.S. Department of Energy  under grant number DE-SC 0016013. Some computing for this project was performed at the High-Performance Computing Center at Oklahoma State University, supported in part through the National Science Foundation grant OAC-1531128.
\appendix
\section{Neutral Current Interaction\label{app:Lag}}
The coefficients of neutral current interaction of $Z_1$ with SM fermions as defined in Eq.~\eqref{eq:LZ1} are
\begin{equation}
\begin{aligned}
C_{L_1}&=\cos^2\theta_W\left[\cos\xi\left(T_{3L}-\frac{g_Y^2}{g_L^2}\frac{Y_{f_L}}{2}\right)-\frac{\sin\xi}{\sqrt{\cos 2\theta_W}}\left(-\frac{g_Y^2}{g_L^2}\frac{Y_{f_L}}{2}\right)\right], \\
\widetilde{C}_{L_1}&=-\cos^2\theta_W\left[\cos\xi\left(T_{3L}-\frac{g_Y^2}{g_L^2}\frac{Y_{f_L}-Y_F}{2}\right)+\frac{\sin\xi}{\sqrt{\cos 2\theta_W}}\left(\frac{g_Y^2}{g_L^2}\frac{Y_{f_L}-Y_F}{2}\right)\right]\times R_{ij},\\ 
C_{R_1}&=\cos^2\theta_W\left[\cos\xi\left(-\frac{g_Y^2}{g_L^2}\frac{Y_{f_R}}{2}\right)-\frac{\sin\xi}{\sqrt{\cos 2\theta_W}}\left(T_{3R}-\frac{g_Y^2}{g_L^2}\frac{Y_{f_R}}{2}\right)\right],\\
\widetilde{C}_{R_1}&=\cos^2\theta_W\frac{\sin\xi}{\sqrt{\cos 2\theta_W}}\left(T_{3R}\right) \times \frac{\kappa_R^2}{\kappa_L^2}R_{ij},
\end{aligned}
\end{equation}
whereas those of $Z_2$ with SM fermions as in Eq.~\eqref{eq:LZ2} are as follows:
\begin{equation}
\begin{aligned}
C_{L_2}&=\cos^2\theta_W\left[\frac{\cos\xi}{\sqrt{\cos 2\theta_W}}\left(-\frac{g_Y^2}{g_L^2}\frac{Y_{f_L}}{2}\right)+\sin\xi\left(T_{3L}-\frac{g_Y^2}{g_L^2}\frac{Y_{f_L}}{2}\right)\right], \\
\widetilde{C}_{L_2}&=\cos^2\theta_W\left[\frac{\cos\xi}{\sqrt{\cos 2\theta_W}}\left(\frac{g_Y^2}{g_L^2}\frac{Y_{f_L}-Y_F}{2}\right)-\sin\xi\left(T_{3L}-\frac{g_Y^2}{g_L^2}\frac{Y_{f_L}-Y_F}{2}\right)\right]\times R_{ij},\\ 
C_{R_2}&=\cos^2\theta_W\left[\frac{\cos\xi}{\sqrt{\cos 2\theta_W}}\left(T_{3R}-\frac{g_Y^2}{g_L^2}\frac{Y_{f_R}}{2}\right)+\sin\xi\left(-\frac{g_Y^2}{g_L^2}\frac{Y_{f_R}}{2}\right)\right],\\
\widetilde{C}_{R_2}&=-\cos^2\theta_W\frac{\cos\xi}{\sqrt{\cos 2\theta_W}}\left(T_{3R}\right)\times \frac{\kappa_R^2}{\kappa_L^2}R_{ij}.
\end{aligned}
\end{equation}

\section{Input Parameters for Meson Mixing\label{1}}
Here, we provide various input parameters useful in computing the meson mixing mass difference. The strong coupling strengths at high scales~\cite{Bethke:2009jm}, $\alpha_s(\Lambda)$, are given in Table~\ref{tab:alphas}.
\begin{table}[]
\footnotesize
	{\renewcommand{\arraystretch}{1.0}
		\begin{center}
			\begin{tabular}{|c|c|c|c|c|c|c|c|}
			\hline
				\textbf{Scale} & $M_{Z_1}$&  $M_{Z_2}$&  $M_{W_L}$&  $M_{W_R}$&  $M_{h}$&  $M_{H}$ &  $M_{t}$ \\
				\hline
				$\Lambda$ & 91.187 GeV & 5 TeV & 80.379 GeV & 4.219 TeV & 125.10 GeV & 10 TeV & 172.9 GeV \\
				\hline
			$\boldsymbol{\alpha_s}$ & 0.1183 & 0.0824 & 0.1206 & 0.0837 & 0.1129 & 0.0774 & 0.1079 \\
				\hline
				\end{tabular}
			\caption{Values of the strong coupling constant at different energy scales.}
   \label{tab:alphas}
		\end{center}}
\end{table}

\subsection{\texorpdfstring{$K-\overline{K}$ Mixing}{kaonmixingvalues}}
\begin{table}[]
\footnotesize
	{\renewcommand{\arraystretch}{1.2}%
		\begin{center}
			\begin{tabular}{|c|c|c|c|c|c|}
				\hline
				\textbf{Constants} & $\mu$ & $f_K$ & $M_K$ & $m_s$ & $m_d$\\
				\hline
				\textbf{Values}  & $2 \text{ GeV}$ 	& $159.8 \text{ MeV}$ & $467.611\text{ MeV}$  & $93 \text{ MeV}$  & $4.67\text{ MeV}$\\
                \hline
                \hline
                \textbf{Constants} &  $R_1$ & $R_2$ & $R_3$ & $R_4$ & $R_5$\\
				\hline
				\textbf{Values}  &  1 & -12.9 & 3.98 & 20.8 & 5.2\\
                \hline
				\end{tabular}
			\caption{Values of input parameters used in $K-\overline{K}$ mixing.}\label{tab:KKconst}
		\end{center}}
\end{table}
The meson mixing mass difference is given by $
    \Delta M_{K} = 2\text{Re}\bra{K}\mathcal{H}\ket{\overline{K}}$. The matrix element at low energy scale is obtained as: 
\begin{equation}
\bra{\overline{K}}\mathcal{H}_\text{eff}\ket{K}_i=\sum_{j=1}^5\sum_{r=1}^5(b_j^{(r,i)}+\eta c_j^{(r,i)})\eta^{a_j}C_i(\Lambda))R_r\bra{\overline{K}}\mathcal{Q}_1\ket{K},\label{Kmix}
\end{equation}

with, $\langle \mathcal{O}_1\rangle=\frac{1}{3}M_{K}f^2_{K}B_1(\mu)$. The non-vanishing entries of the magic numbers are~\cite{Ciuchini:1998ix}:
    \begin{equation*}
        a_i=(0.29,-0.69,0.79,-1.1,0.14)
    \end{equation*}

\begin{minipage}{0.4\textwidth}
\begin{equation*}
\begin{aligned}
b_i^{(11)}&=(0.82,0,0,0,0),\\
b_i^{(22)}&=(0,2.4,0.011,0,0),\\
b_i^{(23)}&=(0,-0.63,0.17,0,0),\\
b_i^{(32)}&=(0,-0.019,0.028,0,0),\\
b_i^{(33)}&=(0,0.0049,0.43,0,0),\\
b_i^{(44)}&=(0,0,0,4.4,0),\\
b_i^{(45)}&=(0,0,0,1.5,-0.17),\\
b_i^{(54)}&=(0,0,0,0.18,0),\\
b_i^{(55)}&=(0,0,0,0.061,0.82),
\end{aligned}
\end{equation*}
\end{minipage}
\begin{minipage}{0.5\textwidth}
\begin{equation}
\begin{aligned}
c_i^{(11)}&=(-0.016,0,0,0,0),\\
c_i^{(22)}&=(0,-0.23,-0.002,0,0),\\
c_i^{(23)}&=(0,-0.018,0.0049,0,0),\\
c_i^{(32)}&=(0,0.0028,-0.0093,0,0),\\
c_i^{(33)}&=(0,0.00021,0.023,0,0),\\
c_i^{(44)}&=(0,0,0,-0.68,0.0055),\\
c_i^{(45)}&=(0,0,0,-0.35,-0.0062),\\
c_i^{(54)}&=(0,0,0,-0.026,-0.016),\\
c_i^{(55)}&=(0,0,0,-0.013,0.018).\\
\end{aligned}
\end{equation}
\end{minipage}
The coefficients and input parameters relevant to this calculation are given in Table~\ref{tab:KKconst}.

\subsection{\texorpdfstring{$D-\overline{D}$ Mixing}{Dmesonmixingvalues} }

\begin{table}[]
\footnotesize
	{\renewcommand{\arraystretch}{1.0}%
		\begin{center}
			\begin{tabular}{|c|c|c|c|c|c|}
				\hline
				\textbf{Constants} & $\mu$ & $f_D$ & $M_D$ & $m_u$ & $m_c$ \\
				\hline
				\textbf{Values}  & $2.8 \text{ GeV}$ 	& $201 \text{ MeV}$ & $1.864\text{ GeV}$  & $2.01 \text{ MeV}$  & $1.01\text{ GeV}$ \\
                \hline
				\hline
				\textbf{Constants} &  $B_1$ & $B_2$ & $B_3$ & $B_4$ & $B_5$\\
				\hline
				\textbf{Values}  &  0.865 & 0.82 & 1.07 & 1.08 & 1.455\\
                \hline
				\end{tabular}
			\caption{Values of input parameters used in $D-\overline{D}$ mixing.}~\label{tab:DDconst}
		\end{center}}
\end{table}
The mass difference is given by $\Delta M_{D} = 2|\bra{D}\mathcal{H}\ket{\overline{D}}|$ with the renormalised operators being:

\begin{equation}
	\begin{aligned}
		 \langle \mathcal{O}_1\rangle&=\frac{1}{3}M_{D}f^2_{D}B_1(\mu),\\
		 \langle \mathcal{O}_2\rangle&=-\frac{5}{24}\left(\frac{M_{D}}{m_u(\mu)+m_c(\mu)}\right)^2M_Df^2_{D}B_2(\mu),\\
		 \langle \mathcal{O}_3\rangle&=\frac{1}{24}\left(\frac{M_{D}}{m_u(\mu)+m_c(\mu)}\right)^2M_Df^2_{D}B_3(\mu),\\
		\langle \mathcal{O}_4\rangle&=\frac{1}{4}\left(\frac{M_{D}}{m_u(\mu)+m_c(\mu)}\right)^2M_Df^2_{D}B_4(\mu),\\
		 \langle \mathcal{O}_5\rangle&=\frac{1}{12}\left(\frac{M_{D}}{m_u(\mu)+m_c(\mu)}\right)^2M_Df^2_{D}B_5(\mu).
		 \end{aligned}\label{Bpar}
\end{equation}
The coefficients and input parameter values are listed in Table~\ref{tab:DDconst} and the non-vanishing entries of the magic numbers are~\cite{Bona:2007vz}:
    \begin{equation*}
        a_i=(0.286,-0.692,0.787,-1.143,0.143)
    \end{equation*}

\begin{minipage}{0.4\textwidth}
\begin{equation*}
\begin{aligned}
b_i^{(11)}&=(0.837,0,0,0,0),\\
b_i^{(22)}&=(0,2.163,0.012,0,0),\\
b_i^{(23)}&=(0,-0.567,0.176,0,0),\\
b_i^{(32)}&=(0,-0.032,0.031,0,0),\\
b_i^{(33)}&=(0,0.008,0.474,0,0),\\
b_i^{(44)}&=(0,0,0,3.63,0),\\
b_i^{(45)}&=(0,0,0,1.21,-0.19),\\
b_i^{(54)}&=(0,0,0,0.14,0),\\
b_i^{(55)}&=(0,0,0,0.045,0.839),
\end{aligned}
\end{equation*}
\end{minipage}
\begin{minipage}{0.5\textwidth}
\begin{equation}
\begin{aligned}
c_i^{(11)}&=(-0.016,0,0,0,0),\\
c_i^{(22)}&=(0,-0.20,-0.002,0,0),\\
c_i^{(23)}&=(0,-0.016,0.006,0,0),\\
c_i^{(32)}&=(0,-0.004,-0.010,0,0),\\
c_i^{(33)}&=(0,0,0.025,0,0),\\
c_i^{(44)}&=(0,0,0,-0.56,0.006),\\
c_i^{(45)}&=(0,0,0,-0.29,-0.006),\\
c_i^{(54)}&=(0,0,0,-0.019,-0.016),\\
c_i^{(55)}&=(0,0,0,-0.009,0.018).
\end{aligned}
\end{equation}
\end{minipage}

\subsection{\texorpdfstring{$B_q-\overline{B}_q$ Mixing}{Bqmesonmixingvalues}\label{app:Bqmix}}

The operators used in evaluating $\Delta M_{B_q} = 2|\bra{B_q}\mathcal{H}\ket{\overline{B}_q}|$ (in Sec.~\ref{sec:mass diff}) are 
\begin{equation}
	\begin{aligned}
		 \langle \mathcal{O}_1\rangle&=f^2_{B_q}M_{B_q}\frac{8}{3}B_1(\mu_b),\\
		\langle \mathcal{O}_2  \rangle&=f_{B_q}^2M_{B_q}\frac{-5M_{B_q}^2}{3(\overline{m}_b(\mu_b)+\overline{m}_q(\mu_b))^2}B_2(\mu_b),\\
		\langle \mathcal{O}_3 \rangle&=f_{B_q}^2M_{B_q}\frac{M_{B_q}^2}{3(\overline{m}_b(\mu_b)+\overline{m}_q(\mu_b))^2}B_3(\mu_b),\\
		\langle \mathcal{O}_4 \rangle&=f_{B_q}^2M_{B_q}\left[\frac{2M_{B_q}^2}{(\overline{m}_b(\mu_b)+\overline{m}_q(\mu_b))^2}+\frac{1}{3}\right]B_4(\mu_b),\\
		\langle \mathcal{O}_5 \rangle&=f_{B_q}^2M_{B_q}\left[\frac{2M_{B_q}^2}{3(\overline{m}_b(\mu_b)+\overline{m}_q(\mu_b))^2}+1\right]B_5(\mu_b).
	\end{aligned}
\end{equation}

The central values of the combination of $f_B^2B_i(\mu_b=4.18\text{ GeV})$ used are as follows \cite{DiLuzio:2019jyq}:

\begin{minipage}{0.45\textwidth}
\begin{equation*}
	\begin{aligned}
		f_{B_s}^2B_1^s(\mu_b)&=&0.0452 \text{ GeV}^2,\\
		f_{B_s}^2B_2^d(\mu_b)&=&0.0441 \text{ GeV}^2,\\
		f_{B_s}^2B_3^d(\mu_b)&=&0.0454 \text{ GeV}^2,\\
		f_{B_s}^2B_4^s(\mu_b)&=&0.0544 \text{ GeV}^2,\\
		f_{B_s}^2B_5^s(\mu_b)&=&0.0507 \text{ GeV}^2,
	\end{aligned}
\end{equation*}
\end{minipage}
\begin{minipage}{0.45\textwidth}
\begin{equation}
	\begin{aligned}
		f_{B_d}^2B_1^d(\mu_b)&=&0.0305 \text{ GeV}^2,\\
		f_{B_d}^2B_2^d(\mu_b)&=&0.0288 \text{ GeV}^2,\\
		f_{B_d}^2B_3^d(\mu_b)&=&0.0281 \text{ GeV}^2,\\
		f_{B_d}^2B_4^d(\mu_b)&=&0.0387 \text{ GeV}^2,\\
		f_{B_d}^2B_5^d(\mu_b)&=&0.0361 \text{ GeV}^2.
	\end{aligned}
\end{equation}
\end{minipage}
\vspace{2mm}
The masses of quarks at $\mu_b$ are $m_s=77.9 \text{ MeV}$, $m_d=3.94\text{ MeV}$ and $m_b=4.18\text{ GeV}$.
\vspace{4mm}

In Sec.~\ref{sec:constrainthiggs}, $B_d-\overline{B_d}$ mass difference was calculated using magic numbers. The renormalized operators in terms of $B_i(\mu)$ parameters are same as in Eq.~\eqref{Bpar} with \{$M_D \to M_{B_d}$, $f_D \to f_{B_d}$, $m_u\to m_b$, $ m_c \to m_d$\}. Other relevant input parameters are given in Table~\ref{tab:constantsBd} and the non-vanishing entries of the magic numbers are \cite{Becirevic:2001jj}:
    \begin{equation}
        a_i=(0.286,-0.692,0.787,-1.143,0.143)
    \end{equation}

\begin{minipage}{0.4\textwidth}
\begin{equation*}
\begin{aligned}
b_i^{(11)}&=(0.865,0,0,0,0),\\
b_i^{(22)}&=(0,1.879,0.012,0,0),\\
b_i^{(23)}&=(0,-0.493,0.18,0,0),\\
b_i^{(32)}&=(0,-0.044,0.035,0,0),\\
b_i^{(33)}&=(0,0.011,0.54,0,0),\\
b_i^{(44)}&=(0,0,0,2.87,0),\\
b_i^{(45)}&=(0,0,0,0.961,-0.22),\\
b_i^{(54)}&=(0,0,0,0.09,0),\\
b_i^{(55)}&=(0,0,0,0.029,0.863),
\end{aligned}
\end{equation*}
\end{minipage}
\begin{minipage}{0.5\textwidth}
\begin{equation}
\begin{aligned}
c_i^{(11)}&=(-0.017,0,0,0,0),\\
c_i^{(22)}&=(0,-0.18,-0.003,0,0),\\
c_i^{(23)}&=(0,-0.014,0.008,0,0),\\
c_i^{(32)}&=(0,-0.005,-0.012,0,0),\\
c_i^{(33)}&=(0,0,0.028,0,0),\\
c_i^{(44)}&=(0,0,0,-0.48,0.005),\\
c_i^{(45)}&=(0,0,0,-0.25,-0.006),\\
c_i^{(54)}&=(0,0,0,-0.013,-0.016),\\
c_i^{(55)}&=(0,0,0,-0.007,0.019),
\end{aligned}
\end{equation}
\end{minipage}

\begin{table}[]
\footnotesize
	{\renewcommand{\arraystretch}{1.0}
		\begin{center}
			\begin{tabular}{|c|c|c|c|c|c|}
				\hline
				\textbf{Constants} & $\mu$ & $f_{B_d}$ & $M_{B_d}$ & $m_b$ & $m_d$\\
				\hline
				\textbf{Values}  & $4.6 \text{ GeV}$ 	& $200 \text{ MeV}$ & $5.279\text{ GeV}$  & $4.61 \text{ GeV}$  & $5.4\text{ MeV}$\\
                \hline
                \hline
				\textbf{Constants} &  $B_1$ & $B_2$ & $B_3$ & $B_4$ & $B_5$\\
				\hline
				\textbf{Values}  & 0.87 & 0.82 & 1.02 & 1.16 & 1.91\\
                \hline
				\end{tabular}
			\caption{Values of input parameters used in $B_d-\overline{B}_d$ mixing.}\label{tab:constantsBd}
		\end{center}}
\end{table}

\section{Form Factors for Meson Decay}\label{app1}
The form factor that appears in Eq.~\eqref{eq:semilepmesondecay} describing the decay width of meson to lighter meson and charged leptons takes the form~\cite{Melikhov:2000yu} 
\begin{equation}
	f_+(q^2)=f_+(0)/(1-\frac{q^2}{M_V^2}).
\end{equation} 
The values of the form factors and the vector meson masses are given in Table~\ref{Input Parameters}.

\begin{table}[]
	{\renewcommand{\arraystretch}{1.5}
		\begin{center}
			\begin{tabular}{|c|c|c|c|}
				\hline
				\textbf{Transition} & $f_+(0)$ &$M_V$(GeV)\\
				\hline
				$K \to\pi$ & 0.9709~\cite{Carrasco:2016kpy} & 0.892\\
				\hline
				$B \to\pi$ & 0.29 & 5.32\\
				\hline
				$B \to K$ & 0.36 & 5.42\\
				\hline
				$D \to \pi$ & 0.69 & 2.01\\
				\hline
				$D_s \to K$ & 0.72 & 2.01\\
				\hline
			\end{tabular}
			\caption{Parameter values used in calculation semileptonic decays of heavy mesons.}
			\label{Input Parameters}
	\end{center}}
\end{table}

\section{Kaon Mixing Box Diagram Expressions\label{app3}}
 The $\lambda$ couplings contributing to kaon mixing box diagrams mediated by $W_{L,\, R}$ in Sec.~\ref{sec:boxWLR} take the form 
\begin{equation}
    \begin{aligned}
    \lambda^{LL}_{i=(u,c,t)}&=(\mathcal{V}^*_{i,1}-\delta\mathcal{V}^\dagger_{i,1})(\mathcal{V}_{i,2}-\delta\mathcal{V}_{i,2}),\\
    \lambda^{LR}_{i=(u,c,t)}&=(\mathcal{V}^*_{i,1}-\delta\mathcal{V}^\dagger_{i,1})(\mathcal{V}_{i,2}-\frac{\kappa_R^2}{\kappa_L^2}\delta\mathcal{V}_{i,2}),\\
     \lambda^{LR}_{i=(u,c,t)}&=(\mathcal{V}^*_{i,1}-\frac{\kappa_R^2}{\kappa_L^2}\delta\mathcal{V}^\dagger_{i,1})(\mathcal{V}_{i,2}-\delta\mathcal{V}_{i,2}),\\
      \lambda^{LL}_{i=(U,C,T)}&=(V_{L_d}\rho_{L_U}^\dagger)_{1,i}(\rho_{L_U}V_{L_d}^\dagger)_{i,2},\\
    \lambda^{LR}_{i=(U,C,T)}&=\lambda^{RL}_{i=(U,C,T)}=\frac{\kappa_L}{\kappa_R}\lambda^{LL}_{i=(U,C,T)}.\\
    \end{aligned}
\end{equation}
From Eq.~\eqref{eq:W_X current}, the script $\mathcal{V}=V_{X_u}V_{X_d}^\dagger$ is interpreted as the $V_{CKM}$ matrix elements, and $\delta\mathcal{V}=\frac{1}{2}(V_{X_u}\rho_{X_D}^\dagger \rho_{X_D} V_{X_d}^\dagger +V_{X_u}\rho_{X_U}^\dagger \rho_{X_U} V_{X_d}^\dagger)$.

\bibliographystyle{style.bst}
\bibliography{refmain}
\end{sloppypar}
\end{document}